\documentclass[a4paper,UKenglish,cleveref, autoref, thm-restate, final]{lipics-v2021}

\nolinenumbers

\usepackage[svgnames,dvipsnames]{xcolor}

\usepackage{xspace}
\usepackage{tikz}
\usetikzlibrary{automata,positioning,shapes,arrows,decorations,decorations.pathmorphing,decorations.markings,bending}
\usetikzlibrary{shapes.misc}

\tikzset{every loop/.style={looseness=7}, >=latex}
\tikzset{every picture/.style={>=latex}}
\tikzstyle{PlayerMin}=[draw,circle,minimum size=6mm,inner sep=1.5pt]
\tikzstyle{PlayerMax}=[draw,rectangle,minimum size=6mm,inner sep=1.5pt]
\tikzstyle{PlayerMinsmall}=[draw,circle,minimum size=5mm,inner sep=1.5pt]
\tikzstyle{PlayerMaxsmall}=[draw,rectangle,minimum size=3.5mm,inner sep=1.5pt]
\tikzstyle{Player}=[draw,diamond,minimum size=6mm,inner sep=1.5pt]
\tikzstyle{target}=[circle, minimum size=1mm,inner sep=-2pt]
\tikzstyle{targetBouded}=[draw, circle, accepting, minimum size=7mm,inner sep=1.5pt]
\tikzstyle{PlayerMinmin}=[draw,circle, minimum size=1.5mm,inner sep=0pt]
\tikzstyle{PlayerMaxmin}=[draw,rectangle,minimum size=1.5mm,inner sep=0pt]
\tikzstyle{proba}=[draw,circle,minimum height=0pt,inner sep=2pt,minimum width=0pt,fill=black]
\tikzstyle{leaf}=[draw,diamond,minimum size=7mm,inner sep=1.5pt]

\tikzstyle{strat} =[minimum width=0.1cm,line width=0.01mm,draw=none]
\tikzstyle{vecArrow} = [decoration={markings,mark=at position
	1 with {\arrow[scale=.5,>=latex]{>}}}%,postaction={decorate}
]

\usepackage[obeyFinal]{todonotes}

\usepackage{bbm}
\usepackage{amsfonts,amssymb,amsmath}
\usepackage{amsthm}
\usepackage{wasysym}
\usepackage{stmaryrd}
\usepackage{mathtools}

\DeclareMathOperator*{\arginf}{\arg\!\inf}
\DeclareMathOperator*{\argsup}{\arg\!\sup}

\newcommand{\arginfepsilon}[1][]{\ifthenelse{\equal{#1}{}}{{\arginf}^{\varepsilon}}{{\arginf}^{#1}}}
\newcommand{\argsupepsilon}[1][]{\ifthenelse{\equal{#1}{}}{{\argsup}^{\varepsilon}}{{\argsup}^{#1}}}
\newcommand{\tuple}[1]{\ensuremath{\langle #1 \rangle}\xspace}
\newcommand{\powerset}[1]{\ensuremath{2^{#1}}\xspace}

\newcommand{\last}{\textsf{last}}

\newcommand{\proj}{\ensuremath{\mathsf{proj}}\xspace}

\newcommand{\inj}{\ensuremath{\mathsf{inj}}\xspace}

\newcommand{\R}{\ensuremath{\mathbb{R}}\xspace}
\newcommand{\Rplus}{\ensuremath{\R_{\geq 0}}\xspace}
\newcommand{\Rbar}{\ensuremath{\R_\infty}\xspace}
\newcommand{\Z}{\ensuremath{\mathbb{Z}}\xspace}
\newcommand{\N}{\ensuremath{\mathbb{N}}\xspace}
\newcommand{\Q}{\ensuremath{\mathbb{Q}}\xspace}

\newcommand{\MinPl}{\ensuremath{\mathsf{Min}}\xspace}
\newcommand{\MaxPl}{\ensuremath{\mathsf{Max}}\xspace}

\newcommand{\WTG}{WTG\xspace}

\newcommand{\game}{\ensuremath{\mathcal G}\xspace}
\newcommand{\gameEx}{\ensuremath{\tuple{\LocsMin, \allowbreak\LocsMax, \allowbreak\LocsT, 
			\allowbreak\Cl, \allowbreak\Trans,\weight}}\xspace}

\newcommand{\clockx}{x}

\newcommand{\delay}{\ensuremath{t}\xspace}
\newcommand{\Cl}{\ensuremath{\mathcal{X}}\xspace}
\newcommand{\clockbound}{\ensuremath{\mathsf{M}}\xspace}
\newcommand{\val}{\ensuremath{\nu}\xspace}
\newcommand{\guard}{\ensuremath{g}\xspace}
\newcommand{\Guards}[1][]{\ifthenelse{\equal{#1}{}}
	{\ensuremath{\mathsf{Guards}(\Cl)}\xspace}
	{\ensuremath{\mathsf{Guards}_{#1}(\Cl)\xspace}}}
\newcommand{\reset}{\ensuremath{Y}\xspace}

\newcommand{\loc}{\ensuremath{\ell}\xspace}
\newcommand{\Locs}{\ensuremath{L}\xspace}
\newcommand{\LocsUrg}{\ensuremath{\Locs_{u}}\xspace}
\newcommand{\LocsMin}{\ensuremath{\Locs_{\MinPl}}\xspace}
\newcommand{\LocsMax}{\ensuremath{\Locs_{\MaxPl}}\xspace}

\newcommand{\LocsT}{\ensuremath{\Locs_T}\xspace}
\newcommand{\Trans}{\ensuremath{\Delta}\xspace}
\newcommand{\trans}{\ensuremath{\delta}\xspace}

\newcommand{\RTgame}{\ensuremath{\mathcal{A}}\xspace}

\newcommand{\rRTgame}[1][]{\ifthenelse{\equal{#1}{}}
	{\ensuremath{\mathcal{R}(\RTgame)}\xspace}
	{\ensuremath{\mathcal{R}^{#1}(\RTgame)}\xspace}}

\newcommand{\QLocs}{Q}

\newcommand{\weight}{\ensuremath{\mathsf{wt}}\xspace}
\newcommand{\weightP}{\weight}
\newcommand{\weightT}{\ensuremath{\weight_T}\xspace}
\newcommand{\weightC}{\ensuremath{\weight_\Sigma}\xspace}

\newcommand{\maxWeightLoc}{W_{\mathsf{loc}}}
\newcommand{\maxWeightTrans}{W_{\mathsf{tr}}}
\newcommand{\maxWeightEdge}{W_{\mathsf{e}}}

\newcommand{\robminstrategy}{\ensuremath{\chi}\xspace}
\newcommand{\robmaxstrategy}{\ensuremath{\zeta}\xspace}

\newcommand{\Strat}{\ensuremath{\mathsf{Strat}}\xspace}

\newcommand{\StratMin}[1][]{\ifthenelse{\equal{#1}{}}
{\ensuremath{\Strat_{\MinPl}}}{\ensuremath{\Strat_{\MinPl}(#1)}}\xspace}
\newcommand{\StratMax}[1][]{\ifthenelse{\equal{#1}{}}
{\ensuremath{\Strat_{\MaxPl}}}{\ensuremath{\Strat_{\MaxPl}(#1)}}\xspace}
\newcommand{\dStrat}[1][]{\ifthenelse{\equal{#1}{}}
{\ensuremath{\mathsf{dStrat}}\xspace}{\ensuremath{\mathsf{dStrat}(#1)}\xspace}}
\newcommand{\dStratMin}[1][]{\ifthenelse{\equal{#1}{}}
{\ensuremath{\dStrat_{\MinPl}}\xspace}{\ensuremath{\dStrat_{\MinPl}(#1)}\xspace}}
\newcommand{\dStratMax}[1][]{\ifthenelse{\equal{#1}{}}
{\ensuremath{\dStrat_{\MaxPl}}\xspace}{\ensuremath{\dStrat_{\MaxPl}(#1)}\xspace}}
\newcommand{\mStrat}{\used to compute the value of a divergent \WTGensuremath{\mathsf{mStrat}}\xspace}
\newcommand{\mStratMin}[1][]{\ifthenelse{\equal{#1}{}}
{\ensuremath{\mStrat_{\MinPl}}}{\ensuremath{\mStrat_{\MinPl}(#1)}}\xspace}
\newcommand{\mStratMax}[1][]{\ifthenelse{\equal{#1}{}}
{\ensuremath{\mStrat_{\MaxPl}}}{\ensuremath{\mStrat_{\MaxPl}(#1)}}\xspace}
\newcommand{\rStrat}[1]{\ensuremath{\mathsf{Strat}^{#1}}\xspace}
\newcommand{\rStratMax}[2][]{\ifthenelse{\equal{#1}{}}
{\ensuremath{\rStrat{#2}_{\MaxPl}}}{\ensuremath{\rStrat{#2}_{\MaxPl}(#1)}}\xspace}
\newcommand{\rStratMin}[2][]{\ifthenelse{\equal{#1}{}}
{\ensuremath{\rStrat{#2}_{\MinPl}}}{\ensuremath{\rStrat{#2}_{\MinPl}(#1)}}\xspace}

\newcommand{\dValue}{\ensuremath{\mathsf{rVal}^0}\xspace}

\newcommand{\rValue}[1][]{\ifthenelse{\equal{#1}{}}
	{\ensuremath{\mathsf{rVal}^{\perturbation}}}
	{\ensuremath{\mathsf{rVal}^{\perturbation,#1}}}}
\newcommand{\rupperValue}[1][]{\ifthenelse{\equal{#1}{}}
	{\ensuremath{\overline{\mathsf{rVal}}^{\perturbation}}}
	{\ensuremath{\overline{\mathsf{rVal}}^{\perturbation,#1}}}}
\newcommand{\rValueP}[1][]{\ifthenelse{\equal{#1}{}}
	{\ensuremath{\mathsf{rVal}^{\perturbation'}}}
	{\ensuremath{\mathsf{rVal}^{\perturbation',#1}}}}
\newcommand{\rValueL}[1][]{\ifthenelse{\equal{#1}{}}
	{\ensuremath{\overline{\mathsf{rVal}}}}
	{\ensuremath{\overline{\mathsf{rVal}}^{#1}}}}
\newcommand{\bValue}[1][]{\ifthenelse{\equal{#1}{}}
{\ensuremath{\overline{\frac{1}{2}\mathsf{rVal}}^{\perturbation}}}
{\ensuremath{\overline{\frac{1}{2}\mathsf{rVal}}^{\perturbation,#1}}}}
\newcommand{\bValueL}[1][]{\ifthenelse{\equal{#1}{}}
{\ensuremath{\overline{\frac{1}{2}\mathsf{rVal}}}}
{\ensuremath{\overline{\frac{1}{2}\mathsf{rVal}}^{#1}}}}
\newcommand{\rValueS}[1][]{\ifthenelse{\equal{#1}{}}
{\ensuremath{\mathsf{rVal}^{\perturbation}}}
{\ensuremath{\mathsf{rVal}^{\perturbation,#1}}}}

\newcommand{\regions}[2]{\ensuremath{\mathsf{Reg}(#1,#2)}\xspace}
\newcommand{\rgame}[1][]{\ifthenelse{\equal{#1}{}}
{\ensuremath{\mathcal{R}(\game)}\xspace}
{\ensuremath{\mathcal{R}^{#1}(\game)}\xspace}}

\newcommand{\moveto}[1]{\ensuremath{\xrightarrow{#1}}\xspace}
\newcommand{\move}[1][]{\ifthenelse{\equal{#1}{}}
{\ensuremath{\trans, \delay}\xspace}
{\ensuremath{\trans_{#1}, \delay_{#1}}\xspace}}
\newcommand{\moveP}[1][]{\ifthenelse{\equal{#1}{}}
{\ensuremath{\trans', \delay'}\xspace}
{\ensuremath{\trans'_{#1}, \delay'_{#1}}\xspace}}
\newcommand{\umove}[1][]{\ifthenelse{\equal{#1}{}}
{\ensuremath{\utrans, \delay}\xspace}
{\ensuremath{\utrans_{#1}, \delay_{#1}}\xspace}}

\newcommand{\play}{\ensuremath{\rho}\xspace}

\newcommand{\Plays}{\ensuremath{\mathsf{Plays}}\xspace}

\newcommand{\FPlays}[1][]{\ifthenelse{\equal{#1}{}}
{\ensuremath{\mathsf{FPlays}}\xspace}{\ensuremath{\mathsf{FPlays}^{#1}}\xspace}}
\newcommand{\FPlaysMin}[1][]{\ensuremath{\FPlays[#1]_\MinPl}\xspace}
\newcommand{\FPlaysMax}[1][]{\ensuremath{\FPlays[#1]_\MaxPl}\xspace}

\newcommand{\outcomes}{\ensuremath{\mathsf{Play}}\xspace}

\newcommand{\FPlaysG}[1][]{\ensuremath{\FPlays[#1]_{\game}}}

\newcommand{\rpath}{\ensuremath{\pi}\xspace}
\newcommand{\ppath}{\ensuremath{\pi}\xspace}

\newcommand{\PVF}{PVF\xspace}
\newcommand{\F}{\ensuremath{\mathcal F}\xspace}
\newcommand{\Hfunction}{\ensuremath{\mathcal H}\xspace}
\newcommand{\V}{\ensuremath{\mathbb{V}}\xspace}

\newcommand{\Eq}{\ensuremath{\mathcal{E}}\xspace}
\newcommand{\unreset}[1]{\ensuremath{\mathsf{Unreset}_{#1}}\xspace}
\newcommand{\fguard}[1]{\ensuremath{\mathsf{Guard}_{#1}}\xspace}
\newcommand{\pre}[1]{\ensuremath{\mathsf{Pre}_{#1}}\xspace}
\newcommand{\perturbs}[1]{\ensuremath{\mathsf{Perturb}^{\perturbation}_{#1}}\xspace}

\newcommand{\EXP}{\ensuremath{\mathsf{EXPTIME}}\xspace}
\newcommand{\PSPACE}{\ensuremath{\mathsf{PSPACE}}\xspace}

\newcommand{\perturbation}{\ensuremath{p}\xspace}
\newcommand{\perturbationBound}{\ensuremath{\eta}\xspace}
\newcommand{\perturbationParam}{\ensuremath{\mathfrak{p}}\xspace}

\newcommand{\rob}{\mathsf{rob}}
\newcommand{\exact}{\ensuremath{0}\xspace}
\newcommand{\excess}{\ensuremath{\mathsf{excess}\xspace}}
\newcommand{\cons}{\ensuremath{\mathsf{cons}\xspace}}

\newcommand{\pertInt}[1][]{\ifthenelse{\equal{#1}{}}
{\ensuremath{[0, 2\perturbationBound]}}
{\ensuremath{[0, 2#1]}\xspace}}
\newcommand{\sem}[2][]{\ifthenelse{\equal{#1}{}}
{\ensuremath{\llbracket#2\rrbracket}}
{\ensuremath{\llbracket#2\rrbracket^{#1}}}\xspace}

\newcommand{\States}{\ensuremath{S}}
\newcommand{\StatesMin}{\ensuremath{\States_{\MinPl}}\xspace}
\newcommand{\StatesMax}{\ensuremath{\States_{\MaxPl}}\xspace}
\newcommand{\StatesT}{\ensuremath{\States_{T}}\xspace}
\newcommand{\states}{\ensuremath{s}}
\newcommand{\Edges}{\ensuremath{E}}
\newcommand{\EdgesMin}{\ensuremath{\Edges_{\MinPl}}\xspace}
\newcommand{\EdgesMax}{\ensuremath{\Edges_{\MaxPl}}\xspace}
\newcommand{\EdgesRob}{\ensuremath{\Edges_{\textsf{rob}}}\xspace}
\newcommand{\edge}{\ensuremath{e}}

\newcommand{\Conf}[1][]{\ifthenelse{\equal{#1}{}}
{\ensuremath{\mathsf{Conf}}\xspace}
{\ensuremath{\mathsf{Conf}}^{#1}\xspace}}
\newcommand{\ConfMin}[1][]{\ensuremath{\Conf[#1]_{\MinPl}}\xspace}
\newcommand{\ConfT}[1][]{\ensuremath{\Conf[#1]_{T}}\xspace}
\newcommand{\ConfMax}[1][]{\ensuremath{\Conf[#1]_{\MaxPl}}\xspace}

\newcommand{\gamedelta}{\game^{\perturbationBound}}

\newcommand{\LocsMindelta}{\LocsMin}

\newcommand{\FPlaysdelta}[1][]{\ifthenelse{\equal{#1}{}}
{\FPlays_{\gamedelta}}{\FPlays^{#1}_{\gamedelta}}}

\newcommand{\gameexcess}{\game^{\mathsf{e}}}

\newcommand{\Locsexcess}{\Locs^{\mathsf{e}}}

\newcommand{\LocsMaxexcess}{\LocsMax^{\mathsf{e}}}

\newcommand{\Transexcess}{\Trans^{\mathsf{e}}}

\newcommand{\weightexcess}{\weight^{\mathsf{e}}}

\newcommand{\FPlaysexcessa}[1][]{\ifthenelse{\equal{#1}{}}
{\FPlays_{\gameexcess}}
{\FPlays^{#1}_{\gameexcess}}}

\title{Synthesis of Robust Optimal Real-Time Systems}

\author{Benjamin Monmege}{Aix Marseille Univ, CNRS, LIS, Marseille,
	France}{benjamin.monmege@univ-amu.fr}{https://orcid.org/0000-0002-4717-9955}{}%

\author{Julie Parreaux}{University of Warsaw, Poland}{j.parreaux@uw.edu.pl}{}{}%

\author{Pierre-Alain Reynier}{Aix Marseille Univ, CNRS, LIS, Marseille,
	France}{pierre-alain.reynier@univ-amu.fr}{}{}

\authorrunning{B. Monmege, J. Parreaux, and P.-A. Reynier}

\Copyright{Benjamin Monmege, Julie Parreaux, and Pierre-Alain Reynier} 

\ccsdesc{Software and its engineering~Formal software verification}
\ccsdesc{Theory of computation~Algorithmic game theory}

\keywords{Weighted timed games, Algorithmic game theory, Robustness}

\hideLIPIcs

\EventEditors{John Q. Open and Joan R. Access}
\EventNoEds{2}
\EventLongTitle{42nd Conference on Very Important Topics (CVIT 2016)}
\EventShortTitle{CVIT 2016}
\EventAcronym{CVIT}
\EventYear{2016}
\EventDate{December 24--27, 2016}
\EventLocation{Little Whinging, United Kingdom}
\EventLogo{}
\SeriesVolume{42}
\ArticleNo{23}
\begin{document}

\maketitle

\begin{abstract}
	Weighted Timed Games (\WTG{s} for short) are widely 
	used to describe real-time controller synthesis problems, 
	but they rely on an
	unrealistic  perfect measure
	of time elapse.
%	The synthesized strategies rely on a , which is not realistic in practice. 
	In order to produce strategies 
	tolerant to timing imprecisions, we consider  
	a notion of robustness, expressed as a parametric semantics, 
	first introduced for timed automata. 
	%More precisely, 
	\WTG{s} are two-player zero-sum games played 
	in a weighted timed automaton %equipped with integer weights 
	in which one of the players, that we call \MinPl, 
	wants to reach a target location while minimising the cumulated weight. 
	The opponent player, in addition to controlling some of the locations,
	can perturb delays chosen by \MinPl.
%	In this work, we equip the underlying timed automaton with a 
%	semantics depending on some parameter (representing the maximal possible perturbation) in which the opponent of \MinPl 
%	can in addition perturb delays chosen by \MinPl.
	%, using a deviation upper-bounded
	%by the parameter value. 
The robust value problem asks, given
some threshold, whether there exists a positive perturbation and
a strategy for \MinPl ensuring to reach the target,
with an accumulated weight below the threshold, whatever
the opponent does.

We provide in this article the
first decidability result for this robust value problem.
More precisely, we show that we can compute the robust value function, in a parametric way,
%this problem is decidable
for the class of divergent \WTG{s} (this class has been introduced
previously to obtain decidability of the (classical) value problem
in \WTG{s} without bounding the number of clocks). 
To this end, we show that the robust value is the fixpoint of some 
operators, as is classically done for value iteration algorithms.
We then combine in a very careful way two representations:
piecewise affine functions introduced
in~\cite{AlurBernadskyMadhusudan-04} to analyse \WTG{s}, 
and shrunk Difference Bound Matrices (shrunk DBMs for short)
considered in~\cite{SankurBM-11} to analyse robustness in timed automata.
The crux of our result consists in showing that using this representation, the operator
 of value iteration can be computed for infinitesimally small perturbations.
Last, we also study qualitative decision problems and close 
an open problem on robust reachability, showing
it is \EXP-complete for general \WTG{s}.

%	Under this semantics, we analyse \WTG with two classes of problems: 
%	the fixed-perturbation robustness problem where we supposed that 
%	the upper-bound over perturbations is known and fixed, and the existential 
%	robustness problem when we ask about the existence of such upper-bound. 
%	In particular, we prove that both kind of robustness problem is \EXP-complete
%	for the qualitative issues (when we only focus on the reachability 
%	objective of \MinPl), and decidable for the quantitative 
%	issues when we restrict ourselves to a subclass of \WTG called divergent.
\end{abstract}

\section{Introduction}

The design and synthesis of real-time systems have long been paramount challenges, given the critical need for dependable and efficient systems in a variety of applications.
% ranging from embedded systems in automotive control to real-time trading platforms.
In particular, the pursuit of robustness and reliability in these systems has led researchers to explore innovative methods and formalisms to address the complexities inherent in real-time environments.
In this work, we focus on game-based models, and more precisely on the game extension of
timed automata~\cite{AlurDill-94}, a.k.a.~timed games,
which provide an elegant framework for capturing the interplay between system components, environment dynamics, and strategic decision-making. More precisely, in
this model, locations
of a timed automaton are split amongst the two players, which play in turn in the
infinite-state space of the automaton.

Regarding robustness, prior studies have primarily focused on areas such as fault tolerance, adaptive control, and formal methods. In this work, we follow a series of works based on game theory. The objective is to fill the gap between mathematical models such as timed automata,
often used for model-checking purposes, and implementation constraints, in which
clocks only have finite precision, and actions are not instantaneous. To that end, 
a parametric semantics has been considered in~\cite{Puri98}, which consists in allowing the
delays to be perturbed by some limited amount. The uncertainty of the system,
i.e.~the perturbation of the delays, is modelled by an adversarial environment.
Two kinds of problems can then be considered: first, the analysis may be done for a fixed perturbation bound (we call it a \emph{fixed-perturbation robustness problem}); second, 
in order to abstract the precise settings of the implementation, and as the exact value
of the perturbation bound may be unknown, one can try to determine whether
 there exists a perturbation bound under which the system
is reliable (we call it an \emph{existential robustness problem}). 
By monotonicity of the semantics w.r.t.~the perturbation bound, if one manages to prove the reliability against
some perturbation bound, then it still holds for smaller perturbations.
Initially introduced for model-checking purposes~\cite{WulfDMR04,BouyerMR06}, this approach has been lifted to automatic
synthesis, yielding the so-called \emph{conservative semantics}, studied for instance 
in~\cite{Sankur2013,OualhadjReynierSankur-14,Busatto-GastonM19}. 
In these works, a player named Controller aims at satisfying a liveness objective while its opponent
may perturb the delays.

In the present work, we aim to go beyond qualitative objectives, and tackle quantitative aspects.
%Traditional qualitative synthesis techniques have been successful in producing functional systems, but they often fall short when it comes to optimizing quantitative objectives. 
In real-time systems and critical applications, quantitative aspects such as resource utilization and performance metrics hold important significance. 
%By infusing game-based models with quantitative elements, we not only enhance our ability to synthesize systems that meet timing requirements but also gain a deeper understanding of how to optimize them with respect to quantitative objectives. 
This has led to the model of \emph{weighted
timed games} (\WTG for short), which has been widely studied during the last two decades.
When considering a reachability objective, Controller (a.k.a.~player \MinPl)
aims at reaching a set of target locations while minimizing the accumulated weight.
One is then interested in the \emph{value problem}, which consists in deciding, given some threshold, whether 
a strategy for \MinPl exists to reach some target location while keeping the accumulated cost below
this threshold.
While this problem is undecidable in general~\cite{BrihayeBruyereRaskin-05,BouyerBrihayeMarkey-06,BrihayeGeeraertsNarayananKrishnaManasaMonmegeTrivedi-14}, 
several subclasses have been identified
that allow one to regain decidability. 
Amongst recent works, we can cite 
the class of divergent \WTG{s}~\cite{BusattoGastonMR23} which
generalize to arbitrary costs the class of strictly non-Zeno costs introduced in~\cite{BouyerCassezFleuryLarsen-04}, 
or the class of one-clock \WTG{s}~\cite{MonPar22}.

The core objective of this research is to explore the synthesis of real-time systems that not only meet timing constraints but also optimize performance with respect to specified weight objectives
and are robust against timing imprecisions.
To that end, we aim to study the setting of timed games extended
with both robustness issues and quantitative aspects. We focus on the conservative
semantics and on reachability objectives. In this setting, under a fixed perturbation, the player 
\MinPl aims 
at reaching a set of target locations while minimizing the accumulated weight, and 
resisting delay perturbations. This leads to a notion of \emph{robust value under
a fixed perturbation}: this is simply the best value \MinPl can achieve. The associated
fixed perturbation robust value problem aims at comparing this value with a given threshold.
When turning to the existential robustness decision problem, 
one considers the notion of \emph{robust value}, defined as the limit of
robust values for arbitrarily small perturbation values. We prove that this limit exists,
and study the associated decision problem, which we simply call \emph{robust value problem}, and which can be defined
as follows:
given a threshold, determine whether there exists a positive perturbation, and
a winning strategy for \MinPl ensuring that the accumulated weight until the target is below
the threshold.%, whatever the behaviour of the antagonist player.
%considering arbitrarily small, but positive, perturbations.

This problem is highly challenging as it combines difficulties coming
from the introduction of weights, with those due to the analysis of
an existential problem for the parametric semantics of robustness.
Unsurprisingly, it has been shown to be undecidable~\cite{Sankur-13}.
To highlight the challenges we face, already in the
qualitative setting (w/o weights), the setting of two-player has not been addressed yet
for the conservative semantics, hence
the existential robust reachability problem was left open in~\cite{Sankur-13}.
Indeed, in~\cite{Busatto-GastonM19,Sankur2013}, only the one-player case is handled (a partial extension is considered 
in~\cite{OualhadjReynierSankur-14})
 and 
 %, where the player controlling the perturbation also
%resolves non-determinism. The 
the existential robustness
reachability problem for the two-player setting
has only been solved for the excessive semantics
(an alternative to the conservative semantics) in~\cite{BouyerMarkeySankur-15}.
Regarding the quantitative setting, very few works have addressed robustness issues.
The fixed-perturbation
robust value problem is shown to be decidable for one-clock weighted timed games in~\cite{GuhaKrishnaManasaTrivedi-15}, 
with non-negative weights only, and for the excessive 
semantics.
In~\cite{BouyerMarkeySankur-13}, the authors consider the one-player case and prove that the 
robust value problem is \PSPACE-complete.

Our contributions are as follows: first, regarding the qualitative setting, 
we close the case of existential robust reachability in two-player timed games
for the conservative semantics, and show that this problem is \EXP-complete.
To do so, we introduce a construction which allows us to reduce the problem to the
excessive semantics solved in~\cite{BouyerMarkeySankur-15}. As a corollary, we deduce an upper bound on the length
of paths to the target.

Then, we turn to the quantitative setting and show that for the
class of divergent \WTG{s} (one of the largest classes of \WTG{s}
for which the decidability of the value problem is known), the robust value problem is decidable. 
We proceed as~follows:
\begin{enumerate}
\item We characterize the robust value for a fixed perturbation as the 
fixpoint of some operator.
\item We show that for acyclic \WTG{s}, this fixpoint can be obtained as a finite iteration 
of this operator, which we decompose using four simpler operators.
\item We introduce a symbolic parametric approach for the computation of this operator, for arbitrarily small values
of the perturbation. This requires carefully combining the representation of
value functions using piecewise affine functions introduced in~\cite{AlurBernadskyMadhusudan-04} with the notion of shrunk DBMs, used in~\cite{SankurBM-11} to analyse robustness issues in timed automata.
This yields the decidability of the robust value problem for the class of acyclic \WTG{s}.
\item By combining this with the upper bound deduced from the qualitative analysis, we show
the decidability of the robust value problem for the whole class of divergent \WTG{s}.
\end{enumerate}

\iffalse
\newcommand{\ours}[1]{\textcolor{green!40!black}{#1}}

\begin{table}
\center
\begin{tabular}{c|c|c|c||c|c|c}
      Class           & \multicolumn{3}{c||}{Qualitative Problems} & \multicolumn{3}{c}{Quantitative Problems}\\
   of WTG  & w/o perturb. & Fixed $p$ & $\exists p >0$ & w/o perturb. & Fixed $p$ & $\exists p >0$ \\
\hline
Acyclic & \EXP & Dec. & Dec. &  \\
Divergent & & bla \\
General & \EXP & \ours{\EXP-c} & \ours{\EXP-c} & Undec. & Undec. & Undec.  \\

\end{tabular}
\caption{Summary of our contributions}
\end{table}
\fi
\begin{comment}
\todo{À enlever pour gagner de la place?}
As other related work, apart from \WTG{s}, energy timed automata have also considered to synthesise real-time systems. 
Intuitively, this model can be understood
as weighted timed automata extended with an energy variable. 
Hence, it is thus a one-player setting, and the 
problems considered often concern infinite runs. 
Robustness of
the synthesized strategies has for instance been tackled
in~\cite{BacciBFLMR21}. Yet another approach 
has been followed in~\cite{LarsenLTW14}
in which the authors consider the formalism of interfaces, that
they lift to the real-time setting, and in which they incorporate
 robustness aspects. However, they only
address the fixed-parameter~case.
\end{comment}
%This article is organized as follows. 
In Section~\ref{sec:prelim}, we 
introduce \WTG{s}, under the prism
of robustness. We describe in Section~\ref{sec:value-val}
the robustness problems 
we consider, present our contributions for qualitative ones, and
state that we can solve the quantitative one for
acyclic \WTG{s}. Sections~\ref{sec:valIt-robust-sem-F} and \ref{sec:encoding} 
detail how to prove this result, following steps \textsf{1.-3.} described 
above. Last, Section~\ref{sec:divergent}
extends this positive result to the class of divergent WTG{s}.
Omitted proofs can be found in the Appendix.

\section{Robustness in weighted timed games}
\label{sec:prelim}

%\noindent\textbf{Clocks, guards and regions.}
We let \Cl be a finite set of variables called clocks. A valuation is
a mapping $\val\colon \Cl\to \Rplus$. For a valuation $\val$, a delay
$t\in\Rplus$ and a subset $Y\subseteq \Cl$ of clocks, we define the
valuation $\val+t$ as $(\val+t)(x)=\val(x)+t$, for all $x\in \Cl$, and
the valuation $\val[Y:=0]$ as $(\val[Y:=0])(x)=0$ if $x\in Y$, and
$(\val[Y:=0])(x)=\val(x)$ otherwise. 
A (non-diagonal) guard on clocks of~\Cl is a
conjunction of atomic constraints of the form $x\bowtie c$, where
${\bowtie}\in\{{\leq},<,=,>,{\geq}\}$ and $c\in \N$. A valuation
$\val\colon \Cl\to \Rplus$ satisfies an atomic constraint $x\bowtie c$
if $\val(x)\bowtie c$. The satisfaction relation is extended to all
guards~$g$ naturally, and denoted by $\val\models g$. We let $\Guards$
denote the set of guards over \Cl.

\begin{figure}
	\centering
	\begin{tikzpicture}[xscale=.8,yscale=.8,every node/.style={font=\scriptsize},
	every label/.style={font=\scriptsize}]
	\node[PlayerMin,label={left:$\loc_0$}] at (0, 0) (s1) {$\mathbf{1}$};
	\node[PlayerMin,label={above:$\loc_4$}] at (3, 1) (s2) {$\mathbf{-1}$};
	\node[PlayerMax,label={below:$\loc_3$}] at (3, -1) (s4) {$\mathbf{0}$};
	\node[PlayerMin,label={below:$\loc_2$}] at (8, -1) (s3) {$\mathbf{1}$};
	\node[PlayerMin,label={right:$\loc_1$}] at (11, 0) (s5) {$\mathbf{1}$};
	\node[target] at (8, 1)  (s0) {\LARGE $\smiley$};
	
	% Connect the states with arrows
	\draw[->]
	(s1) edge node[above,xshift=-.7cm] 
	{$\begin{array}{c}
		1 < x_1 <2 \\ 
		x_1 \coloneqq 0 
		\end{array}$}
	node[below] {$\mathbf{1}$} (s2)
	(s1) edge node[below,xshift=-.7cm] {$1 \leq x_1 \leq 2$} (s4)
	(s2) edge node[above]
	{$\begin{array}{c}
		1 \leq x_1 < 2\land 
		x_2 < 2
		\end{array}$} (s0)
	(s3) edge node[below,xshift=.7cm]
	{$\begin{array}{c}
		1 \leq x_2 \leq 2 \\
		x_2 \coloneqq 0
		\end{array}$}
	node[above,xshift=-.25cm] {$\mathbf{1}$} (s5)
	(s4) edge node[below]
	{$\begin{array}{c}
		1 < x_1 < 2\land x_2 < 1 \\ x_1 \coloneqq 0
		\end{array}$}
	node[above] {$\mathbf{-2}$} (s3)
	(s5) edge node[above,xshift=.25cm] {$0 < x_2$} (s0);
	\end{tikzpicture}
	\caption{An acyclic \WTG with two clocks.}
	\label{fig:rob-valIt_ex-3}
\end{figure}

%\paragraph*{Weighted timed games}
\begin{definition}
	A \emph{weighted timed game} (\WTG) is a tuple $\game=\gameEx$
	where $\LocsMin$, $\LocsMax$, $\LocsT$ are finite disjoint subsets
	of \MinPl locations, \MaxPl locations, and target locations,
	respectively (we let $\Locs=\LocsMin\uplus\LocsMax\uplus\LocsT$),
	$\Cl$ is a finite set of clocks, 
	$\Trans\subseteq \Locs\times\Guards\times \powerset \Cl \times
	\Locs$ is a finite set of transitions, and
	$\weight\colon \Trans\uplus\Locs \to \Z$ is the weight~function.
\end{definition}
%Without loss of generality, we suppose the absence of deadlocks except
%on target locations, i.e.~for each location
%$\loc\in \Locs\backslash\LocsT$ and valuation $\val$, there exists
%$(\loc,g,Y,\loc')\in \Trans$ such that $\val\models g$, and no
%transitions start in \LocsT.
%
The usual semantics, called \emph{exact semantics}, of a \WTG $\game$ 
is defined in terms of a game played
on an infinite transition system whose vertices are configurations of
the \WTG denoted by $\Conf = \ConfMin \uplus \ConfMax \uplus \ConfT$. 
A configuration is a pair $(\loc,\val)$ with a location and
a valuation of the clocks. 
A configuration is final (resp.~belongs to \MinPl,
or \MaxPl), and belongs to $\ConfT$ (resp.~to $\ConfMin$,
or $\ConfMax$)
if its location is a target location of $\LocsT$ (resp.~of $\LocsMin$, or $\LocsMax$). 
The alphabet of the transition system
is given by $\Trans\times\Rplus$: a pair $(\move)$ encodes the
delay $\delay$ that a player wants to spend in the current location, before
firing transition $\trans$. An
example of \WTG is depicted in Figure~\ref{fig:rob-valIt_ex-3}.

In this article, we consider an alternative semantics to model the robustness, traditionnally called 
the \emph{conservative semantics}. It is defined in a WTG~$\game$ according to a fixed
parameter $\perturbation > 0$.
This semantics allows \MaxPl to slightly perturb the delays chosen by \MinPl with an amplitude bounded by $\perturbation$. 
From the modelling perspective, the perturbations model
the small errors of physical systems on the real value of clocks.
Conservative means that the delays proposed by \MinPl must remain 
feasible after applying all possible perturbations.
In particular, the conservative semantics does not add new edges with respect to the exact one.

\begin{definition}
	\label{def:sem-squelette}
	Let $\game = \gameEx$ be a \WTG. For $p\geq 0$, we let
	$\sem[\perturbation]{\game} = \tuple{\States, \Edges, \weight}$
	with $\States = \StatesMin \uplus \StatesMax \uplus \StatesT$ the set
	of \emph{states} with $\StatesMin = \ConfMin$,
	$\StatesT = \ConfT$ and $\StatesMax = \ConfMax \cup
	(\ConfMin \times \Rplus \times \Trans)$;
	$\Edges = \EdgesMin \uplus \EdgesMax \uplus \EdgesRob$ the
	set of \emph{edges} with%\todo{BM: je pense que c'est quand même plus clair en laissant ces formules centrées, même si ça prend un peu de place}
	% $\EdgesMax = 
	% \big\{\big((\loc, \val) \moveto{\move}
	% (\loc', \val')\big) ~|~ \loc \in \LocsMax, \val + \delay \models \guard 
	% \text{ and } \val' = (\val + \delay)[\reset \coloneqq 0] \big\}$, 
	% $\EdgesMin = \big\{\big((\loc, \val) \moveto{\move}
	% (\loc, \val, \move)\big) ~|~ \loc \in \LocsMin, \val + \delay \models \guard 
	% \text{ and } \val + \delay + 2\perturbation \models \guard \big\}$, and
	% $\EdgesRob = \big\{\big((\loc, \val, \move)
	% \moveto{\trans, \varepsilon} (\loc', \val')\big) ~|~
	% \varepsilon \in [0, 2\perturbation] 
	% \text{ and } \val' = (\val + \delay + \varepsilon)[\reset \coloneqq 0] \big\}$
	\begin{align*}
	\EdgesMax &= 
	\big\{\big((\loc, \val) \moveto{\move}
	(\loc', \val')\big) ~|~ \loc \in \LocsMax, \val + \delay \models \guard 
	\text{ and } \val' = (\val + \delay)[\reset \coloneqq 0] \big\} \\
	\EdgesMin &= \big\{\big((\loc, \val) \moveto{\move}
	(\loc, \val, \move)\big) ~|~ \loc \in \LocsMin, \val + \delay \models \guard 
	\text{ and } \val + \delay + 2\perturbation \models \guard \big\} \\
	\EdgesRob &= \big\{\big((\loc, \val, \move)
	\moveto{\trans, \varepsilon} (\loc', \val')\big) ~|~
	\varepsilon \in [0, 2\perturbation] 
	\text{ and } \val' = (\val + \delay + \varepsilon)[\reset \coloneqq 0] \big\}
	\end{align*}
	where $\trans = (\loc, \guard, \reset, \loc') \in \Trans$; and 
	$\weight \colon \States \cup \Edges \to \Z$ the weight
	function such that for all states $\states \in \States$ with
	$\states = (\loc, \val)$ or $\states = (\loc, \val, \move)$, 
	$\weight(\states) = \weight(\loc)$, and all edges $\edge \in \Edges$, 
	$\weight(\edge) = \weight(\trans)$ if $\edge = (\states \moveto{\move} \states')$ 
	with $\states \in \Conf$, or $\weight(\edge) = 0$ otherwise.
\end{definition}

When $p=0$, the infinite transition system $\sem[0]{\game}$ describes the exact semantics of the game, the usual semantics where each step of the player $\MinPl$ is cut into the true step, followed by a useless edge $(\loc, \val, \delta, t) \xrightarrow{\delta, 0} (\loc', \val')$ where $\MaxPl$ has no choice. When $p>0$, the infinite transition system $\sem[\perturbation]{\game}$ describes the conservative semantics of the game: states
$(\loc, \val, \move) \in \ConfMin \times \Rplus \times \Trans$
where \MaxPl must choose the perturbation in the interval $[0, 2\perturbation]$ are called \emph{perturbed states}.

Let $s$ be a state of $\sem[\perturbation]{\game}$, we denote 
by $\Edges(s)$ the set of possible outgoing edges of $\sem[\perturbation]{\game}$ from $s$. 
We extend this notation to locations to denote 
the set of outgoing transitions in~$\game$.
A state (resp.~location) $s$ is a \emph{deadlock} when $\Edges(s) = \emptyset$. 
We note that the conservative semantics may introduce deadlock in configurations of \MinPl 
(even if an outgoing edge exists in the exact semantics). 
Thus, unlike in the literature~\cite{AlurBernadskyMadhusudan-04,BusattoGastonMR23}, 
we allow state \emph{and} location deadlocks.

A \emph{finite play} of $\game$
w.r.t. the conservative semantics with parameter $\perturbation$ 
is a sequence of edges in the transition system
$\sem[\perturbation]{\game}$
starting in a configuration of $\game$.
We denote by $|\play|$ the number of edges of $\play$, and by $\last(\play)$ its last state. 
The concatenation of two finite plays $\play_1$ and
$\play_2$, such that $\play_1$ ends in the same state as
$\play_2$ starts, is denoted by $\play_1\play_2$.
Moreover, for modelling reasons, we only consider finite plays (starting and)
ending in a configuration of $\game$.
%In particular, we say that a finite play $\play$ is \emph{well-formed}
%when $\last(\play) \in \Conf$ (i.e.~not a perturbed state),
%otherwise $\play$ is said \emph{bad-formed}\todo{BM: est-ce qu'on utilise vraiment ces termes ?}.
Since a finite play is always defined regarding a parameter $\perturbation$ for the conservative semantics, 
we denote by $\FPlays[\perturbation]$
this set of finite plays.
Moreover, we denote by $\FPlaysMin[\perturbation]$
(resp.~$\FPlaysMax[\perturbation]$) the subset of these finite plays
ending in a state of \MinPl (resp.~\MaxPl).
A \emph{maximal play} 
is then a maximal sequence of consecutive edges: 
it is either a finite play reaching a deadlock (not necessary in \LocsT), 
or an infinite sequence such that all its prefixes are finite plays.

%We call \emph{path} a finite or infinite sequence $\ppath$ of
%transitions of~$\game$. Each play~$\play$ of $\game$ is associated
%with a unique path $\ppath$ (by projecting away everything but the
%transitions): we say that $\play$ \emph{follows} the path $\ppath$. A
%\emph{target path} is a finite path ending in the target set
%$\LocsT$. We denote by $\TPaths$ the set of target paths. We let
%$\TPaths_\play$ (resp. $\TPaths^n_\play$) the subset of target paths
%that start from the last location of the finite play $\play$
%(resp. containing $n$ transitions). A path is said to be \emph{maximal} if
%it is infinite or if it is a target path.

The objective of \MinPl is to reach a target configuration, while
minimising the cumulated weight up to the target. Hence, we
associate to every finite play
$\play= \states_0 \moveto{\move[0]} \states_1
\moveto{\move[1]} \cdots \states_k$ (some edges are in $\EdgesRob$, others are not) 
its cumulated weight, taking into account both discrete and continuous costs:
$\weightC(\play)=\sum_{i=0}^{k-1} [t_i\times \weight(\states_i) + \weight(\trans_i)]$. 
Then, the weight of a maximal play $\play$,
denoted by $\weight(\play)$, is defined by $+\infty$ if \play does not reach $\LocsT$ 
(because it is infinite or reaches another deadlock), 
and $\weightC(\play)$ if it ends in $(\loc,\val)$ with $\loc\in \LocsT$.

%\paragraph*{Robust strategies under the conservative semantics}

A strategy for \MinPl (resp.~\MaxPl)
is a mapping from finite plays ending in a state of \MinPl
(resp.~\MaxPl) to a decision in $(\move)$ labelling an edge of $\sem[\perturbation]{\game}$ from the last state of the~play.
% \ state\footnote{Under a conservative semantics, the set of states of 
%	the semantics that belongs to \MaxPl, i.e. $\StatesMax$,
%	are the configurations of \MaxPl \emph{and} the perturbed states.}
%Even if we suppose that deadlocks are not allowed,  
Since plays could reach a deadlock state of \MinPl, we consider strategies of \MinPl to be partial mappings. 
For instance, in the \WTG depicted in \figurename{~\ref{fig:rob-valIt_ex-3}} 
and a perturbation $\perturbation$, a strategy for \MinPl in all plays ending in $(\loc_2, \val)$ 
can be defined only when $\val(x_2) \leq 2 - 2\perturbation$ since, otherwise, there are no outgoing edges 
in $\sem[\perturbation]{\game}$ from this state. 
Symmetrically, we ask for \MaxPl to always propose a move if we are not in a deadlock state. 
More formally, a strategy for \MinPl, denoted $\robminstrategy$, 
is a (possibly partial) mapping $\robminstrategy \colon \FPlaysMin[\perturbation] \to \Edges$ 
such that $\robminstrategy(\play) \in \Edges(\last(\play))$. \todo{Exemple ici}
A strategy for \MaxPl, denoted $\robmaxstrategy$, 
is a (possibly partial) 
mapping $\robmaxstrategy \colon \FPlaysMax[\perturbation] \to \Edges$ 
such that for all $\play$, if $\Edges(\last(\play))\neq \emptyset$, then $\robminstrategy(\play)$ is defined, 
and in this case belongs to $\Edges(\last(\play))$.
The set of strategies of \MinPl (resp.~\MaxPl) with the perturbation $\perturbation$ is denoted by $\rStratMin{\perturbation}$
(resp.~$\rStratMax{\perturbation}$).

% We explain the asymmetric definition between the strategies of \MinPl
% and \MaxPl by the objective of both players.
% Indeed, if the strategy of \MinPl does not choose a robust decision,
% i.e.\ the strategy stops the play without reaching a target
% location, then \MinPl loses and obtains $+\infty$ as a payoff which is
% the worst situation for \MinPl.
% However, it is in the interest of \MaxPl to choose to play no (robust) decision
% even if there exist feasible ones, since in this case, \MaxPl wins by obtaining $+\infty$
% as a payoff (that is the best situation for \MaxPl).
%\begin{example}
%	Consider the \WTG depicted in \figurename{~\ref{fig:rob-valIt_ex-3}}
%	and a perturbation $\perturbation$.
%	As an example, let us define a particular strategy $\robminstrategy$ 
%	for \MinPl as a function defined on plays $\play$ ending in $(\loc_2, \val)$ with 
%	$\val(x_2) \leq 2 - 2\perturbation$ by
%	$\robminstrategy(\play) = (\trans, 2 - 2\perturbation - \val(x_2))$ where 
%	$\trans$ is the unique transition between $\loc_2$ and $\loc_1$ (we do not define here what happens in plays ending in $\loc_0$ to simplify). 
%	Notice that $\robminstrategy(\play)$ cannot be defined if the last state is 
%	$(\loc_2, \val)$ with $\val(x_2) > 2 - 2\perturbation$, since there are no 
%	outgoing edges from such states in $\sem[\perturbation]{\game}$: indeed \MinPl 
%	should propose a delay such that all possible perturbations later chosen by 
%	\MaxPl still satisfy the guard on the transition.
%\end{example}

A play or finite play
$\play= \states_0 \moveto{\move[0]} \states_1
\moveto{\move[1]} \cdots$ \emph{conforms} to a
strategy $\robminstrategy$ of $\MinPl$ (resp.~$\MaxPl$)
if for all $k$ such that $\states_k$ belongs to $\MinPl$
(resp.~$\MaxPl$), we have that
$(\move[k]) =
\robminstrategy(\states_0 \moveto{\move[0]} \cdots \states_k)$. 
For all strategies $\robminstrategy$
and $\robmaxstrategy$ of players \MinPl and \MaxPl, respectively, and
for all configurations~$(\loc_0,\val_0)$, we let
$\outcomes((\loc_0,\val_0),\robminstrategy,\robmaxstrategy)$ be
the outcome of $\robminstrategy$ and $\robmaxstrategy$, defined as the
unique maximal play conforming to $\robminstrategy$ and
$\robmaxstrategy$ and starting in~$(\loc_0,\val_0)$.

The semantics $\sem[\perturbation]{\game}$ is monotonic with respect to the perturbation $\perturbation$ in the sense that $\MinPl$ has more strategies when $p$ decreases, while $\MaxPl$ can obtain, against a fixed strategy of $\MinPl$, a smaller weight when $p$ decreases. Formally, we have:
%(see Appendix~\ref{app:sem-strat_MinMax} for the detailed proof),

\begin{restatable}{lemma}{lemSemstratMinMax}
	\label{lem:sem-strat_MinMax}
	Let $\game$ be a \WTG, and  $\perturbation > \perturbation' \geq 0$ be two perturbations. Then
	\begin{enumerate}
		\item\label{itm:sem-strat_min}
		$\rStratMin{\perturbation} \subseteq
		\rStratMin{\perturbation'}$;
		
		\item\label{itm:sem-strat_max}
		for all $\robminstrategy \in \rStratMin{\perturbation}$,
		$\sup_{\robmaxstrategy \in \rStratMax{\perturbation}}
		\weight(\outcomes((\loc, \val), \robminstrategy, \robmaxstrategy)) \geq
		\sup_{\robmaxstrategy \in \rStratMax{\perturbation'}}
		\weight(\outcomes((\loc, \val), \robminstrategy, \robmaxstrategy))$. 
	\end{enumerate}
\end{restatable}

\section{Deciding the robustness in weighted timed games }
\label{sec:value-val}

We aim to study what \MinPl can guarantee, qualitatively and then quantitatively, 
in the conservative semantics of weighted timed games whatever \MaxPl does.

\smallskip
\noindent\textbf{Qualitative robustness problems.}
%\subparagraph*{Qualitative robustness problems.}
%We start by studying the qualitative question of whether \MinPl can guarantee to reach a target location from a given configuration. 
Formally, given a WTG $\game$ and a perturbation
$\perturbation$, we say a strategy  $\robminstrategy$ of \MinPl
is \emph{winning} in $\sem[\perturbation]{\game}$ from configuration $(\loc,\val)$ if for all
strategies $\robmaxstrategy$ of \MaxPl, $\outcomes((\loc, \val),
\robminstrategy, \robmaxstrategy)$ is a finite play ending in a location of
$\LocsT$.

There are two possible questions, whether the perturbation $\perturbation$ is fixed, or if we should consider it to be infinitesimally small: 
\begin{itemize}
	\item \emph{fixed-perturbation robust reachability problem}: given a WTG $\game$, a configuration
	$(\loc,\val)$ and
	a perturbation $\perturbation >0$, decide whether \MinPl has a
	winning strategy $\robminstrategy$ from $(\loc,\val)$ in 
	$\sem[\perturbation]{\game}$;
	\item \emph{existential robust reachability problem}: given a WTG $\game$ and a
	configuration $(\loc,\val)$, decide whether
	there exists $\perturbation >0$ such that 
	\MinPl has a winning strategy $\robminstrategy$ from $(\loc,\val)$ in
	$\sem[\perturbation]{\game}$. Notice that by Lemma~\ref{lem:sem-strat_MinMax}, if \MinPl has a winning strategy $\robminstrategy$ from $(\loc,\val)$ in
	$\sem[\perturbation]{\game}$, then he has one in $\sem[\perturbation']{\game}$ for all $\perturbation'\leq \perturbation$.
\end{itemize}

When the perturbation $\perturbation$ is fixed, we can encode in a \WTG the conservative semantics described in $\sem[\perturbation]{\game}$, by adding new locations for \MaxPl to choose a perturbation, and by modifying the guards that will now use the perturbation $\perturbation$. Solving the fixed-perturbation robust reachability problem then amounts to solving a reachability problem in the modified 
\WTG\footnote{By transforming the \WTG, its guards use rational instead of natural numbers (due to 
	$\perturbation$). To fit the classical definition of \WTG, we can apply a scaling factor (i.e. 
	$1/\perturbation$) to all constants appearing in this \WTG. We note that this operation preserves the 
	set of winning strategies for the reachability objective (here weights are irrelevant) by applying the 
	scaling operations on strategies too.}
which can be performed in \EXP \cite{AsarinMaler-99} (here weights are useless). Since the reachability problem in timed games is already \EXP-complete \cite{JurdzinskiTrivedi-07}, we obtain:
\begin{proposition}
	The fixed-perturbation robust reachability problem is \EXP-complete.
\end{proposition}

\begin{figure*}
	\centering
	\vspace{-1cm}
	\begin{tikzpicture}[xscale=.8,every node/.style={font=\scriptsize},
	every label/.style={font=\scriptsize}]
	% Draw the states
	\node[target] at (1.75, 1) (cons) {conservative semantics};

	\node[PlayerMin, label={above:$\loc$}] at (0, 0) (s1) {$\mathbf{w_0}$};
	\node[Player, label={above:$\loc'$}] at (3.5, 0)  (s0) {$\mathbf{w_1}$};
	%	\node at (7.5, 0) (s) {$\rightsquigarrow$};
	
	% Connect the states with arrows
	\draw[->] (s1) edge node[above] {$\guard, \reset$} node[below] {$\mathbf{w}$} (s0);
	\draw[->, decorate, decoration=snake] (4.5, 0) -- (6.5, 0);
	
	\begin{scope}[xshift=7.5cm]
	\node[PlayerMin, label={above:$\loc$}] at (0, 0) (s1) {$\mathbf{w_0}$};
	\node[PlayerMax, label={above:$\loc^{\trans}$}] at (4, 0) (s2) {$\mathbf{0}$};
	\node[Player, label={above:$\loc'$}] at (8, 0) (s0) {$\mathbf{w_1}$};
	\node[target] at (4, -1) (s3) {\Large $\frownie$};
	\node[target] at (4, 1) (excess) {excessive semantics};
	
	% Connect the states with arrows
	\draw[->]
	(s1) edge node[above] {$x^{\mathsf{e}} := 0$} node[below] {$\mathbf{w}$} (s2)
	(s2) edge node[above] {$\guard \wedge x^{\mathsf{e}} = 0, \reset$} node[below] {$\mathbf{0}$} (s0)
	(s2) edge node[right] {$\neg\guard \wedge x^{\mathsf{e}} = 0$} node[left] {$\mathbf{0}$} (s3)
	;
	\end{scope}
	\end{tikzpicture}
	\caption{Gadget used to encode the conservative semantics into the excessive one. Each transition
	$\trans =(\loc, \guard, \reset, \loc')$ with $\loc\in \LocsMin$ is replaced by the gadget. 
	Symbols $w, w_0, w_1$ denote weights from~$\game$. The new location $\loc^\trans$
	of \MaxPl uses a fresh clock $x^{\mathsf{e}}$ to test the guard after the perturbation 
	(as in the conservative semantics). The new location $\frownie$ is a deadlock (thus winning for \MaxPl).}
	\label{fig:valPb-cons2excess_gadget}
\end{figure*}

% In timed automata, the robust reachability problem 
% is not harder to solve than the reachability problem
% in timed automata (under the exact semantics), that is \PSPACE-complete~\cite{AlurDill-94}.
% Indeed, the region abstraction with only open regions works alike
% since the unique player cannot see the perturbation of the environment.
% In \WTG{s}, the problem is still open.
We now turn our attention to the existential robust reachability problem.
% with a non fixed perturbation. 
This problem was left open for the conservative semantics 
(see~\cite{Sankur-13}, Table 1.2 page 17), while
it has been solved in~\cite{BouyerMarkeySankur-15}
for an alternative semantics of robustness, known as 
the \emph{excessive semantics}. 
Intuitively, while the conservative semantics requires that the delay, 
after perturbation, satisfies the guard, the excessive semantics 
only requires that the delay, \emph{without perturbation}, 
satisfies the guard. 
We present %in Appendix~\ref{app:rob-val_reach-reduction} 
a 
reduction from the conservative semantics
to the excessive one, allowing us to solve the
existential robust reachability problem
for the conservative semantics.
Intuitively, the construction (depicted on \figurename{~\ref{fig:valPb-cons2excess_gadget}})
adds a new location (for \MaxPl) for each transitions of \MinPl 
to test the delay chosen by \MinPl after the perturbation:
\begin{restatable}{proposition}{propReachCons}
	\label{prop:reach-cons}
	The existential robust reachability problem is \EXP-complete.
\end{restatable}

\smallskip
\noindent\textbf{Quantitative fixed-perturbation robustness problem.}
%\subparagraph*{Quantitative fixed-perturbation robustness problem.}
We are also interested in the minimal weight that \MinPl can guarantee while reaching the target  
whatever \MaxPl does:
to do that we define \emph{robust values}.
First, we define the \emph{fixed-perturbation robust value}: 
for all configurations $(\loc, \val)$
of $\game$ (and not for all states of the semantics), we let
$\rupperValue(\loc, \val) =
\inf_{\robminstrategy \in \rStratMin{\perturbation}}~
\sup_{\robmaxstrategy \in \rStratMax{\perturbation}}
\weight(\outcomes((\loc, \val), \robminstrategy, \robmaxstrategy))$.

Since a fixed-perturbation conservative semantics defines a
\emph{quantitative reachability game}\footnote{A quantitative reachability game introduced
	in~\cite{BrihayeGeeraertHaddadLefaucheuxMonmege-15} is an abstract model to formally
	define the semantics of quantitative (infinite) games.},
where configurations of \MaxPl also contain the robust states, we obtain that
the fixed-perturbation robust value is determined, by applying~\cite[Theorem 2.2]{brihaye2021oneclock}, i.e.~that 
$\rupperValue(\loc, \val) = 
\sup_{\robmaxstrategy \in \rStratMax{\perturbation}} ~\inf_{\robminstrategy \in \rStratMin{\perturbation}}
\weight(\outcomes((\loc, \val), \robminstrategy, \robmaxstrategy))$. 
We therefore denote $\rValue$ this~value.

\begin{remark}
	In~\cite{BouyerMarkeySankur-13,BouyerMarkeySankur-15, GuhaKrishnaManasaTrivedi-15}, 
	the set of possible perturbations for \MaxPl is $[-\perturbation, \perturbation]$.
	For technical reasons, we use a (equivalent) perturbation 
	with a shift of the delay proposed by $\MinPl$ by $\perturbation$.
%	(the proof that both definitions are equivalent w.r.t. to the robust value
%	is given in 
%	Appendix~\ref{app:valPb-fixed-rob-div_translated}).
\end{remark}

When $\perturbation = 0$, $\dValue$
defines the \emph{(exact) value} 
that is used to study the value problem in \WTG{s}. 
By Lemma~\ref{lem:sem-strat_MinMax}, we can deduce that 
the fixed-perturbation robust value is monotonic with respect to the 
perturbation $p$ and is always an upper-bound for the (exact) value.
% (see Appendix~\ref{app:rVal-monotony} for the detailed proof).
\begin{restatable}{lemma}{lemRValMonotony}
	\label{lem:rVal-monotony}
	Let $\game$ be a \WTG, and  $\perturbation > \perturbation' \geq 0$ be two perturbations. 
	Then, for all configurations $(\loc, \val)$, 
	$\rValue(\loc, \val) \geq \rValueP(\loc, \val)$.
\end{restatable}

As in the qualitative case, when the perturbation $\perturbation$ is fixed, we
can encode in polynomial time in a \WTG the conservative semantics described in
$\sem[\perturbation]{\game}$. Unfortunately, the value of \WTG{s} is not always
computable since the associated decision problems (in particular the
\emph{value problem} that requires to decide if the value of a given
configuration is below a given threshold) are undecidable
\cite{BrihayeBruyereRaskin-05, BouyerBrihayeBruyereRaskin-07,
BrihayeGeeraertsNarayananKrishnaManasaMonmegeTrivedi-14}. However, in
subclasses of \WTG{s} where the value function can be computed, like acyclic
\WTG{s} (where every path in the graph of the \WTG{s} is acyclic, decidable in
2-\EXP~\cite{Bus19}) or divergent \WTG{s} (that we recall in
Section~\ref{sec:divergent}, in 3-\EXP~\cite{BouyerCassezFleuryLarsen-04,BusattoGastonMonmegeReynier-17}),
the fixed-perturbation robust value is also computable (if the modified game
falls in the subclass). In particular, we obtain:

\begin{proposition}
	\label{prop:fixed-val}
	We can compute the fixed-perturbation robust value of a \WTG that is acyclic (in 2-\EXP) or divergent (in 3-\EXP), for all possible initial configurations. 
\end{proposition}

On top of computing the robust values, the previous works also allow one to synthesize almost-optimal (\emph{i.e.} arbitrarily close from the value) strategies for both players. 

%\begin{proposition}
%	\label{prop:val-det}
	% Let $\game$ be a \WTG under a conservative semantics $\cons_{\perturbationBound}$
	% of parameter $\perturbationBound > 0$.
	% Then, the fixed-perturbation robust value, denoted by $\rValue_{\game}$, is determined, i.e.
	% \begin{displaymath}
	% \rupperValue(\loc, \val) =
	% \sup_{\robmaxstrategy \in \rStratMax{\perturbationBound}}~
	% \inf_{\robminstrategy \in \rStratMin{\perturbationBound}}~
	% \weightP(\outcomes((\loc, \val), \robminstrategy, \robmaxstrategy)) \,.
	% \end{displaymath}
%\end{proposition}

%Moreover, we say that a robust strategy $\robminstrategy \in \rStratMin{\cons'_\perturbationBound}$
%of \MinPl is \emph{$\varepsilon$-optimal} with respect to $\rValue[\cons'_\perturbationBound]$ if,
%for all configurations $(\loc, \val)$,
%\begin{displaymath}
%\sup_{\robmaxstrategy \in \rStratMax{\cons'_\perturbationBound}(\game)}
%\weightP(\outcomes((\loc, \val), \robminstrategy, \robmaxstrategy)) \leq
%\rValue[\cons'_\perturbationBound](\loc ,\val) + \varepsilon\,.
%\end{displaymath}

\smallskip
\noindent\textbf{Quantitative robustness problem.} 
%\subparagraph*{Quantitative robustness problem.} 
Now, we consider the existential version of this problem by considering an infinitesimal perturbation.
We thus want to know what \MinPl can guarantee as a value if \MaxPl plays infinitesimally 
small perturbations. To define properly the problem, we introduce a new value: given a \WTG \game,
the \emph{robust value} is defined,
for all configurations $(\loc, \val)$ of $\game$, by
$\rValueL(\loc, \val) =
\lim_{\perturbation \to 0, \perturbation > 0} \rValue(\loc, \val)$.
This value is defined as a limit of functions
(the fixed-perturbation robust values), which can be proved to always exist 
as the limit of a non-increasing sequence of functions
(see Lemma~\ref{lem:rVal-monotony}). 
%
%
%Moreover, in Section~\ref{subsec:val-comp_opt-lim},
%we study the link between all robust values according to a class of robust semantics
%and a permutation bound.
%In particular, by letting $\game$ be a \WTG,
%for all classes of robust semantics $(\cons'_\perturbationBound)_\perturbationBound$
%and parameter $\perturbationBound > 0$, we~have
%\begin{equation*}
%\bValueL[\rob]_{\game} \underbrace{=}_{\text{Prop.~\ref{prop:val-comp_opt-lim}}}
%\rValueL[\rob]_{\game} \underbrace{\leq}_{\text{Lem.~\ref{lem:val-comp_opt-lim-1}}}
%\bValue[\rob]_{\game} \underbrace{\leq}_{\text{Lem.~\ref{lem:val-comp_opt-lim-2}}}
%\rValue[\cons'_\perturbationBound]_{\game} \,.
%\end{equation*}
%
%Finally, we say that a robust strategy $\robminstrategy \in \rStratMin{\cons'_\perturbationBound}$
%of \MinPl is \emph{$\varepsilon$-optimal} with respect to $\rValueL[\rob]$ if,
%for all configurations $(\loc, \val)$,
%\begin{displaymath}
%\sup_{\robmaxstrategy \in \rStratMax{\cons'_\perturbationBound}(\game)}~
%\weightP(\outcomes((\loc, \val), \robminstrategy, \robmaxstrategy)) \leq
%\rValueL[\rob](\loc ,\val) + \varepsilon\,.
%\end{displaymath}
%
The decision problem associated to this robust value is: given a \WTG $\game$,
an initial configuration $(\loc, \val)$,
and a threshold $\lambda \in \Q \cup \{-\infty, +\infty\}$, decide if
$\rValueL(\loc, \val) \leq \lambda$. We call it the \emph{robust value problem}. 

Unsurprisingly, this problem is undecidable~\cite[Theorem 4]{BouyerMarkeySankur-13}. 
We will thus consider some restrictions over \WTG{s}.
In particular, we consider classes of \WTG{s} where the (non robust) value problem is 
known as decidable for the exact semantics:
acyclic \WTG{s}~\cite{AlurBernadskyMadhusudan-04}, and 
%one-clock \WTG{s}~\cite{BouyerLarsenMarkeyRasmussen-06, MonPar22} and 
divergent \WTG{s}~\cite{BouyerCassezFleuryLarsen-04,BusattoGastonMonmegeReynier-17}. 
% We want to check if the decidability remains for the robustness in these classes of \WTG{s}. 
%
Our first main contribution concerns the acyclic case:
\begin{theorem}
	\label{thm:main}
	The robust value problem is decidable over the subclass of acyclic \WTG{s}.
\end{theorem}

The next two sections sketch the proof of this theorem via an adaptation of the value iteration 
algorithm~\cite{AlurBernadskyMadhusudan-04} used to compute the value function in (non-robust) acyclic \WTG{s}: 
it consists in computing iteratively the best thing that both players can guarantee in a bounded number of steps, 
that increases step by step. It is best described by a mapping $\F$ that explains how a value function gets 
modified by allowing one more step for the players to play. 
The adaptation we propose consists in taking the robustness into account by using shrunk DBM techniques 
introduced in~\cite{SankurBM-11}: instead of inequalities on the difference of two clock values of the 
form $x-y \leq c$ involving rational constants $c$, the constants $c$ will now be of the form 
$a - b\perturbation$, with $a$ a rational and $b$ a positive integer, $p$ being an infinitesimal perturbation. 
This will allow us to compute a description of the fixed-perturbation values for all initial configurations, 
for all perturbations $\perturbation$ smaller than an upperbound that we will compute. 
The robust value will then be obtained by taking the limit of this parametric representation 
of the fixed-perturbation values when $\perturbation$ tends to~$0$. 
Our algorithm will also compute an upperbound on the biggest allowed perturbation $p$. 
%(we illustrate this in Appendix~\ref{app:valIt-ex-rob}). 
As in previous works, once the value is computed, we can also synthesize almost-optimal strategies.

Section~\ref{sec:valIt-robust-sem-F} will describe the mapping $\F_{\perturbation}$, with a known perturbation $\perturbation$: the iteration of this operator will be shown to converge towards the fixed-perturbation value $\rValue$. The robust value functions will be shown to always be piecewise affine functions with polytope pieces. Section~\ref{sec:encoding} describes the parametric representation of these functions, where the perturbation is no longer fixed but is a formal parameter $\perturbationParam$. We then explain how the mapping $\F_{\perturbation}$ can be computed for all small enough values of $\perturbation$ at once, allowing us to conclude.

% To compute the iterations of the operator, in the exact case, 
% authors of~\cite{AlurBernadskyMadhusudan-04} use 
% \emph{piecewise affine functions} described by a set of \emph{cells} 
% and a set of affine equations to represent the manipulated value functions.
% The main idea of our algorithm is the modification of this data structure 
% through the computation of the new operator:
% we will use parametric (in $\perturbationBound$) piecewise affine functions on 
% $\Rplus^{\Cl} + \perturbationBound\,\Q$ 
% (where the parameter appears in cells and affine equation), 
% and an upper bound constraint over $\perturbationBound$.
% The formal definition of this data structure is given in Section~\ref{sec:encoding}.
% Thus, we conclude the proof by proving that our data structure is preserved by 
% applying $\F^{\perturbationBound}$ on them (see Section~\ref{sec:rob-valIt_robust-sem-op}).

% In particular, our algorithm allows us to conclude that 
% \begin{proposition}
% 	\label{prop:rob-valIt_robust-sem}
% 	Let $\game$ be an acyclic \WTG under a class of conservative
% 	semantics $(\cons_\perturbationBound)_{\perturbationBound > 0}$.
% 	Then, we can compute a parameter $\perturbationBound > 0$
% 	and the mapping $\perturbationBound' \mapsto
% 	\rValueP$ for all
% 	$\perturbationBound' \leq \perturbationBound$.
% \end{proposition}

% In the rest of this paper, we present the algorithm to obtain 
% Proposition~\ref{prop:rob-valIt_robust-sem}.

\section{Operator $\F_{\perturbation}$ to compute the fixed-perturbation value}
\label{sec:valIt-robust-sem-F}

The first step of the proof is the definition of the new operator adapted from the operator $\F$ of \cite{AlurBernadskyMadhusudan-04}. 
We thus fix a perturbation $\perturbation>0$, and we define an operator $\F_\perturbation$ taking as input a mapping 
$X \colon \Locs \times \Rplus^{\Cl} \to \Rbar$, computing a mapping $\F_\perturbation(X) \colon \Locs \times \Rplus^{\Cl} \to \Rbar$ 
defined for all configurations $(\loc, \val)$ by $\F_{\perturbation}(X)(\loc, \val)$ equal to
\begin{displaymath}
\begin{cases}
0 \qquad\;\;\;\;\, \text{if $\loc \in \LocsT$} \\
+\infty \qquad \text{if $\loc \in \LocsMax$ and~$E(\loc, \val) = \emptyset$}  \quad \text{(if \MaxPl reaches a deadlock, he wins)}\\
\sup_{(\loc, \val) \moveto{\move} (\loc', \val') \in \sem[\perturbation]{\game}}
\big[\delay\,\weight(\loc) + \weight(\trans) + X(\loc', \val')\big] \qquad \text{if $\loc \in \LocsMax$ and $E(\loc, \val) \neq \emptyset$} \\
\inf_{(\loc, \val) \moveto{\move} (\loc, \val, \move) \in \sem[\perturbation]{\game}}\!
\sup_{(\loc, \val, \move) \moveto{\trans, \varepsilon} (\loc, \val') \in \sem[\perturbation]{\game}} \!\!
\big[(\delay{+}\varepsilon)\,\weight(\loc) {+} \weight(\trans) {+} X(\loc', \val')\big] \;\; \text{if $\loc \in \LocsMin$}
\end{cases}
\end{displaymath}
In the following, we let $\V^0$ be the mapping 
$\Locs \times \Rplus^{\Cl} \to \Rbar$ defined
by $\V^0(\loc, \val) = 0$ if $\loc \in \LocsT$ and
$\V^0(\loc, \val) = +\infty$ otherwise. 
%An adaptation (see Appendix~\ref{app:valIt-robust-sem-F}) 
%of the proof in the non-robust setting allows us to obtain:
By adapting the proof of the non-robust setting, we show:

\begin{restatable}{lemma}{lemValItF}
	\label{lem:valIt-robust-sem-F}
	Let $\game$ be an acyclic \WTG, $\perturbation> 0$, $D$ is the depth of $\game$, 
	i.e.~the length of a longest path in~$\game$.
	Then, $\rValue$ is a fixpoint of $\F_{\perturbation}$, and
	$\rValue = \F_{\perturbation}^D(\V^0)$.
\end{restatable}

We can thus compute the fixed-perturbation robust value of an acyclic \WTG by repeatedly computing $\F_{\perturbation}$. We will see in the next section that this computation can be made for all small enough $\perturbation$ by using a parametric representation of the mappings. It will be easier to split the computation of $\F_{\perturbation}$ in several steps (as done in the non robust case \cite{AlurBernadskyMadhusudan-04,BusattoGastonMR23}). Each of the four operators takes as input a mapping $V\colon \Rplus^{\Cl} \to \Rbar$ (where the location $\loc$ has been fixed, with respect to mappings $\Locs \times \Rplus^{\Cl} \to \Rbar$), and computes a mapping of the same type. 
% Based on the case of the exact semantics,
% we split this computation into intermediate operators~\cite{AlurBernadskyMadhusudan-04,Bus19}.
% In particular, we extend the set of operators used in this previous case:
% we extend the data structure with a parameter $\perturbationBound$
% (i.e.\ operators work on function $\Vals \to \Rbar + \perturbationBound \, \Q$)
% and we add a new operator to model the perturbation chosen by \MaxPl. 
% In particular, we define intermediate operations over robust value functions such that 
% by letting $\loc \in \Locs$ be a location and $\val \in \Vals$ be a valuation,
% we consider the following operations over $\V_{\loc}$ applying to $\val$.
\begin{itemize}
	\item The operator $\unreset{Y}$, with $Y \subseteq \Cl$ a subset of clocks, is such that for all $\val\in \Rplus^{\Cl}$, 
	$\unreset{Y}(V)(\val) = V(\val[Y \coloneqq 0])$.
	
	\item The operator $\fguard{\delta}$, with $\delta=(\loc,g,Y,\loc')$ a transition of $\Trans$, is such that for all $\val\in \Rplus^{\Cl}$, 
	if $\val \models \guard$, then $\fguard{\delta}(V)(\val) = V(\val)$; otherwise, 
	$\fguard{\delta}(V)(\val)$ is equal to $-\infty$ if $\loc \in \LocsMax$, and $+\infty$ if $\loc \in \LocsMin$.
	
	\item The operator $\pre{\loc}$, with $\loc \in \Locs$, 
	is such that for all $\val\in \Rplus^{\Cl}$, $\pre{\loc}(V)(\val)$ is equal 
	to $\sup_{\delay\geq 0} [\delay \, \weight(\loc) + V(\val + \delay)]$ 
	if $\loc \in \LocsMax$, and $\inf_{\delay\geq 0} [\delay \, \weight(\loc) + V(\val + \delay)]$
	if $\loc \in \LocsMin$.
	
	\item The operator $\perturbs{\loc}$, with perturbation $\perturbation>0$
	and $\loc \in \LocsMin$, is such that for all $\val\in \Rplus^{\Cl}$, 
	$\perturbs{\loc}(V)(\val) =
	\sup_{\varepsilon \in [0, 2\perturbation]}
	[\varepsilon \, \weight(\loc) + V(\val + \varepsilon)]$.
\end{itemize}

Though the situation is not symmetrical for $\MinPl$ and $\MaxPl$ in $\F_p$, in particular for the choice of delay $t$, the definition of $\pre{\loc}$ does not differentiate the two players with respect to their choice of delay. However, the correctness of the decomposition comes from the combination of this operator with $\fguard\trans$ (that clearly penalises $\MinPl$ if he chooses a delay such that the translated valuation does not satisfy the guard) and $\perturbs{\loc}$ that allows \MaxPl to select a legal perturbation not satisfying the guard, leading to a value $+\infty$ afterwards. 

For a mapping $\V\colon \Locs \times \Rplus^{\Cl} \to \Rbar$ and a location $\loc$, we can extract the submapping for the location $\loc$, that we denote by $\V_\loc\colon \Rplus^{\Cl} \to \Rbar$, defined for all $\val\in \Rplus^{\Cl}$ by $\V_\loc(\val) = \V(\loc, \val)$. Mappings $\Rplus^{\Cl} \to \Rbar$ can be compared, by using a pointwise comparison: in particular the maximum or minimum of two such mappings is defined pointwisely. The previous operators indeed allow us to split the computation of $\F_{\perturbation}$ (see the proof 
in Appendix~\ref{app:rob-valIt_Frob-cons}). We also rely on the classical notion of \emph{regions}, 
as introduced in the seminal work on timed automata~\cite{AlurDill-94}. Indeed, for a given location $\ell$, the set of deadlock valuations $\val$ where $E(\ell,\val)=\emptyset$ is a union of regions that we denote $R_\ell$ in the following, and that can easily be computed.

\begin{restatable}{lemma}{lemValItFcons}
	\label{lem:rob-valIt_Frob-cons}
	For all 
	$\V \colon \Locs\times \Rplus^{\Cl} \to \Rbar$, 
 	$\loc\in \Locs$, and $p>0$, 
	$\F_{\perturbation}(\V)(\loc)$ equals
	\begin{displaymath}
	\begin{cases}
		\val\mapsto 0 & \!\!\!\!\text{if $\loc \in \LocsT$}\\
		\left(\val \mapsto \begin{cases} 
			+\infty & \!\!\!\text{if $\val\in R_\ell$}\\
		 \big(
		 \displaystyle{\max_{\trans = (\loc, \guard, Y, \loc') \in \Trans}} \big[\weight(\trans) +
	\pre{\loc}(\fguard{\delta}(\unreset{Y}(\V_{\loc'})))\big]\big) (\val) & \!\!\!\text{if $\val\notin R_\ell$}\end{cases}\right) 
	& \!\!\!\!\text{if $\loc \in \LocsMax$}\\ 
	\displaystyle{\min_{\trans = (\loc, \guard, Y, \loc') \in \Trans}} \big[\weight(\trans) +
	\pre{\loc}(\perturbs{\loc}(\fguard{\delta}(\unreset{Y}(\V_{\loc'}))))\big]  & \!\!\!\!\text{if $\loc \in \LocsMin$}
	\end{cases}
	\end{displaymath}
\end{restatable}

\section{Encoding parametric piecewise affine functions}
\label{sec:encoding}

We now explain how to encode the mappings that the operators defined in the previous section take as input, to compute $\F_\perturbation$ for all perturbation bounds $\perturbation>0$ at once. 
We adapt the formalism used in~\cite{AlurBernadskyMadhusudan-04,BusattoGastonMR23} to incorporate the perturbation $\perturbation$. 
This formalism relies on the remark that $\V^0$ is a piecewise affine function, and that if $\V$ is piecewise affine, 
so is $\F_{\perturbation}(\V)$: thus we only have to manipulate such piecewise affine functions. 

To model the robustness, that depends on the perturbation bound
$\perturbation$, and maintain a parametric description of all the value
functions for infinitesimally small values of $\perturbation$, we consider
piecewise affine functions that depend on a formal parameter
$\perturbationParam$ describing the perturbation. The pieces over which the
function is affine, that we call cells in the following, are polytopes
described by a conjunction of affine equalities and inequalities involving
$\perturbationParam$. Some of our computations will only hold for small enough
values of the parameter $\perturbationParam$ and we will thus also maintain an
upperbound for this parameter.

\begin{definition}
We call \emph{parametric affine expression} an expression $E$ of the form $\sum_{x\in \Cl} \alpha_x \, x + \beta + \gamma\perturbationParam$ with $\alpha_x\in \Q$ for all $x\in\Cl$, $\beta\in \Q\cup\{-\infty,+\infty\}$, and $\gamma\in \Q$. The semantics of such an expression is given for a particular perturbation $p$ as a mapping $\sem E_p\colon \Rplus^\Cl \to \Rbar$ defined for all $\val \in \Rplus^\Cl$ by $\sem E_p(\val) = \sum_{x\in \Cl} \alpha_x \, \val(x) + \beta + \gamma p$. 
\end{definition}

A partition of $\Rplus^{\Cl}$ into cells is described by a set 
$\Eq = \{E_1, \ldots, E_m\}$ of parametric affine expressions.  
 Every expression can be turned into an equation or inequation by comparing it to $0$ with the symbol $=$, $<$ or $>$. The partition of $\Rplus^\Cl$ is obtained by considering all the combinations of equations and inequations for each $1\leq i \leq m$: such a combination is described by a tuple $(\bowtie_i)_{1\leq i \leq m}$ of symbols in $\{=,<,>\}$. For a given perturbation $\perturbation$, we let $\sem{\Eq,(\bowtie_i)_{1\leq i \leq m}}_p$ be the set of valuations $\val$ such that for all $i\in\{1, \ldots, m\}$, $\sem{E_i}_p \bowtie_i 0$.

\newcommand{\cells}{\mathcal{C}}

We call \emph{cell} every such combination such that for $\perturbation$ that tends to $0$, while being positive, 
the set $\sem{\Eq,(\bowtie_i)_{1\leq i \leq m}}_p$ is non empty. 
We let $\cells(\Eq)$ be the set of cells of $\Eq$. 
Notice that it can be decided (in at most exponential time) 
if a combination $(\bowtie_i)_{1\leq i \leq m}$ is a cell, by encoding the semantics in the first order theory of the reals,
and deciding if there exists an upperbound $\perturbationBound>0$ 
such that for all $0<\perturbation\leq \perturbationBound$, $\sem{\Eq,(\bowtie_i)_{1\leq i \leq m}}_p$ is non empty. 
Moreover, we can compute the biggest such upperbound $\perturbationBound$ if it exists. 
The upperbound of the partition
$\Eq = \{E_1, \ldots, E_m\}$ is then defined as the minimum such upperbound over all cells 
(there are at most $3^m$ cells), and denoted by $\perturbationBound(\Eq)$ in the following. 
On the left of \figurename{~\ref{fig:valIt-cell}}, we depict the partition of $\Rplus^{\Cl}$ defined from
$\Eq = \{x_2 - 2\perturbationParam, 
2x_1 + x_2 - 2 + \perturbationParam, 
2x_1 - x_2 + 1/2\}$, with a fixed value of the perturbation.
In blue, we color the cell defined by 
$(>, <, >)$.
We note that this cell is non-empty when $\perturbation \leq 1/2$.
By considering other cells, we obtain that $\perturbationBound(\Eq)=1/2$.

In the following, we may need to record a smaller upperbound than $\perturbationBound(\Eq)$, in order to keep the tightest constraint seen so far in the computation. We thus call \emph{parametric partition} a pair $\langle\Eq, \perturbationBound\rangle$ given by a set of equations and a perturbation $\perturbationBound>0$ that is at most~$\perturbationBound(\Eq)$.

For a cell $c\in \cells(\Eq)$, an expression $E$ of $\Eq$ is said to be on the \emph{border} of $c$ if the removal of $E$ from the set $\Eq$ 
of expressions forbids one to obtain the set of valuations $\sem{c}_p$ with the resulting cells for all small enough values of $p>0$: more precisely, we require that no cell $c'\in \cells(\Eq\setminus\{E\})$ is such that for some 
$p\leq \perturbationBound(\Eq)$, $\sem c_p = \sem{c'}_p$. Because of the definition of $\perturbationBound(\Eq)$, this definition does not depend on the actual value of $p$ that we consider (and we could thus replace "for some" by "for all" above).
On the left of \figurename{~\ref{fig:valIt-cell}}, all expressions are on the border for the blue cell, but only two of them are on the border of the orange cell.

\begin{figure*}
	\centering
	\begin{tikzpicture}[scale=1.5,every node/.style={font=\scriptsize},
	every label/.style={font=\scriptsize}]
	\draw[->] (0,0) -- (0,1.5);
	\draw[->] (0,0) -- (1.25,0);
	
	\node at (1.25,-.15) {\scriptsize$x_1$};
	\node at (-.15,1.5) {\scriptsize$x_2$};
	
	\node at (1,-.15) {\scriptsize$1$};
	
	\node at (-.15,0) {\scriptsize$0$};
	\node at (-.15,1) {\scriptsize$1$};
	
	\draw[dashed,opacity=0.35] (1,0) -- (1,1.25);
	
	\draw[dashed,opacity=0.35] (0,1) -- (1.25,1);
	
	\draw[fill=LightBlue,draw=none,opacity=0.75]
	(0, 0.25) -- (0, 0.5) -- (0.3425, 1.1856) -- (0.805, 0.25) -- (0, 0.25);
	\draw[fill=Peach,draw=none,opacity=0.75]
	(0, 0.25) -- (0, 0) -- (0.9375, 0) -- (0.805, 0.25) -- (0, 0.25);
	
	\draw[Purple,very thick] (0.235,1.4) -- (0.9355,0); %y = 2 - p - 2x
	\draw[Purple,very thick] (0,.5) -- (.45,1.4); %y = 2x + .5
	\draw[Purple,very thick] (0,.25) -- (1,0.25); %y = 2p
	
	\node[Purple,draw=none] at (1.35,.25) {$x_2 - 2\perturbationParam$};
	\node[Purple,draw=none] at (1.35,.7) {$2\, x_1 + x_2 - 2 + \perturbationParam$};
	\node[Purple,draw=none] at (0.9,1.5) {$2\,x_1 - x_2 + \frac{1}{2}$};
	
	\begin{scope}[xshift=3cm]
	\draw[->] (0,0) -- (0,1.5);
	\draw[->] (0,0) -- (1.25,0);
	
	\node at (1.25,-.15) {\scriptsize$x_1$};
	\node at (-.15,1.5) {\scriptsize$x_2$};
	
	\node at (1,-.15) {\scriptsize$1$};
	
	\node at (-.15,0) {\scriptsize$0$};
	\node at (-.15,1) {\scriptsize$1$};
	
	\draw[dashed,opacity=0.35] (1,0) -- (1,1.25);
	
	\draw[dashed,opacity=0.35] (0,1) -- (1.25,1);
	
	\draw[fill=ForestGreen,draw=none,opacity=0.75]
	(0, 0.25) -- (0, 0.5) -- (0.46, .96) -- (0.54, 0.79) -- (0, 0.25);

	\draw[fill=YellowGreen,draw=none,opacity=0.75]
	(0, 0.5) -- (0.3425, 1.1856) -- (0.46, .96) -- (0, 0.5);
	
	\draw[fill=LightBlue,draw=none,opacity=0.75]
	(0, 0.25) -- (0.54, 0.79) -- (0.8125, 0.25) -- (0, 0.25);

	\draw[Purple,very thick] (0.235,1.4) -- (0.9355,0); %y = 2 - p - 2x
	\draw[Purple,very thick] (0,.5) -- (.45,1.4); %y = 2x + .5
	\draw[Purple,very thick] (0,.25) -- (1,0.25); %y = 2p
	
	\draw[red,very thick] (0,.25) -- (1,1.25); % diagonal from (0, 2p) -> y = x + 2p
	\draw[red,very thick] (0.5625,0) -- (1, 0.4375); % diagonal from (1-3p/2, 2p) -> y = x - 1 + 7p/2

	\node[red,draw=none] at (1.8,0.4) {$x_1 - x_2 - 1 + 7\perturbationParam/2$};

	\node[red,draw=none] at (1.6,1.2) {$x_1 - x_2 + 2\perturbationParam$};

	\node[red,draw=none] at (1.5,1.4) {$x_1 - x_2 + 1/2$};

	\node[red,draw=none] at (1.3,1.65) {$x_1 - x_2 + 7/8-\perturbationParam/4$};

	\draw[red,very thick] (0,.5) -- (.9,1.4); %diagonal from (0, 0.5) -> y = x + .5
	\draw[red,very thick] (0,.844) -- (0.7,1.544); %diagonal from (3/4-p/4, 5/4-p/2) -> y = x + 7/8 - p/4
	\end{scope}
	\end{tikzpicture}
	\caption{On the left, we depict the partition defined from
		$\Eq = \{x_2 - 2\perturbationParam, 2 x_1 + x_2 - 2 + \perturbationParam, 
		2x_1 - x_2 + 1/2 \}$, for a small enough value of $\perturbationParam$.
		On the right, we depict the atomic partition induced by $\Eq$, and draw in red the added 
		parametric affine expressions.}
	\label{fig:valIt-cell}
\end{figure*}

The proofs that follow (in particular time delaying that requires to move along diagonal lines) requires to adapt the notion of \emph{atomicity} of a parametric partition, originally introduced in the non-robust setting \cite{AlurBernadskyMadhusudan-04,BusattoGastonMR23}.
A parametric affine expression $E = \sum_{x\in \Cl} \alpha_x \, x 
+ \beta + \gamma\perturbation$ is said to be \emph{diagonal}
if $\sum_{x\in \Cl} \alpha_x = 0$: indeed, for all $p>0$, $\sem{E}_p(\val)= \sem{E}_p(\val + t)$ for all $t \in \R$. A parametric partition is said to be \emph{atomic} if for all cells $c\in \cells(\Eq)$, there are at most two non-diagonal parametric affine expressions on the border of $c$: intuitively, one border is reachable from every valuation by letting time elapse, and the other border is such that by letting time elapse from it we can reach all valuations of the cell. An atomic partition decomposes the space into tubes whose borders are only diagonal, each tube being then sliced by using only non-diagonal expressions. In particular, each cell $c$ of an atomic partition has a finite set of cells that it can reach by time elapsing (and dually a set of cells that can reach $c$ by time elapsing), and this set does not depend on the value of the parameter $p$, nor the starting valuation in $\sem c_p$. On the right of \figurename{~\ref{fig:valIt-cell}}, we depict
the atomic partition associated with the same set of expressions used on the left. The diagonal expressions are depicted in red. 
We note that the cell colored in blue on the left
is split into five cells that are non-empty when $\perturbation \leq 3/7$. We can describe the new parametric partition as 
$(\{x_2 - 2\perturbationParam, 
2x_1 + x_2 - 2 + \perturbationParam, 
2x_1 - x_2 + 1/2, x_1 - x_2 - 1 + 7\perturbationParam/2, 
x_1 - x_2 + 2 \perturbationParam, x_1 - x_2 + 1/2, x_1 - x_2 + 7/8 - \perturbationParam/4\}, 3/7)$. As we will see below, a parametric partition can always be made atomic, by adding some diagonal parametric affine expressions. 

A \emph{parametric value function} (\PVF for short) is a tuple $F = \langle
\Eq, \perturbationBound, (f_c)_{c\in \cells(\Eq)}\rangle$ where $\langle \Eq,
\perturbationBound\rangle$ is a partition and, for all cells $c\in \cells(\Eq)$,
$f_c$ is a parametric affine expression. For a perturbation $0<p\leq
\perturbationBound$, the semantics $\sem{F}_p$ of this tuple is a mapping
$\Rplus^\Cl \to \Rbar$ defined for all valuations $\val$ by $\sem{F}_p(\val) =
\sem{f_c}_p(\val)$ where $c$ is the unique cell such that $\val\in \sem c_p$. 
A \PVF is said to be \emph{atomic} if its parametric partition is atomic. 
As announced above, we can always refine a \PVF so that it becomes atomic. 
%(the proof is given in Appendix~\ref{app:rob-valIt_computation-atomic}).

\begin{restatable}{lemma}{lemComputationAtomic}
	\label{lem:rob-valIt_computation-atomic}
	If $F=\tuple{\Eq,\perturbationBound, (f_c)_{c\in \cells(\Eq)}}$ is a \PVF, 
	we can compute an atomic \PVF $F'=\tuple{\Eq',\perturbationBound', (f'_c)_{c\in \cells(\Eq')}}$
	such that 
	$\perturbationBound' \leq \perturbationBound$, 
	and $\sem{F}_p = \sem{F'}_p$ for all $p\leq \perturbationBound'$.
\end{restatable}

% \section{Operations over parametric value functions}
% \label{sec:rob-valIt_robust-sem-op}

To conclude the proof of Theorem~\ref{thm:main}, 
we need to compute one application of $\F_{\perturbation}$ 
over a mapping $X \colon \Locs\times \Rplus^\Cl \to \Rbar$ 
that is stored by a \PVF for 
each location. Moreover, our computations must be done 
for all small enough values of $\perturbation$ simultaneously.
%(the detailed proof is given in Appendix~\ref{app:Fp_parametric}). 

\begin{restatable}{proposition}{propFpParametric}
	\label{prop:Fp_parametric}
	Let $F = \tuple{\Eq, \perturbationBound, (f_c)_{c\in \cells(\Eq)}}$ 
	be a \PVF.
	We can compute a \PVF
	$F' = \tuple{\Eq', \perturbationBound', (f'_c)_{c\in \cells(\Eq')}}$
	with $\perturbationBound' \leq \perturbationBound$, and $\sem{F'}_p = \F_\perturbation(\sem{F}_p)$
	for all $p \leq \perturbationBound'$.
\end{restatable}
\begin{proof}[Sketch of the proof]
	By using Lemma~\ref{lem:rob-valIt_Frob-cons}, it suffices to perform the proof independently for the four operators, as well as maximum or minimum operations. Proofs from \cite{AlurBernadskyMadhusudan-04,BusattoGastonMR23} can be adapted 
	for the maximum/minimum operations as well as the operators $\fguard{\trans}$ and $\unreset{Y}$ 
	that exist also in the non-robust setting. In the case of $\MaxPl$, the two cases depend only on the regions and we thus only apply the various operators for starting valuations not in $R_\ell$. 
	
	For the operator $\pre{\loc}$ (and similarly $\perturbs{\loc}$), the adaptation is more subtle. 
	The key ingredient, for instance to compute it over an atomic partition for a location $\loc$ of \MaxPl, is to transform the computation of 
	$(\sem{F}_p)(\val) = \sup_{\delay\geq 0}
	[\delay \, \weight(\loc) + \sem{F}_p(\val + \delay)]$ involving a supremum (for a fixed valuation $\val$, and a fixed perturbation $p\leq \perturbationBound$), by using a maximum over a finite set of interesting delays. 
	First, we remark that for every delay $\delay>0$, the valuation $\val + \delay$ belongs
	to the open diagonal half-line from $\val$, which crosses some of the semantics 
	$\sem{c'}_p$ for certain cells $c'$. 
	Moreover, this finite (since there are anyway only a finite number of cells in the partition) subset  
	of crossing cells neither depends on the choice of $\val$ in a given starting cell $c$, nor on the 
	perturbation $p$ as long as it is at most $\perturbationBound$ (by atomicity of the partition).
	Since the function $\sem{F}_p$ is affine in each cell, the above supremum 
	over the possible delays is obtained for a value $\delay$ that is either 
	$0$, or tending to $+\infty$, or on one of the two non-diagonal 
	borders of the previous crossing cells, 
	and we thus only have to consider those borders (that do not depend on the choice
	of $\val$ in a given starting cell $c$, nor on the perturbation $p$). 
%	as long as it is at most~$\perturbationBound$).
%
%	For the operator $\perturbs{\loc}$ for a location $\loc\in \LocsMin$, we subtly adapt the construction for $\pre{\loc}$, 
%	carefully taking into account the parametric representation of the perturbation $p$. 
\end{proof}

\section{Divergent weighted timed games}\label{sec:divergent}

From our algorithm to solve acyclic \WTG{s}, we naturally want to extend
the computation of the robust value to other classes of \WTG{s} by 
using an unfolding of the \WTG.
In particular, we consider the natural extension of \emph{divergent \WTG{s}}
(like in~\cite{BusattoGastonMonmegeReynier-17,BusattoGastonMR23}) that define a large class 
of decidable \WTG{s} for the exact semantics, with no limitations 
on the number of~clocks.

%\paragraph*{Definition of divergent \WTG{s}}
As usual in related work 
\cite{AlurBernadskyMadhusudan-04,BouyerCassezFleuryLarsen-04,
	BouyerJaziriMarkey-15,BusattoGastonMR23},
we now assume that all clocks are \emph{bounded} by a constant
$\clockbound\in\N$, i.e.~every transition of the \WTG is equipped with
a guard $g$ such that $\val\models g$ implies
$\val(x)\leq \clockbound$ for all clocks $x\in \Cl$. We denote by
$\maxWeightLoc$ (resp.~$\maxWeightTrans$, $\maxWeightEdge$) the maximal weight in
absolute values of locations (resp.~of transitions, edges) of $\game$,
i.e.~$\maxWeightLoc = \max_{\loc\in \Locs} |\weight(\loc)|$
(resp.~$\maxWeightTrans = \max_{\trans \in \Trans} |\weight(\trans)|$,
$\maxWeightEdge = \clockbound\maxWeightLoc + \maxWeightTrans$).

We use the exact semantics to define the divergence property by 
relying once again on the regions~\cite{AlurDill-94}.
We let $\regions \Cl \clockbound$ be the set of regions when clocks are bounded by $\clockbound$.
A game $\game$ (w.r.t.~the exact semantics)
can be populated with the region information as described formally in
\cite{BusattoGastonMR23}: we obtain the \emph{region game} $\rgame$.
%is thus the \WTG with locations
%$\RStates = \Locs\times \regions \Cl \clockbound$ and all transitions
%$((\loc,r),\guard'',\reset,(\loc',r'))$ with
%$(\loc,\guard,\reset,\loc')\in \Trans$ such that the model of guard
%$\guard''$ (i.e.~all valuations $\val$ such that
%$\val\models \guard''$) is a region $r''$, time successor of $r$ such
%that $r''$ satisfies the guard $\guard$, and $r'$ is the region
%obtained from $r''$ by resetting all clocks of $\reset$. 
%Distribution of locations to players, final locations, and weights are inherited
%from $\game$. 
%
We call \emph{region path} a finite or infinite sequence
of transitions in this game, and we denote by $\rpath$ such
paths. A play~$\play$ in $\game$ can be projected on a region path
$\rpath$:
we say that $\play$ \emph{follows} the region path~$\rpath$. 
It is important to notice that, even if $\rpath$ is a \emph{cycle}
(i.e.~starts and ends in the same location of the region game), there
may exist plays following it in $\game$ that are not cycles, due to
the fact that regions are sets of valuations. 

Divergent \WTG{s} are obtained 
by enforcing a semantical property of divergence 
(originally called \emph{strictly non-Zeno cost} 
when only dealing with non-negative weights~\cite{BouyerCassezFleuryLarsen-04}):
it asks that every play (w.r.t.~the exact semantics) 
following a cyclic region path has weight far from $0$. 
Formally, a cyclic region path~\rpath of~\rgame is said to be a positive 
(resp.~negative) if every finite play~\play following~\rpath satisfies
$\weightC(\play)\geq 1$ (resp.~$\weightC(\play)\leq -1$).

\begin{definition}\label{def:divergent}
	A \WTG is divergent if every cyclic region path is positive or negative.
\end{definition}

Finally, with loss of generality,  we only consider divergent \WTG{s} 
containing no configurations with a value equal to $-\infty$. 
Intuitively, guaranteeing a value $-\infty$ 
resembles a Büchi condition for $\MinPl$, since this means that \MaxPl cannot avoid the iteration of 
negative cycles with his delays.
In the robust settings, testing Büchi condition for automata is already 
non-trivial~\cite{Busatto-GastonM19}, thus we remove this behaviour in our games in this article.
Since the value is a lower bound of the robust value 
(by Lemma~\ref{lem:rVal-monotony}), we obtain that all locations have 
a robust value distinct from $-\infty$. 
Moreover, testing if such a location exists in a divergent \WTG 
can be done in \EXP~\cite{BusattoGastonMR23}. 
Our second contribution is to extend the symbolic algorithm 
used in the case of acyclic \WTG{s} to compute the robust value in 
this subclass of divergent \WTG{s}.

\begin{restatable}{theorem}{thmDiv}
	\label{thm:rob-limit_div}
	The robust value problem is decidable over the subclass 
	of divergent \WTG{s} without configurations of value 
	$-\infty$. 
\end{restatable}
\begin{proof}[Sketch of the proof]
	We compute the robust value by using an adaptation of the 
	algorithm of~\cite{BusattoGastonMR23} used to compute the (exact) 
	value function in divergent \WTG{s}. 
	In particular, its termination is guaranteed by the use
	of an equivalent definition of divergent \WTG requiring that for all strongly 
	connected components (SCC) $S$ of the graph of the region game, 
	either every cyclic region path~$\rpath$ inside $S$ is positive 
	(we say that the SCC is positive) or every cyclic region path 
	~$\rpath$ inside $S$ is negative (we say that the SCC is negative).
	
	We adapt this argument in the case of the computation of 
	the robust value of a divergent \WTG 
	(without configurations with a value equal to $-\infty$). 
	In particular, we observe that if a cyclic region path  is positive 
	(resp.~negative) w.r.t.~the exact semantics, then it is also positive 
	(resp.~negative) w.r.t.~the conservative semantics, as the latter only 
	filters some plays. 
	Thus, the finite convergence of the value iteration algorithm 
	(defined by $\F_\perturbation$ as for acyclic \WTG, 
	i.e.~initialised by the function $\V^0$ defined such that 
	$\V^0(\loc) = 0$ for all target locations, and $\V^0(\loc) = +\infty$ otherwise) 
	is guaranteed by its finite convergence in finite time in each positive 
	(resp.~negative) SCC. 
	Intuitively, in a positive (resp.~negative) SCC, the interest of \MinPl 
	(resp.~\MaxPl) is to quickly reach a target location of $\game$ to minimise 
	the number of positive (resp.~negative) cyclic region paths followed along the play
	that allow us to upperbound the number of iterations needed to obtain the robust value
	of all locations.
	Thus, the number of iterations needed to compute 
	the robust value in a divergent \WTG is defined by the sum of 
	the number of iterations for each SCC along the longest path 
	of the SCC decomposition. 
%	Details of this proof are given in Appendix~\ref{app:rob-limit_div}.
\end{proof}

On top of computing the value, the modified algorithm allows one 
to synthesize almost-optimal strategies 
(we can adapt recent works~\cite{MonmegeParreauxReynier-ICALP21} 
showing that those strategies can be taken among switching strategies 
for \MinPl and memoryless strategies for \MaxPl).

\section{Conclusion}

\todo{BM: en fait, on avait déjà écrit quelque chose pour la complexité... on semblait vachement optimiste ;-)}

%\smallskip
%\noindent
%\textbf{Complexity analysis.}
This article allows one to compute (finite) robust values of weighted timed games 
in classes of games (acyclic and divergent) where the non-robust values are 
indeed computable.
 
As future works, we would like to carefully explore the exact complexity of our 
algorithms. Intuitively, each operator used to describe $\F_\perturbation$ can be 
computed in exponential time with respect to the set of cells and the size of 
$\eta$. By~\cite{BusattoGastonMR23}, the number of cells exponentially grows at each 
application of $\F_\perturbation$ (so it is doubly exponential for the whole 
computation) and the constants in affine expressions polynomially grow, in the 
non-robust setting. We hope that such upperbounds remain in the robust setting. This 
would imply that, for divergent \WTG{s}, our algorithm requires a triply-exponential 
time, since the unfolding is exponential in the size of $\game$. 
\todo{BM: j'ai supprimé (par manque de place) le second paragraphe de conclusion qui parlait d'autoriser les locations de valeur $-\infty$, mais aussi d'approximation dans le cas almost-divergent, et le cas d'une horloge... si on a de la place, on peut remettre le paragraphe}

 As future works, we also suggest to extend the setting to incorporate
 divergent \WTG{s} that contains location with a value equal to 
 $-\infty$. However, fixing it for all divergent \WTG{s} seems to be difficult since, 
 intuitively, the condition to guarantee $-\infty$ looks like a Büchi condition 
 where \MaxPl can avoid the iteration of cycles with his delays.
 Moreover, we would to study almost-divergent weighted timed games (studied in 
 \cite{BusattoGastonMonmegeReynier-18,BusattoGastonMR23}), a class of games 
 undecidable, but with approximable value functions. We wonder if the robust values 
 could also be approximated by similar techniques. Another direction of research 
 would be to consider the fragment of one-clock weighted timed games, another class 
 of games where the value function is known to be computable (for a long time in the 
 non-negative case \cite{BouyerLarsenMarkeyRasmussen-06}, very recently in the 
 general case \cite{MonPar22}). The difficulty here would be that encoding the 
 conservative semantics in an exact semantics, even with fixed-perturbation, requires 
 the addition of a clock, thus exiting the decidable fragment. The question thus 
 becomes a possible adaptation of techniques used previously to solve non robust 
 one-clock \WTG{s} to incorporate directly the robustness.

\bibliographystyle{plainurl}

\newpage

\appendix

\section{Proof of Lemma~\ref{lem:sem-strat_MinMax}}
\label{app:sem-strat_MinMax}

\lemSemstratMinMax*
\begin{proof}
	\begin{enumerate}
		\item Let $\robminstrategy \in \rStratMin{\perturbation}$
		be a strategy of $\MinPl$ in $\game$, and
		$\play \in \FPlays[\perturbation']$ be a finite play in $\game$
		such that $\play \in \FPlays[\perturbation]$.
		Otherwise, $\play \notin \FPlays[\perturbation]$, 
		and $\robminstrategy(\play)$ is not defined.
		Now, by definition of the strategy, we have
		$\robminstrategy(\play) = (\move)$ 
		such that $(\loc, \val) \xrightarrow{\move} (\loc, \val, \move) \in 
		\sem[\perturbation]{\game}$ where $\last(\play) = (\loc, \val)$, 
		i.e. $\val + \delay \models \guard$ and 
		$\val + \delay + 2\perturbation \models \guard$ 
		where $\guard$ is the guard of $\trans$.
		Since valuations that satisfies a guard defined an interval, we conclude that
		$\val + \delay \models \guard$ and 
		$\val + \delay + 2\perturbation' \models \guard$ by 
		$\val + \delay \leq \val + \delay + 2\perturbation' < \val + \delay + 2\perturbation$.
		Thus, $(\loc, \val) \xrightarrow{\move} (\loc, \val, \move) \in 
		\sem[\perturbation']{\game}$, i.e. $\robminstrategy \in \rStratMin{\perturbation'}$.
		
		\item The proof consists in proving the following claim:
		\begin{equation}
		\label{eq:rob-val_strat-Max}
		\forall \robmaxstrategy' \in \rStratMax{\perturbation'}~
		\exists \robmaxstrategy \in \rStratMax{\perturbation},~
		\outcomes((\loc, \val), \robminstrategy, \robmaxstrategy) =
		\outcomes((\loc, \val), \robminstrategy, \robmaxstrategy')
		\tag*{$(\star)$}
		\end{equation}
		Indeed, we can conclude since,
		for all $\robmaxstrategy' \in \rStratMax{\perturbation'}$, we~have
		\begin{displaymath}
		\sup_{\robmaxstrategy \in \rStratMax{\perturbation}}
		\weightP(\outcomes((\loc, \val), \robminstrategy, \robmaxstrategy)) \geq
		\weightP(\outcomes((\loc, \val), \robminstrategy, \robmaxstrategy)) =
		\weightP(\outcomes((\loc, \val), \robminstrategy, \robmaxstrategy')) \,.
		\end{displaymath}
		Now, by letting $\robmaxstrategy' \in \rStratMax{\perturbation'}$,
		we prove~\ref{eq:rob-val_strat-Max} by defining
		$\robmaxstrategy \in \rStratMax{\perturbation}$ as an extension
		of $\robmaxstrategy'$, i.e.\ for all finite play
		$\play \in \FPlaysMax[\perturbation]$,
		we~fix
		\begin{displaymath}
		\robmaxstrategy(\play) =
		\begin{cases}
		\robmaxstrategy'(\play) & \text{if $\play \in \FPlaysMax[\perturbation']$;} \\
		\edge & \text{where $\edge \in \Edges(\last(\play))$, otherwise;}
		\end{cases}
		\end{displaymath}
		where $\Edges(\last(\play))$ is the set of edges of $\sem[\perturbation]{\game}$ 
		from $\last(\play)$.
		In particular, we conclude by induction on the length of the play induced by
		$\robminstrategy$ and $\robmaxstrategy$.
		The property is trivial when $\play = (\loc, \val)$.
		Otherwise, we fix $\play = \play_1 \moveto{\move} \states \in \FPlays[\perturbation]$
		be the play conforming to $\robminstrategy$ and $\robmaxstrategy$ of length $i$
		from $(\loc, \val)$, and
		$\play' = \play_1' \moveto{\moveP} \states \in \FPlays[\perturbation']$ be the play
		conforming to $\robminstrategy$ and $\robmaxstrategy'$ of length $i$ from $(\loc, \val)$
		(we note that such a play exists since
		$\robminstrategy \in \rStratMin{\perturbation'}$, by the previous property).
		Thus, by the hypothesis of induction, $\play_1 = \play_1'$, thus we need to
		prove that $(\move) = (\moveP)$.
		In particular, if $\last(\play_1) \in \ConfMin$, then the property holds since we apply the
		same strategy in both cases.
		Otherwise, since $\play_1  = \play_1' \in \FPlaysMax[\perturbation']$,
		we conclude by definition of $\robmaxstrategy$.
		\qedhere
	\end{enumerate}
\end{proof}

\section{Proof of Proposition~\ref{prop:reach-cons}}
%Encoding the conservative semantics in the excessive semantics}
\label{app:rob-val_reach-reduction}

\propReachCons*

Before to provide the proof of this proposition, we first formally recall the
definition of the excessive semantics, where we reuse the sets 
$\StatesMin$, $\StatesMax$, $\StatesT$, $\EdgesMax$, $\EdgesRob$ 
and $\weight$ of Definition~\ref{def:sem-squelette}:

\begin{definition}
	\label{def:sem-excess}
	Let $\game = \gameEx$ be a \WTG.
	The \emph{excessive semantics} of $\game$ of perturbation $\perturbation > 0$ 
	is the transition system
	$\sem[\perturbation]{\game}_{\excess} = \tuple{\States, \Edges, \weight}$ only differing from $\sem[\perturbation]{\game}$ by the set 
	$\EdgesMin' = \left\{\Big((\loc, \val) \moveto{\move}
	(\loc, \val, \move)\Big) ~|~ \loc \in \LocsMin, 
	\val + \delay \models \guard  \right\}$
	where $\trans = (\loc, \guard, \reset, \loc') \in \Trans$.
\end{definition}

\begin{figure*}
	\centering
	\begin{tikzpicture}[xscale=.8,every node/.style={font=\scriptsize},
	every label/.style={font=\scriptsize}]
	% Draw the states
	\node[target] at (1.75, 1) (cons) {$\game$: conservative semantics};

	\node[PlayerMin, label={above:$\loc$}] at (0, 0) (s1) {$\mathbf{w_0}$};
	\node[Player, label={above:$\loc'$}] at (3.5, 0)  (s0) {$\mathbf{w_1}$};
	%	\node at (7.5, 0) (s) {$\rightsquigarrow$};
	
	% Connect the states with arrows
	\draw[->] (s1) edge node[above] {$\trans =(\loc, \guard, \reset, \loc')$} node[below] {$\mathbf{w}$} (s0);
	\draw[->, decorate, decoration=snake] (4.5, 0) -- (6.5, 0);
	
	\begin{scope}[xshift=7.5cm]
	\node[PlayerMin, label={above:$\loc$}] at (0, 0) (s1) {$\mathbf{w_0}$};
	\node[PlayerMax, label={above:$\loc^{\trans}$}] at (4, 0) (s2) {\textsf{urgent}};
	\node[Player, label={above:$\loc'$}] at (8, 0) (s0) {$\mathbf{w_1}$};
	\node[target] at (4, -1.5) (s3) {\Large $\frownie$};
	\node[target] at (4, 1) (excess) {$\gameexcess$: excessive semantics};
	
	% Connect the states with arrows
	\draw[->]
	(s1) edge node[above] {$(\loc, \text{true},\emptyset,\loc^{\trans})$} node[below] {$\mathbf{w}$} (s2)
	(s2) edge node[above] {$(\loc^{\trans},\guard, \reset,\loc')$} node[below] {$\mathbf{0}$} (s0)
	(s2) edge node[right] {$(\loc^{\trans}, \neg\guard, \emptyset, \frownie)$} node[left] {$\mathbf{0}$} (s3)
	;
	\end{scope}
	\end{tikzpicture}
	\caption{Gadget used to define $\gameexcess$ from $\game$. We detail how a transition
		$\trans =(\loc, \guard, \reset, \loc')$ starting in a location belonging to \MinPl is transformed.
		Symbols  $w, w_0, w_1$ denote weights from~$\game$. The new location $\loc^\trans$
		of \MaxPl is urgent and tests the guard after the perturbation 
		(as in the conservative semantics).}
	\label{fig:valPb-cons2excess_gadget2}
\end{figure*}

As explain before, the proof consists on a reduction from 
the conservative semantics to the excessive one.
Intuitively, the conservative semantics imposes that the guards on transitions
of \MinPl are tested after applying the perturbation.
The reduction exploits this idea: all transitions of \MinPl in
$\game$ will be modified by adding a new 
\emph{urgent location}\footnote{An urgent location is a location where time elapsing is not allowed. 
	We can model it with a new clock that is reset before entering the urgent location 
	and will be equal to $0$ in all guards of its outgoing transitions. 
	In particular, we use the urgent locations in this reduction to simplify its proof of correctness. 
	Indeed, with urgent locations, $\game$ and $\gameexcess$  have the same number of clocks, 
	i.e. valuations in both games are the same dimension.} 
of \MaxPl (to forbid modification of the delay) dedicated to check the delay
chosen by \MinPl (after perturbation).
This gadget is depicted on \figurename{~\ref{fig:valPb-cons2excess_gadget2}}.
If \MinPl plays a decision that is not robust w.r.t. the conservative semantics
(that is possible since we ask for a robust decision w.r.t. the excessive one),
then \MaxPl wins by reaching $\frownie$.
Otherwise, \MinPl chooses a robust decision w.r.t. the conservative semantics and
\MaxPl must apply the original transition.
Formally,

\begin{proposition}
	\label{prop:valPb-cons2excess}
	Let $\game$ be a \WTG.
	We can build in polynomial time a \WTG $\gameexcess$ 
	such that for all configurations $(\loc, \val)$, 
	and all perturbation perturbation $\perturbation > 0$,
	\MinPl has a winning strategy from $(\loc, \val)$ in 
	$\sem[\perturbation]{\game}$
	iff she has a winning strategy from $(\loc, \val)$ in 
	$\sem[\perturbation]{\gameexcess}_{\excess}$.
\end{proposition}

Using this reduction, we can solve the existential robust reachability problem (for the
conservative semantics), and obtain the following result that directly imply Proposition~\ref{prop:reach-cons}:
\begin{proposition}
	\label{prop:reach-cons2}
	The existential robust reachability problem is \EXP-complete. 
	
	In addition, if the problem has a positive answer, then there exists a bound $A$
	exponential in the size of the \WTG such that for every small enough
	perturbation $\perturbation>0$,
	\MinPl has a winning strategy in $\sem[\perturbation]{\game}$ ensuring 
	to reach the target within at most $A$ steps.
\end{proposition}
\begin{proof}
	Let $\game$ be a \WTG. The \EXP membership directly follows from 
	Proposition~\ref{prop:valPb-cons2excess} 
	and~\cite{BouyerMarkeySankur-15} that shows the decidability of the existential
	robust reachability problem for the excessive semantics in exponential time.
	To prove the \EXP hardness, we apply the reduction from the halting problem of
	linear-bounded alternating Turing machines from~\cite{OualhadjReynierSankur-14}.
	
	Regarding the second statement of the Proposition, 
	\cite{BouyerMarkeySankur-15} also shows that
	if the existential robust reachability problem for a \WTG $\game'$
	under excessive semantics admits a positive answer, then there
	exists some positive parameter and a strategy
	for \MinPl which ensures to reach the target within at most
	$A$ steps, where $A$ is exponential in the size of $\game'$.
	For details, see the proof of Proposition 4.1 of~\cite{BouyerMarkeySankur-15}.
	To conclude, we observe that the proof of Proposition~\ref{prop:valPb-cons2excess} shows,
	given some perturbation $\perturbation$ and some bound $A$,
	that if \MinPl has a winning strategy ensuring to reach the target within at most $A$ steps in 
	$\sem[\perturbation]{\gameexcess}_{\excess}$, then so does she in 
	$\sem[\perturbation]{\game}$.
\end{proof}

In the rest of this Appendix, we give the proof of Proposition~\ref{prop:valPb-cons2excess}. 
In particular, we provide the formal construction of the game $\gameexcess$ from a game $\game$. 
To do it, we need to slightly modify the definition of a \WTG by allowing \emph{urgent locations}
under the excessive semantics. 
Formally, a \emph{\WTG with urgent locations} is a tuple 
$\game=\tuple{\LocsMin, \LocsMax, \LocsT, \LocsUrg,	\Cl, \Transexcess, \weightexcess}$
where $\gameEx$ is a \WTG and $\LocsUrg \subseteq \Locs$ is a set of urgent locations.
The usage of urgent location modifies the set of edges of the semantics of the \WTG. 

\begin{definition}
	Let $\game = \tuple{\LocsMin, \LocsMax, \LocsT, \LocsUrg, \Cl, \Transexcess, \weightexcess}$ 
	be a \WTG with urgent locations. 
	For $\perturbation \geq 0$, we define a transition system
	$\sem[\perturbation]{\game}_{\excess} = \tuple{\States, \Edges, \weight}$
	only differing from the excessive semantics of $\gameEx$ by the set 
	$\Edges$ defined such that, when $\loc \in \LocsUrg$, then for all 
	valuations $\val$, $(\loc, \val) \moveto{\move} (\loc', \val') \in \Edges(\loc, \val)$ 
	if and only if $\delay = 0$.
\end{definition}

Now, we have equipped to lead to the following definition of $\gameexcess$:
\begin{definition}
	\label{def:valPb-gameexcess}
	Let $\game = \gameEx$ be a \WTG and
	$\Trans_\MinPl = \{\trans \mid \trans = (\loc, \guard, \reset, \loc')
	\in \Trans \text{ with } \loc \in \LocsMin\}$.
	We let $\gameexcess =\tuple{\LocsMin^{\mathsf{e}}, \LocsMax^{\mathsf{e}}, \LocsT, 
		\LocsUrg^{\mathsf{e}}, \Cl, \Transexcess, \weightexcess}$ be a \WTG with urgent locations where:
	$\LocsMin^{\mathsf{e}}   = \LocsMin \cup \{\frownie\}$,
	$\LocsMax^{\mathsf{e}}  = \LocsMax \cup \{\loc^{\trans} \mid \trans \in \Trans_\MinPl\}$,
	$\LocsUrg^{\mathsf{e}}  = \{\loc^{\trans} \mid \trans \in \Trans_\MinPl\}$, and 
	$\Transexcess  =  \Transexcess_{\MinPl} \uplus \Transexcess_{\MaxPl}
	\uplus \Transexcess_{\cons'_\perturbation} \uplus \left\{(\frownie, \text{true}, \emptyset, \frownie) \right\}$
	such that 
	\begin{align*}
	\Transexcess_{\MaxPl}
	= & \left\{(\loc, \guard, \reset, \loc')
	\mid (\loc, \guard, \reset, \loc') \in \Trans \text{ with }
	\loc \in \LocsMax \right\} \\
	\Transexcess_{\MinPl}
	= & \left\{(\loc, \text{true}, \emptyset, \loc^{\trans}) \mid
	\trans = (\loc, \guard, \reset, \loc') \in \Trans_\MinPl \right\}\\
	\Transexcess_{\rob}
	= & \left\{(\loc^{\trans}, \guard, \reset, \loc') \mid
	\trans = (\loc, \guard, \reset, \loc') \in
	\Trans_\MinPl \right\} \cup \left\{(\loc^{\trans}, \neg\guard, \emptyset, {\frownie}) \mid
	\trans = (\loc, \guard, \reset, \loc') \in \Trans_\MinPl \right\}
	\end{align*}
	where $\neg\guard$ is the complement
	\footnote{Formally, the complement of a guard is not a guard, as a guard
		is a conjunction of atomic constraints. However, it can be represented
		as a disjunction of guards, which can be dealt with using multiple
		transitions.}
	%	\footnote{We recall that a guard
	%		is a conjunction of atomic constraints of the form $x \bowtie a$ with
	%		$\bowtie~\in \{\leq, \allowbreak < , >, \geq\}$.
	%		Thus, the complement of a guard is defined by the disjunction
	%		of the negation of each atomic constraint that defines the guards.
	%		However, the complement of a guard is not a guard
	%		(since it is a conjunction of atomic constraints).
	%		To avoid this pitfall, the complementary of a guard defines a
	%		set of (at most $2|\Cl|$) transitions
	%		defined by each atomic constraint of the disjunction.
	%		For instance, the transition with the guard $a \leq x < b$
	%		introduces two complement transitions with the guard
	%		$x < a$ and $b \leq x$.}
	of $\guard$, and $\weightexcess: \Locsexcess \cup \Transexcess \to \Z$ be the weight
	function defined for all locations $\loc \in \Locsexcess$ and
	for all transitions $\trans = (\loc, \guard, \reset, \loc') \in \Transexcess$ by
	$\weightexcess(\loc) =\weight(\loc)$ if $\loc \in \Locs$, and $0$ otherwise, and
	by: 
	\begin{displaymath}
	\weightexcess(\trans) =
	\begin{cases}
	\weight(\trans)
	& \text{if $\loc \in \LocsMax$} \\
	\weight(\trans') & \text{if $\loc \in \LocsMindelta$ and $\loc' = \loc^{\trans'}$} \\
	0 & \text{otherwise.}
	\end{cases}
	\end{displaymath}
\end{definition}

This construction can be done in polynomial time.
We now turn to its correction.
Consider  a perturbation $\perturbation > 0$
and a configuration $(\loc, \val)$. We  have to prove that the following 
properties are equivalent:
	\begin{enumerate}
		\item\label{itm:valPb-cons2excess_game} 
		\MinPl has a winning strategy in $\sem[\perturbation]{\game}$ from $(\loc, \val)$;
		\item\label{itm:valPb-cons2excess_gameexcass} 
		\MinPl has a winning strategy in $\sem[\perturbation]{\gameexcess}_{\excess}$
		from $(\loc, \val)$.
	\end{enumerate}

\text{}

In this proof, we will consider two different semantics, the conservative
and the excessive one. 
We naturally extend all notions of plays and strategies introduced for
the conservative semantics to the excessive one. 
In particular, we denote by $\FPlays[\excess_\perturbation]$ the set of plays 
and $\rStrat{\excess_\perturbation}$ the 
set of strategies w.r.t. the excessive semantics. 
Recall that we
denote by $\FPlays[\perturbation]$ the set of plays 
and $\rStrat{\perturbation}$ the 
set of strategies w.r.t. the conservative semantics.

We will also say that a decision (i.e. a transition) is robust for $\cons_\perturbation$
(resp. $\excess_{\perturbation}$) if it is well-defined in the corresponding semantics.
The following property directly follows from the definition of the conservative
and excessive semantics, as the last one is more restrictive than the former one.

\begin{lemma}
	\label{lem:valPb-translated-cons2excess}
	Let $\game$ be a \WTG and a robust decision $(\move)$ for $\cons_\perturbation$ 
	with $\perturbation > 0$, from a configuration $(\loc, \val)$.
	Then, $(\move)$ is robust for the $\excess_{\perturbation}$ from $(\loc, \val)$.
\end{lemma}

\paragraph*{First implication: a winning strategy in $\sem[\perturbation]{\game}$ 
	implies a winning strategy in $\sem[\perturbation]{\gameexcess}_\excess$}
To prove this implication, we 
 consider $\robminstrategy \in \rStratMin{\perturbation}$ 
be a winning strategy for \MinPl in $\sem[\perturbation]{\game}$, and we define
$\robminstrategy^{\mathsf{e}} \in \rStratMin{\excess_\perturbation}$ 
be a strategy for \MinPl in $\sem[\perturbation]{\gameexcess}_\excess$. 
In particular, we prove that $\robminstrategy^{\mathsf{e}}$ is winning.
To define this strategy, we use the (partial) function
$\proj \colon \FPlays[\excess_{\perturbation}]_{\gameexcess} \to
\FPlays[\perturbation]_{\game}$ on finite plays in
$\FPlays[\excess_{\perturbation}]_{\gameexcess}$ starting from $(\loc, \val)$
and ending in a state of $\sem[\perturbation]{\game}$.
In particular, $\proj$ is defined on finite plays that do not reach the location $\frownie$.
By construction of $\gameexcess$, except configurations around $\loc^\trans$,
edges along $\play$ are ones of $\sem[\perturbation]{\game}$:
all decisions applied in $\play$ are robust for $\cons_\perturbation$,
or feasible for \MaxPl.
Thus, the projection function consists in removing the configurations around $\loc^\trans$.
Formally, for all finite plays $\play \in \FPlays[\excess_{\perturbation}]_{\gameexcess}$ 
%that end in a configuration of $\game$
starting in  $(\loc, \val)$ and ending
in a state of $\sem[\perturbation]{\game} = \tuple{\States, \Edges}$,
we inductively define:
\begin{displaymath}
\proj(\play) =
\begin{cases}
(\loc, \val) & \text{if $\play = (\loc, \val)$;} \\
\proj(\play_1) \moveto{\move} \states &
\text{if $\play = \play_1 \moveto{\move} \states$ with $\last(\play_1) \in \States$;} \\
\proj(\play_1) \moveto{\trans, \perturbation} (\loc', \val') &
\text{if $\play = \play_1 \moveto{\trans', \perturbation}
	(\loc^{\trans}, \val) \moveto{\trans, 0} (\loc', \val')$.}
\end{cases}
\end{displaymath}
Moreover, by definition of $\proj$, 
we remark that $\proj$ fulfils the following property:
\begin{lemma}
	\label{lem:valPb-cons-proj}
	Let $\play \in \FPlays[\excess_{\perturbation}]_{\gameexcess}$
	be a finite play ending in a state of $\sem[\perturbation]{\game}$.
	Then, $\proj(\play) \in \FPlays[\perturbation]_{\game}$
	and $\last(\play) = \last(\proj(\play))$.
\end{lemma}

Now, we have tools to define $\robminstrategy^{\mathsf{e}}$:
for all finite plays $\play \in \FPlays[\excess_{\perturbation}]_{\game}$
starting in $(\loc, \val)$ and ending in a configuration of \MinPl, we let
\begin{displaymath}
\robminstrategy^{\mathsf{e}}(\play) =
(\trans', \delay) \qquad
\text{if $\robminstrategy(\proj(\play)) = (\move)$} \,.
\end{displaymath}
We note that this strategy chooses robust decisions for $\excess_\perturbation$
since guards in transitions starting in a location of \MinPl in $\gameexcess$ are equal
to $\text{true}$.
Moreover, since the definition of $\robminstrategy^{\mathsf{e}}$
relies on $\proj$, it is no surprise that:
\begin{lemma}
	\label{lem:valPb-cons-proj-conf}
	Let $\play \in \FPlays[\excess_{\perturbation}]_{\gameexcess}$
	conforming to $\robminstrategy^{\mathsf{e}}$. Then,
	\begin{enumerate}
		\item\label{itm:valPb-cons-proj-conf1} $\play$ can not reach $\frownie$;
		\item\label{itm:valPb-cons-proj-conf2} if $\proj(\play)$ is defined (i.e. $\last(\play) \in S$),
		then $\proj(\play)$ is conforming to $\robminstrategy$.
	\end{enumerate} 
\end{lemma}
\begin{proof}
	We reason by induction on the length of finite plays
	$\play \in \FPlays[\excess_{\perturbation}]_{\gameexcess}$
	conforming to starting in $(\loc, \val)$. 
	If $\play = (\loc, \val)$ then $\proj(\play) = (\loc, \val)$, and both properties trivially hold.
	Otherwise, we suppose that $\play = \play_1 \moveto{\move} \states$ and we prove both properties. 
	\begin{enumerate}
		\item By contradiction, we suppose that $\states = (\frownie, \val)$ for a valuation $\val$.
		In particular, by definition of $\gameexcess$, $\last(\play_1) = (\loc^\trans_1, \val')$ such that 
		$\trans = (\loc_1, \guard, \reset, \loc_1') \in \Trans_\MinPl$ and $\val' \not\models g$. 
		Now, since $\play$ start a configuration of $\game$, we can write 
		$\play_1 = \play_2 \moveto{\trans', \delay_1} (\loc, \val, \trans', \delay_1) 
		\moveto{\trans', \varepsilon} (\loc^\trans_1, \val')$ where 
		$\last(\play_2) = (\loc_2, \val_2)  \in \ConfMin[\game]$ (by definition of $\gameexcess$).
		Moreover, as $\play_1$ is conforming to $\robminstrategy^{\mathsf{e}}$ and does not reach $\frownie$, 
		we know that $\robminstrategy^{\mathsf{e}}(\play_2) = (\trans', \delay_1)$, 
		i.e. $\robminstrategy(\proj(\play_2)) = (\trans, \delay_1)$.
		Since $\robminstrategy$ is a strategy for \MinPl in $\game$ (under the conservative semantics), 
		then for all perturbations $\varepsilon \in [0, 2\perturbation]$, 
		$\val_2 + \delay + \varepsilon \models \guard$, i.e. $\val' \models \guard$.
		Thus, we obtain a contradiction.

		\item We distinguish two cases according the last state of $\play_1$. 
		First, if $\last(\play_1) \in S$, then $\proj(\play_1)$ is defined.
		Since $\proj(\play_1)$ is conforming to $\robminstrategy$
		(by the hypothesis of induction) and $\last(\play_1) =\last(\proj(\play_1))$
		(by Lemma~\ref{lem:valPb-cons-proj},
		we conclude that $\proj(\play)$ is conforming to $\robminstrategy$
		when $\last(\play_1) \notin \ConfMin[\game]$.
		In particular, we suppose that $\last(\play_1) \in \ConfMin[\game]$,
		and $\proj(\play) = \proj(\play_1) \moveto{\move} \states$.
		We note that by definition of $\robminstrategy^{\mathsf{e}}$,
		the decision chosen by the strategy $\robminstrategy$ for $\proj(\play)$ is $(\move)$.
		Thus, $\proj(\play)$ is conforming to~$\robminstrategy$.
		
		Otherwise, we suppose that $\last(\play_1) \notin S$ and by the previous property, we know that 
		$\last(\play_1) = (\loc^\trans_1, \val')$ such that 
		$\trans = (\loc_1, \guard, \reset, \loc_1') \in \Trans_\MinPl$.
		Now, since $\play$ start a configuration of $\game$, we can write 
		$\play_1 = \play_2 \moveto{\trans', \varepsilon} (\loc^\trans_1, \val')$ where 
		$\last(\play_2) \in S \setminus \Conf[\game]$ (by definition of $\gameexcess$).
		To conclude, we note that $\proj(\play) =\proj(\play_2) \moveto{\trans, \varepsilon} s$	
		with $\last(\proj(\play_2))$ be a state of \MaxPl (in $\sem[\perturbation]{\game}$). 
		Thus, $\proj(\play)$ is conforming to $\robminstrategy$ since $\proj(\play_2)$ is 
		conforming to $\robminstrategy$ by hypothesis of induction. 	
		\qedhere
	\end{enumerate}
\end{proof}

By contradiction, we suppose that $\robminstrategy^{\mathsf{e}}$ is not a winning strategy.
In particular, we suppose that there exists a maximal play $\play$ conforming to 
$\robminstrategy^{\mathsf{e}}$ that does not reach a target location. 
Moreover, by Lemma~\ref{lem:valPb-cons-proj-conf}.\ref{itm:valPb-cons-proj-conf1}, 
we know that $\play$ can neither reached $\frownie$ nor ends in an urgent location of \MaxPl
(they do not introduce new deadlock). 
In particular, if $\play$ is a finite play, its ends in a state of $\sem[\perturbation]{\game}$. 
Thus, by Lemma~\ref{lem:valPb-cons-proj}, we deduce that $\proj(\play)$ 
does not reach a target location (we note that if $\play$ is infinite, $\proj$ is defined on 
all its finite prefix ending in a state of $\sem[\perturbation]{\game}$). 
Since $\proj(\play)$ is conforming to $\robminstrategy$ 
(by Lemma~\ref{lem:valPb-cons-proj-conf}.\ref{itm:valPb-cons-proj-conf2}), 
we deduce that $\robminstrategy$ is not winning.
Thus, we obtain a contradiction.

\paragraph*{Second  implication: a winning strategy in 
$\sem[\perturbation]{\gameexcess}_\excess$ 
	implies a winning strategy in~$\sem[\perturbation]{\game}$}
Conversely, consider 
a (given) winning strategy $\robminstrategy \in \rStratMin{\excess_\perturbation}$
in $\sem[\perturbation]{\gameexcess}_\excess$.
We define a strategy
$\robminstrategy^{\mathsf{c}} \in \rStratMin{\perturbation}$
for \MinPl in $\sem[\perturbation]{\game}$ which 
 is winning.
To do it, we define a function that encodes plays in $\game$ w.r.t.
conservative semantics into plays in $\gameexcess$
w.r.t. excessive semantics.
Since the interval of possible perturbations is the same in both semantics,
all robust decisions for the conservative semantics in $\game$ are
robust decisions for the excessive semantics in $\gamedelta$ where
all guards from a location of \MinPl contain no constraints.
Formally, we define, by induction on the length of finite plays,
$\inj : \FPlays[\perturbation]_{\game} \to
\FPlays[\excess_{\perturbation}]_{\gameexcess}$
such that for all finite plays $\play \in \FPlays[\perturbation]_{\game}$, 
we let
\begin{displaymath}
\inj(\play) = 
\begin{cases}
\states & \text{if $\play = \states$;} \\
\inj(\play_1) \moveto{\move} \states &
\text{if $\play = \play_1 \moveto{\move} \states$ with $\last(\play_1) \in \Conf[\game]$;} \\
\inj(\play_1) \moveto{\trans', \perturbation} (\loc^{\trans}, \val) \moveto{\trans, 0} \states&
\text{if $\play = \play_1 \moveto{\trans, \perturbation} \states$ with
	$\last(\play_1) = (\loc, \val, \move)$.}
\end{cases}
\end{displaymath}
By definition of $\inj$, we remark that $\inj$ fulfils the following property:
\begin{lemma}
	\label{lem:valPb-cons-inj}
	Let $\play \in \FPlaysG[\perturbation]$ be a finite play ending 
	in a configuration of $\game$.
	Then, $\inj(\play) \in \FPlays[\excess_{\perturbation}]_{\game}$
	and $\last(\play) = \last(\inj(\play))$.
\end{lemma}

Now, we have tools to define $\robminstrategy^{\mathsf{c}}$: for all finite plays
$\play \in \FPlays[\perturbation]_{\game}$, we let
\begin{displaymath}
\robminstrategy^{\mathsf{c}}(\play) = (\move)  \qquad
\text{if $\robminstrategy(\inj(\play)) = (\trans', \delay)$}
\end{displaymath}
such that $\trans' = (\last(\play), \text{true}, \emptyset, \loc^{\trans})$
and $(\move)$ is a robust decision for the conservative semantics
from $\last(\play)$.
We note, by definition, that this strategy chooses a robust decision
for the conservative semantics.
Moreover, since the definition of $\robminstrategy^{\mathsf{c}}$
relies on $\inj$, we can show:
\begin{lemma}
	\label{lem:valPb-cons-inj-conf}
	Let $\play \in \FPlays[\perturbation]_{\game}$
	conforming to $\robminstrategy^{\mathsf{c}}$.
	Then, $\inj(\play)$ is conforming to~$\robminstrategy$.
\end{lemma}
\begin{proof}
	We reason by induction on the length of finite plays
	$\play \in \FPlays[\perturbation]_{\game}$.
	If $\play = \states$ then $\inj(\play) = \states$, and the property trivially holds.
	Otherwise, we suppose that $\play = \play_1 \moveto{\move} \states$.
	Since $\proj(\play_1)$ is conforming to $\robminstrategy$
	(by the hypothesis of induction) and $\last(\play_1) = \last(\inj(\play_1))$
	(by Lemma~\ref{lem:valPb-cons-inj}),
	we conclude that $\inj(\play)$ is conforming to $\robminstrategy$
	when $\last(\play_1) \notin \ConfMin[\game]$.
	In particular, we suppose that $\last(\play_1) \in \ConfMin[\game]$,
	$\inj(\play) = \inj(\play_1) \moveto{\move} \states$.
	To conclude, we remark that $(\move)$ is robust for the conservative semantics
	in $\game$ since $\robminstrategy$ is supposed to win.
	Otherwise, \MaxPl can choose a perturbation such that $\frownie$ was reached,
	which contradicts that $\robminstrategy$ is winning.
\end{proof}

By contradiction, we suppose that $\robminstrategy^{\mathsf{c}}$ 
is not a winning strategy.
In particular, we suppose that there exists a maximal play $\play$ conforming to 
$\robminstrategy^{\mathsf{c}}$ that does not reach a target location. 
By Lemma~\ref{lem:valPb-cons-inj}, we deduce that $\inj(\play)$ does not reach a
target location. 
Since $\inj(\play)$ is conforming to $\robminstrategy$ 
(by Lemma~\ref{lem:valPb-cons-inj-conf}), we deduce that $\robminstrategy$ 
is not winning.
Thus, we obtain a contradiction which conclude the proof of 
Proposition~\ref{prop:valPb-cons2excess}.

\section{Conservative semantics is equivalent to the ``classical'' conservative semantics}
\label{app:valPb-fixed-rob-div_translated}

In the literature, the conservative semantics is defined such that 
\MaxPl can choose a perturbation in $[-\perturbation, \perturbation]$ 
instead in $[0, 2\perturbation]$ in our case. 
Even if this semantics change a little the possible choices of \MinPl, 
we prove that their robust values are always equal. 
Formally, we define the ``classical'' conservative semantics, given 
in~\cite{BouyerMarkeySankur-13, BouyerMarkeySankur-15, GuhaKrishnaManasaTrivedi-15} by:

\begin{definition}
	\label{def:sem-cons-classic}
	Let $\game = \gameEx$ be a \WTG.
	The \emph{``classical'' conservative semantics}
	of $\game$ of perturbation $\perturbation > 0$ is a transition system
	$\sem[\perturbation]{\game}_{\cons} = \tuple{\States, \Edges, \weight}$
	only differing from $\sem[\perturbation]{\game}$ by the sets
	\begin{align*}
	\EdgesMin &= \left\{\left((\loc, \val) \moveto{\move}
	(\loc, \val, \move)\right) ~|~ \loc \in \LocsMin, \delay \geq \perturbation,~
	\val + \delay - \perturbation \models \guard \text{ and }
	\val + \delay + \perturbation \models \guard \right\} \\
	\EdgesRob &= \left\{\left((\loc, \val, \move)
	\moveto{\trans, \varepsilon} (\loc', \val')\right) ~|~
	\varepsilon \in [-\perturbation,\perturbation] \text{ and }
	\val' = (\val + \delay + \varepsilon)[\reset \coloneqq 0] \right\}
	\end{align*}
	where $\trans \in \Trans$.
\end{definition}

We note that with respect to this new conservative semantics, 
all delays $\delay$ chosen by \MinPl must satisfy $\delay \geq \perturbation$, 
since, otherwise, \MaxPl can produce a negative delay.
Moreover, we note that $\sem[\perturbation]{\game}_{\cons}$
correspond to a shift by $\perturbation$ of delays used in 
$\sem[\perturbation]{\game}$. 
In particular, by applying the definition of both semantics, 
we obtain the following lemma: 

\begin{lemma}
	\label{lem:valPb-translated-decision}
	Let $\game$ be a \WTG, $\perturbation > 0$ be a perturbation. 
	Then, $(\loc, \val) \xrightarrow{\move} (\loc, \val, \move) \in \sem[\perturbation]{\game}$ 
	is an edge of the conservative semantics if, and only if, 
	$(\loc, \val) \xrightarrow{\move+ \perturbation} (\loc, \val, \move+ \perturbation) 
	\in \sem[\perturbation]{\game}_{\cons}$ is an edge of the ``classical'' 
	conservative semantics.
\end{lemma}

We naturally extend all notions of plays and strategies in this context. 
In particular, we denote by $\FPlays[\perturbation, \cons]$ the set of plays 
and $\rStrat{\perturbation, \cons}$ the set of strategies w.r.t. the ``classical''
conservative semantics.  
Moreover, we defined a new robust value (that is also determined 
by~\cite[Theorem 1]{brihaye2021oneclock}):
\begin{displaymath}
\rValue(\loc, \val) = 
\inf_{\robminstrategy \in \rStratMin{\perturbation, \cons}}~
\sup_{\robmaxstrategy \in \rStratMax{\perturbation, \cons}}
\weight(\outcomes((\loc, \val), \robminstrategy, \robmaxstrategy)) \,,
\end{displaymath}
Now, we prove that

\begin{proposition}
	\label{prop:valPb-translated}
	Let $\game$ be a \WTG and $\perturbation > 0$ be a perturbation.
	Then, for all configurations $(\loc, \val)$,
	we have $\rValue(\loc, \val) = \rValue_{\cons}(\loc, \val)$.
\end{proposition}

Given an initial configuration $(\loc, \val)$, we reason by double inequalities.

\paragraph*{First inequality: $\rValue_{\cons}(\loc, \val) \geq \rValue(\loc, \val)$.}
By considering a strategy $\robminstrategy' \in \rStratMin{\perturbation, \cons}$
for \MinPl, we exhibit a strategy $\robminstrategy \in \rStratMin{\perturbation}$
at least as good as $\robminstrategy'$, i.e. such that
\begin{displaymath}
\sup_{\robmaxstrategy' \in \rStratMax{\perturbation, \cons}}
\weight(\outcomes((\loc, \val), \robminstrategy', \robmaxstrategy'))
\geq
\sup_{\robmaxstrategy \in \rStratMax{\perturbation}}
\weight(\outcomes((\loc, \val), \robminstrategy, \robmaxstrategy)) \,.
\end{displaymath}

The definition of $\robminstrategy$ is induced by a function between plays
$\FPlays[\perturbation]$ to $\FPlays[\perturbation, \cons]$
such that this function preserves the weight of plays.
Intuitively, this function implements the property on robust decision 
given by Lemma~\ref{lem:valPb-translated-decision}.
Formally, we define the function $\proj \colon \FPlays[\perturbation]
\to \FPlays[\perturbation, \cons]$ by induction on the length of plays such that
for all finite plays $\play \in \FPlays[\perturbation]$, we have
\begin{displaymath}
\proj(\play) =
\begin{cases}
(\loc, \val) & \text{if $\play = (\loc, \val)$;} \\
\proj(\play') \moveto{\move} (\loc, \val)
& \text{if $\play = \play' \moveto{\move} (\loc, \val)$ and $\last(\play') \in \ConfMax$;} \\
\proj(\play') \moveto{\trans, \delay + \perturbation} (\loc, \val, \trans, \delay + \perturbation)
& \text{if $\play = \play' \moveto{\move} (\loc, \val, \move)$;} \\
\proj(\play') \moveto{\trans, \varepsilon - \perturbation} (\loc, \val)
& \text{if $\play = \play' \moveto{\trans, \varepsilon} (\loc, \val)$
	and $\last(\play') \notin \Conf$.} \\
\end{cases}
\end{displaymath}
Moreover, $\proj$ fulfils the following property:
\begin{lemma}
	\label{lem:valPb-translate-proj}
	For all finite plays
	$\play \in \FPlays[\perturbation]$, we have:
	\begin{enumerate}
		\item\label{itm:valPb-translate-proj-last}
		$\proj(\play) \in \FPlays[\perturbation, \cons]$ and
		$\last(\play) \in \Conf$ if and only if $\last(\proj(\play)) \in \Conf$.
		Moreover, when $\last(\play) \in \Conf$
		we have $\last(\proj(\play)) = \last(\play)$;
		
		\item\label{itm:valPb-translate-proj-weight}
		if $\play$ ends in a configuration of $\game$, then
		$\weightC(\play) = \weightC(\proj(\play))$.
	\end{enumerate}
\end{lemma}
\begin{proof}
	\begin{enumerate}
		\item By definition of $\proj$.
		
		\item We reason by induction on the length of a $\play$ where $\last(\play) \in \Conf$.
		If $\play = (\loc, \val)$, then the property trivially holds.
		Otherwise, we let $\play = \play' \moveto{\move} \states$.
		Thus, we conclude by distinguishing cases according to~$\last(\play')$.
		First, we remark that $\last(\play') \notin \ConfMin$.
		Otherwise, that will contradict $\last(\play) \in \Conf$.
		
		Now, we suppose that $\last(\play') = (\loc, \val) \in \ConfMax$, and we have
		$\proj(\play) = \proj(\play') \moveto{\move} \states$.
		In particular, by hypothesis of induction applied on $\play'$
		that ends in a configuration of $\game$, we have $\weightC(\proj(\play')) = \weight(\play')$.
		Moreover, since $\last(\play')  = (\loc, \val) = \last(\proj(\play'))$
		(by item~\eqref{itm:valPb-translate-proj-last}), we deduce that
		\begin{align*}
		\weightC(\proj(\play))
		&= \weightC(\proj(\play')) + \delay \, \weight(\loc) + \weight(\trans) \\
		&= \weightC(\play') + \delay \, \weight(\loc) + \weight(\trans)
		= \weightC(\play) \,.
		\end{align*}
		
		Finally, we suppose that $\last(\play') = (\loc, \val, \move') \notin \Conf$,
		and we have $\proj(\play) = \proj(\play') \moveto{\trans, \delay - \perturbation} \states$ such that
		\begin{equation}
		\label{eq:pert-exact_proj_wt1}
		\weightC(\proj(\play))
		= \weightC(\proj(\play')) + (\delay - \perturbation) \,
		\weight(\loc, \val, \move') + 0 = \weightC(\proj(\play')) + (\delay - \perturbation) \,
		\weight(\loc) \,.
		\end{equation}
		Since $\last(\play') \notin \Conf$, we can not apply the
		hypothesis of induction on $\play'$.
		However, we can rewrite
		$\play'= \play'' \moveto{\trans, \delay'} (\loc, \val, \trans, \delay')$
		where $\last(\play'') = (\loc, \val) \in \ConfMin$ (since $\play'$ begins by a configuration).
		In particular, $\proj(\play') = \proj(\play'') 
		\moveto{\trans, \delay' + \perturbation}
		(\loc, \val, \trans, \delay' + \perturbation)$, and we obtain that
		\begin{displaymath}
		\weightC(\proj(\play'))
		= \weightC(\proj(\play'')) + (\delay' + \perturbation) \, 
		\weight(\loc) + \weight(\trans)  
		= \weightC(\play'') + (\delay' + \perturbation) \, \weight(\loc) 
		+ \weight(\trans)
		\end{displaymath}
		since $\weightC(\proj(\play'')) = \weight(\play'')$
		(by hypothesis of induction applied on $\play''$) and
		$\last(\play'')  = (\loc, \val) = \last(\proj(\play''))$
		(by item~\eqref{itm:valPb-translate-proj-last}).
		By combining the previous equation with~\eqref{eq:pert-exact_proj_wt1},
		we conclude~that
		\begin{align*}
		\weightC(\proj(\play))
		&= \weightC(\play'') + (\delay' + \perturbation) \, \weight(\loc) +
		(\delay - \perturbation) \, \weight(\loc)) + \weight(\trans)  \\
		&= \weightC(\play'') + (\delay' + \delay) \ \weight(\loc) + 
		\weight(\trans) = \weightC(\play) \,.
		\qedhere
		\end{align*}
	\end{enumerate}
\end{proof}

Now, we define $\robminstrategy \in \rStratMin{\perturbation}$
such that for all $\play \in \FPlays[\perturbation]$
\begin{displaymath}
\robminstrategy(\play) = (\trans, \delay - \perturbation) 
\qquad \text{if } \robminstrategy'\big(\proj(\play)\big) = (\move) \,.
\end{displaymath}
By Lemma~\ref{lem:valPb-translated-decision}, we note that this choice induced
an edge of $\sem[\perturbation]{\game}$.
Moreover, since the definition of $\robminstrategy$ relies on $\proj$,
it is no surprise that:
\begin{lemma}
	\label{lem:valPb-translate-conf}
	Let $\play \in \FPlays[\perturbation]$ be a play conforming to $\robminstrategy$.
	Then, $\proj(\play)$ is conforming to~$\robminstrategy'$.
\end{lemma}
\begin{proof}
	We reason by induction on the length of $\play$.
	If $\play = (\loc, \val)$ then $\proj(\play) = (\loc, \val)$,
	and the property trivially holds.
	Otherwise, we can write $\play = \play' \moveto{\move} \states$
	such that $\proj(\play')$ is conforming to $\robminstrategy'$
	by the hypothesis of induction.
	In particular, if $\last(\play') \notin \ConfMin$ then
	$\last(\proj(\play')) \notin \ConfMin$
	(by Lemma~\ref{lem:valPb-translate-proj}-\eqref{itm:valPb-translate-proj-last})
	and $\proj(\play)$ is conforming to $\robminstrategy'$.
	
	Now, we suppose that $\last(\play') \in \ConfMin$,
	and $\proj(\play) = \proj(\play') \moveto{\trans, \delay + \perturbation} \states$.
	Since $\play$ is conforming to $\robminstrategy$, then
	$\robminstrategy(\play') = (\move)$.
	We conclude by definition of $\robminstrategy$,
	since $\robminstrategy'(\proj(\play')) = (\trans, \delay + \perturbation)$.
\end{proof}

We consider a finite play $\play \in \FPlays[\perturbation]$
ending a configuration of $\game$ and 
conforming to $\robminstrategy$.
We prove that the play
$\proj(\play) \in \FPlays[\perturbation]$ conforming to $\robminstrategy'$
(by Lemma~\ref{lem:valPb-translate-conf})
satisfies $\weight(\play) \leq \weight(\proj(\play))$.
Indeed, if $\play$ reaches a target location, then $\proj(\play)$ reaches the same target location
by applying Lemma~\ref{lem:valPb-translate-proj}-\eqref{itm:valPb-translate-proj-last}.
In particular, by Lemma~\ref{lem:valPb-translate-proj}-\eqref{itm:valPb-translate-proj-weight},
we conclude that $\weight(\play) \leq \weight(\proj(\play))$.
Otherwise, $\play$ does not reach a target location and $\weight(\play) = +\infty$.
By using Lemma~\ref{lem:valPb-translate-proj}-\eqref{itm:valPb-translate-proj-weight},
we conclude that $\proj(\play)$ does not reach a target location.
Otherwise, $\play$ would reach the same last configuration
(since it ends in a configuration of $\game$), i.e. 
$\weightP(\proj(\play)) = +\infty$.

Finally,we have shown that for all maximal plays $\play$ conforming to $\robminstrategy$,
we can define a play $\proj(\play)$ conforming to $\robminstrategy'$
(by Lemma~\ref{lem:valPb-translate-conf}) such that
$\weightP(\play) \leq \weightP(\proj(\play))$.
In particular, we do not prove that there does not exist infinite maximal 
play conforming to $\robminstrategy$, but if such a play exists, 
it also exists an infinite maximal play conforming to $\robminstrategy'$. 
Indeed, we prove that for all finite plays conforming to $\robminstrategy$, 
there exists a finite play with a weight at least equal to the finite play
conforming to $\robminstrategy'$ such that the last configuration of two plays is equal. 
In particular, if all maximal plays conforming to chi are finite, 
we obtain the inequality over values of $\robminstrategy$ and $\robminstrategy'$. 
Now, if there exists an infinite maximal play conforming to $\robminstrategy$, 
then all its prefixes do not reach the target. 
In particular, we can define an infinite play conforming to $\robminstrategy'$ 
that does not reach the target (since for each prefix, both plays end in the 
same configuration) and the inequality holds.
Thus, we obtain that
\begin{displaymath}
\sup_{\robmaxstrategy' \in \rStratMax{\perturbation, \cons}}
\weightP(\outcomes((\loc, \val), \robminstrategy', \robmaxstrategy'))
\geq
\sup_{\robmaxstrategy \in \rStratMax{\perturbation}}
\weightP(\outcomes((\loc, \val), \robminstrategy, \robmaxstrategy))
\geq \rValue(\loc, \val) \,.
\end{displaymath}
Since, this inequality holds for all $\robminstrategy'$, we obtain the claimed
inequality.

\paragraph*{Second inequality: $\rValue_{\cons}(\loc, \val) \leq \rValue(\loc, \val)$.}
Conversely, by considering a strategy $\robminstrategy \in \rStratMin{\perturbation}$
for \MinPl, we exhibit a strategy $\robminstrategy' \in \rStratMin{\perturbation, \cons}$
at least as good as $\robminstrategy$, i.e.
\begin{displaymath}
\sup_{\robmaxstrategy' \in \rStratMax{\perturbation, \cons}}
\weightP(\outcomes((\loc, \val), \robminstrategy', \robmaxstrategy'))
\leq
\sup_{\robmaxstrategy \in \rStratMax{\perturbation}}
\weightP(\outcomes((\loc, \val), \robminstrategy, \robmaxstrategy)) \,.
\end{displaymath}
In particular, we use the function
$\inj \colon \FPlays[\perturbation, \cons] \to \FPlays[\perturbation]$
defined by induction on the length of plays such that for all finite plays
$\play \in \FPlays[\perturbation, \cons]$, we have
\begin{displaymath}
\inj(\play) =
\begin{cases}
(\loc, \val) & \text{if $\play = (\loc, \val)$} \\
\inj(\play') \moveto{\move} (\loc, \val)
& \text{if $\play = \play' \moveto{\move} (\loc, \val)$ and $\last(\play') \in \ConfMax$} \\
\inj(\play') \moveto{\trans, \delay - \perturbation} (\loc, \val, \trans, \delay- \perturbation)
& \text{if $\play = \play' \moveto{\move} (\loc, \val, \move)$} \\
\inj(\play') \moveto{\trans, \varepsilon + \perturbation} \states
& \text{if $\play = \play' \moveto{\trans, \varepsilon} \states$ and $\last(\play') \notin \Conf$} \\
\end{cases}
\end{displaymath}
Moreover, $\inj$ satisfies the properties:
\begin{lemma}
	\label{lem:valPb-translate-inj}
	For all finite plays
	$\play \in \FPlays[\perturbation, \cons]$, we have:
	\begin{enumerate}
		\item\label{itm:valPb-translate-inj-last}
		$\inj(\play) \in \FPlays[\perturbation]$ and
		$\last(\play) \in \Conf$ if and only if $\last(\inj(\play)) \in \Conf$.
		Moreover, when $\last(\play) \in \Conf$
		we have $\last(\inj(\play)) = \last(\play)$;
		
		\item\label{itm:valPb-translate-inj-bij}
		$\inj$ is the inverse of $\proj$;
		
		\item\label{itm:valPb-translate-inj-weight}
		if $\play$ ends in a configuration of $\game$, i.e. $\last(\play) \in \Conf$, then
		$\weightC(\play) = \weightC(\inj(\play))$.
	\end{enumerate}
\end{lemma}
\begin{proof}
	\begin{enumerate}
		\item By definition of $\proj$.
		
		\item Let $\play \in \FPlays[\perturbation, \cons]$, we prove
		that $\proj(\inj(\play)) = \play$ by induction on the length of $\play$.
		If $\play = (\loc, \val)$, then $\proj(\inj(\play)) = (\loc, \val)$.
		Otherwise, we rewrite by $\play = \play' \moveto{\move} \states$ such that
		$\proj(\inj(\play')) = \play'$ by the hypothesis of induction.
		We conclude by distinguishing cases according to $\last(\play')$.
		If $\last(\play') \in \ConfMax$, then
		\begin{displaymath}
		\proj(\inj(\play)) = \proj\big(\inj(\play')\moveto{\move} \states\big) =
		\proj(\inj(\play')) \moveto{\move} \states = \play'\moveto{\move} \states\,.
		\end{displaymath}
		If $\last(\play') \in \ConfMin$, then
		\begin{displaymath}
		\proj(\inj(\play)) = \proj\big(\inj(\play')\moveto{\move- \perturbation} \states\big) =
		\proj(\inj(\play')) \moveto{\move- \perturbation + \perturbation} \states = \play'\moveto{\move} \states\,.
		\end{displaymath}
		Finally, if $\last(\play') \notin \Conf$, we have
		\begin{displaymath}
		\proj(\inj(\play)) = \proj\big(\inj(\play')\moveto{\move + \perturbation} \states\big) =
		\proj(\inj(\play')) \moveto{\move+ \perturbation - \perturbation} = \play'\moveto{\move} \states\,.
		\end{displaymath}
		
		Conversely, via an analogous proof, we obtain that for
		all $\play \in \FPlays[\perturbation]$
		we have $\inj(\proj(\play)) = \play$ by induction on the length of $\play$.
		
		\item Let $\play \in \FPlays[\perturbation, \cons]$, we have
		$\weightC(\proj(\inj(\play))) = \weightC(\inj(\play))$
		by Lemma~\ref{lem:valPb-translate-proj}-\eqref{itm:valPb-translate-proj-weight}.
		We conclude that $\weightC(\play) = \weightC(\inj(\play))$
		by item~\eqref{itm:valPb-translate-inj-bij}.
		\qedhere
	\end{enumerate}
\end{proof}

Now, we define $\robminstrategy' \in \rStratMin{\perturbation, \cons}$ such that
for all $\play \in \FPlays[\perturbation, \cons]$
\begin{displaymath}
\robminstrategy'(\play) = (\trans, \delay + \perturbation) \qquad \text{if }
\robminstrategy\big(\inj(\play)\big) = (\move) \,.
\end{displaymath}
By Lemma~\ref{lem:valPb-translated-decision}, we note that this choice induced
an edge of $\sem[\perturbation]{\game}_\cons$.
Moreover, since the definition of $\robminstrategy$ relies on $\inj$,
it is no surprise that:
\begin{lemma}
	\label{lem:valPb-translate-conf2}
	Let $\play \in \FPlays[\perturbation, \cons]$ be a play conforming to
	$\robminstrategy'$.
	Then, $\inj(\play)$ is conforming to~$\robminstrategy$.
\end{lemma}
\begin{proof}
	We reason by induction on the length of $\play$ and as in
	Lemma~\ref{lem:valPb-translate-conf}, the interesting case is
	when $\play = \play' \moveto{\move} \states$ where $\last(\play') \in \ConfMin$
	and $\inj(\play')$ is conforming to $\robminstrategy$ by the hypothesis of induction.
	In this case, $\inj(\play) = \inj(\play') \moveto{\trans, \delay - \perturbation} \states$.
	Since $\play$ is conforming to $\robminstrategy'$, then we have
	$\robminstrategy'(\play') = (\move)$.
	We conclude by definition of~$\robminstrategy'$,
	since $\robminstrategy(\inj(\play')) = (\trans, \delay - \perturbation)$.
\end{proof}

We consider a finite play $\play \in \FPlays[\perturbation, \cons]$
ending a configuration of $\game$ and 
conforming to $\robminstrategy'$.
We prove that the play
$\inj(\play) \in \FPlays[\perturbation]$ conforming to $\robminstrategy$
(by Lemma~\ref{lem:valPb-translate-conf2})
satisfies $\weightP(\play) \leq \weightP(\inj(\play))$.
Indeed, if $\play$ reaches a target location, then $\inj(\play)$ reaches the same target location
by applying Lemma~\ref{lem:valPb-translate-inj}-\eqref{itm:valPb-translate-inj-last}.
In particular, by Lemma~\ref{lem:valPb-translate-inj}-\eqref{itm:valPb-translate-inj-weight},
we conclude that $\weightP(\play) \leq \weightP(\inj(\play))$.
Otherwise, $\play$ does not reach a target location and $\inj(\play)$ too
by Lemma~\ref{lem:valPb-translate-inj}-\eqref{itm:valPb-translate-inj-last}.
Thus, $\weightP(\play) = +\infty = \weightP(\inj(\play))$.

Finally, we have shown that for all plays $\play$ conforming to $\robminstrategy$,
we can define a play $\inj(\play)$ conforming to $\robminstrategy'$
(by Lemma~\ref{lem:valPb-translate-conf2}) such that
$\weightP(\play) \leq \weightP(\inj(\play))$.
In particular, we conclude since the following inequality holds
for all $\robminstrategy'$,
\begin{displaymath}
\rValue_{\cons}(\loc, \val) \leq
\sup_{\robmaxstrategy' \in \rStratMax{\perturbation, \cons}}
\weightP(\outcomes((\loc, \val), \robminstrategy', \robmaxstrategy'))
\leq
\sup_{\robmaxstrategy \in \rStratMax{\perturbation}}
\weightP(\outcomes((\loc, \val), \robminstrategy, \robmaxstrategy)) \,.
\end{displaymath}

\section{Proof of Lemma~\ref{lem:rVal-monotony}}
\label{app:rVal-monotony}

\lemRValMonotony*
\begin{proof}
	Let $\robminstrategy \in \rStratMin{\perturbation}$ be a strategy for \MinPl
	such that
	\begin{displaymath}
	\sup_{\robmaxstrategy \in \rStratMax{\perturbation}}
	\weightP(\outcomes((\loc, \val), \robminstrategy, \robmaxstrategy)) \geq
	\sup_{\robmaxstrategy \in \rStratMax{\perturbation'}}
	\weightP(\outcomes((\loc, \val), \robminstrategy, \robmaxstrategy))
	\end{displaymath}
	(by Lemma~\ref{lem:sem-strat_MinMax}-\eqref{itm:sem-strat_max}).
	We conclude, by applying the infimum over strategies in
	$\rStratMin{\perturbation} \subseteq \rStratMin{\perturbation'}$
	(by Lemma~\ref{lem:sem-strat_MinMax}-\eqref{itm:sem-strat_min}) for \MinPl:
	\begin{align*}
	\rValue(\loc, \val)
	&= \inf_{\robminstrategy \in \rStratMin{\perturbation}}~
	\sup_{\robmaxstrategy \in \rStratMax{\perturbation}}~
	\weightP(\outcomes((\loc, \val), \robminstrategy, \robmaxstrategy)) \\
	&\geq \inf_{\robminstrategy \in \rStratMin{\perturbation'}}~
	\sup_{\robmaxstrategy \in \rStratMax{\perturbation'}}
	\weightP(\outcomes((\loc, \val), \robminstrategy, \robmaxstrategy)) = \rValueP(\loc, \val) \,.
	\qedhere
	\end{align*}
\end{proof}

\section{Example of computation of robust values functions}
\label{app:valIt-ex-rob}

We consider the acyclic \WTG depicted in \figurename{~\ref{fig:rob-valIt_ex-3}}
with two clocks $\Cl= \{x_1, x_2\}$ and we compute the fixed-perturbation robust 
value function w.r.t. the conservative semantics
with a parameter $\perturbation > 0$, for all valuations in $[0, 2]^2$.
All cells used throughout this computation are depicted in
\figurename{~\ref{fig:rob-valIt_ex-3_cons-cell}}
(we note that the cells are depicted for a fixed $\perturbation$,
otherwise another dimension is needed).

\text{}

\noindent\textbf{The fixed-perturbation robust value function in $\loc_1$}
Intuitively by the positivity of the weight in $\loc_1$,
\MinPl wants to play with the minimal delay, i.e.\ $\textcolor{magenta}{0}$,
whereas \MaxPl wants to maximise its perturbation,
i.e.\ $\textcolor{blue}{2\,\perturbation}$.
Thus, we have
\begin{displaymath}
\rValue(\loc_1, \val) =
\inf_{\delay}
\sup_{\varepsilon \in [0, 2\perturbation]}
[(\textcolor{magenta}{\delay} + \textcolor{blue}{\varepsilon}) \times 1] =
2 \, \perturbation \,.
\end{displaymath}

\text{}

\noindent\textbf{The fixed-perturbation robust value function in $\loc_2$}
Again, since the weight of $\loc_2$ is positive, 
\MinPl wants to minimise the time spent in this location
whereas \MaxPl wants to maximise the perturbation,
i.e.\ $\textcolor{blue}{2\,\perturbation}$.
In particular, from the guard, we remark that if $\val(x_2) \geq 1$,
\MinPl can directly leave the location and
the fixed-perturbation robust value function is equal to
$1 + (\textcolor{magenta}{0} + \textcolor{blue}{2\,\perturbation}) \times 1 +
2\perturbation = 1 + 4\,\perturbation$;
otherwise, if $\val(x_2) < 1$, \MinPl must wait until the valuation of $x_2$ is equal to $1$
and the fixed-perturbation robust value function is equal to
$1 + (\textcolor{magenta}{1 - \val(x_2)} + \textcolor{blue}{2\,\perturbation}) \times 1 +
2\perturbation = 2 + 4\perturbation - \val(x_2)$.
To summarize, for all valuations in $[0,2]^2$, we~have
\begin{displaymath}
\rValue\big(\loc_2, \val\big) =
\begin{cases}
2 + 4\,\perturbation - \val(x_2) & \text{if \textcolor{Brown}{$\val(x_2) < 1$};} \\
1 + 4\,\perturbation &
\text{if \textcolor{ForestGreen}{$1 \leq \val(x_2) \leq 2 - 2\,\perturbation$};} \\
+\infty & \text{otherwise.}
\end{cases}
\end{displaymath}

\text{}

\noindent\textbf{The fixed-perturbation robust value function in $\loc_3$}
Since the conservative semantics do not perturb the decisions of \MaxPl,
$\loc_2$ can be reached only on the projection of the
brown cell on the affine equation $x_1 = 0$
and the fixed-perturbation robust value function is equal to
\begin{displaymath}
\rValue\big(\loc_3, \val\big) =
\sup_{\substack{\delay \\ 1 < \val(x_1) + \delay < 2 \\ \val(x_2) + \delay < 1}}
\rValue(\loc_2, \val') - 2 \qquad \text{where $\val'(x_1) = 0$ and $\val'(x_2) = \val(x_2) + \delay$.} 
\end{displaymath}
Moreover, \MaxPl wants to play a delay such that the 
valuation of $x_1$ is close enough to $1$
when the transition is applied.
In particular, we distinguish two cases according to the valuation of $x_1$.
\begin{itemize}
	\item We suppose that $1 < \val(x_1)$, and \MaxPl can directly leave the
	location to obtain the weight $2 + 4\,\perturbation - \val(x_2)$ in $\loc_2$.
	In particular, when $\val(x_2) < 1$, we deduce that the 
	fixed-perturbation robust value function is equal to $4\,\perturbation - \val(x_2)$.
	
	\item Otherwise, we suppose that $\val(x_1) \leq 1$ and \MaxPl wants to
	play the delay is $1 - \val(x_1)$.
	We note that this decision is feasible only when $\val(x_2) < \val(x_1)$.
	In this case, the fixed-perturbation robust value function is equal to
	$-2 + 2 + 4\perturbation - (\val(x_2) + 1 - \val(x_1)) =
	4\perturbation + \val(x_1) - \val(x_2) - 1$.
\end{itemize}
Thus, for all valuations in $[0,2]^2$, we conclude that
\begin{displaymath}
\rValue(\loc_3, \val) =
\begin{cases}
4\perturbation + \val(x_1) - \val(x_2) - 1 & \text{if \textcolor{Purple}{$\val(x_1) \leq 1$ and $\val(x_2) < \val(x_1)$};} \\
4\,\perturbation - \val(x_2) & \text{if \textcolor{ForestGreen}{$1 < \val(x_1) < 2$ and $\val(x_2) < 1$};} \\
+\infty & \text{otherwise.}
\end{cases}
\end{displaymath}

\text{}

\noindent\textbf{The fixed-perturbation robust value function in $\loc_4$}
By negativity of the weight of $\loc_4$, \MaxPl wants to minimise the perturbation,
whereas \MinPl wants to maximise the time spent in $\loc_4$,
i.e.\ \MinPl waits until a valuation equal to $2 - 2\,\perturbation$
(since the guard must be satisfied for all perturbations, \MinPl can not wait longer).
Again, we distinguish two cases according to the order over valuations of $x_1$ and $x_2$.
\begin{itemize}
	\item We suppose that $\val(x_2) \leq \val(x_1) \leq 2- 2\,\perturbation$ and
	$\textcolor{magenta}{2 -2\,\perturbation - \val(x_1)}$ is a robust delay for \MinPl.
	Moreover, since \MaxPl does not perturb the delay chosen by \MinPl, 
	the fixed-perturbation robust value is equal to
	$-(\textcolor{magenta}{2 -2\,\perturbation - \val(x_1)})$.
	
	\item We suppose that $\val(x_1) < \val(x_2) \leq 2- 2\,\perturbation$ and, symmetrically,
	$\textcolor{magenta}{2 -2\,\perturbation - \val(x_2)}$ is a robust delay for \MinPl 
	when $\val(x_2) + 2\,\perturbation -1 \leq x_1 \leq \val(x_2)$.
	Moreover, since \MaxPl does not perturb the delay chosen by \MinPl, 
	the fixed-perturbation robust value is equal to
	$-(\textcolor{magenta}{2 -2\,\perturbation - \val(x_2)})$.
\end{itemize}
Thus, for all valuations in $[0,2]^2$, we~have
\begin{displaymath}
\rValue\big(\loc_4, \val\big) =
\begin{cases}
-(2 - 2\,\perturbation - \val(x_1)) & \text{if \textcolor{Brown}{$\val(x_2) \leq \val(x_1) < 2 - 2\,\perturbation$};} \\
-(2 - 2\,\perturbation - \val(x_2))
& \text{if \textcolor{ForestGreen}{$\val(x_2) + 2\,\perturbation - 1 \leq \val(x_1) < \val(x_2) < 2 - 2\,\perturbation$};} \\
+\infty & \text{otherwise.} \\
\end{cases}
\end{displaymath}
We note that $\val(x_2) + 2\perturbation - 1 \leq 2$ (i.e. the green cell is not empty)
if and only if $\perturbation \leq 1/2$.
For the rest of the computation, we suppose that the upper bound over 
$\perturbation$ is $\perturbationBound = 1/2$.

\begin{figure}
	\centering
	\begin{tikzpicture}[scale=0.8,every node/.style={font=\scriptsize},
	every label/.style={font=\scriptsize}]
	\begin{scope}[xshift=0cm,scale=1.25]
	\draw[->] (0,0) -- (0,2.5);
	\draw[->] (0,0) -- (2.5,0);
	
	\draw[dashed] (1,0) -- (1,2);
	\draw[dashed] (2,0) -- (2,2);
	
	\draw[dashed] (0,1) -- (2,1);
	\draw[dashed] (0,2) -- (2,2);

	\node at (2.5,-.25) {$x_1$};
	\node at (-.25,2.5) {$x_2$};
	
	\node at (0,-.25) {$0$};
	\node at (1,-.25) {$1$};
	\node at (2,-.25) {$2$};
	
	\node at (-.25,0) {$0$};
	\node at (-.25,1) {$1$};
	\node at (-.5,1.9) {\textcolor{red}{$2 - 2\,\perturbation$}};
	
	\draw[Brown,fill=Tan,opacity=.7] (0,0) -- (2,0) -- (2,1) -- (0,1) -- (0,0);
	\draw[ForestGreen,fill=LightGreen,opacity=.7] (0,1) -- (2,1) -- (2,1.9) -- (0,1.9) -- (0,1);
	
	\draw[Brown,very thick] (0,0) -- (0,1);
	\draw[Brown,very thick] (2,0) -- (2,1);
	\draw[Brown,very thick] (0,0) -- (2,0);
	\draw[ForestGreen,very thick] (0,1) -- (2,1);
	\draw[ForestGreen,very thick] (0,1.9) -- (2,1.9);
	\draw[ForestGreen,very thick] (0,1) -- (0,1.9);
	\draw[ForestGreen,very thick] (2,1) -- (2,1.9);
	
	\node at (1,2.5) {$\loc_2$};
	\end{scope}
	
	\begin{scope}[xshift=5.5cm,scale=1.25]
	\draw[->] (0,0) -- (0,2.5);
	\draw[->] (0,0) -- (2.5,0);
	
	\draw[dashed] (1,0) -- (1,2);
	\draw[dashed] (2,0) -- (2,2);
	
	\draw[dashed] (0,1) -- (2,1);
	\draw[dashed] (0,2) -- (2,2);

	\node at (2.75,0) {$x_1$};
	\node at (-.25,2.5) {$x_2$};
	
	\node at (0,-.25) {$0$};
	\node at (1,-.25) {$1$};
	\node at (1.9,-.25) {\textcolor{red}{$2 - 2\,\perturbation$}};
	
	\node at (-.85,.9) {\textcolor{red}{$x_1 - 2\,\perturbation + 1$}};
	\node at (-.5,1.9) {\textcolor{red}{$2 - 2\,\perturbation$}};
	
	\draw[Brown,fill=Tan,opacity=.7] (0,0) -- (1.9,1.9) -- (1.9,0) -- (0,0);
	\draw[ForestGreen,fill=LightGreen,opacity=.7] (0,0) -- (0,.9) -- (1,1.9) -- (1.9,1.9) -- (0,0);
	
	\draw[red,very thick] (1.9,1.9) -- (2,2);
	\draw[red,very thick] (1.9,0) -- (1.9,2);
	\draw[red,very thick] (1,1.9) -- (1.1,2);
	\draw[Brown,very thick] (0,0) -- (1.9,1.9);
	\draw[ForestGreen,very thick] (0,.9) -- (1,1.9);
	\draw[ForestGreen,very thick] (0,0) -- (0,.9);
	\draw[Brown,very thick] (0,0) -- (1.9,0);
	\draw[red,very thick] (0,1.9) -- (2,1.9);
	
	\node at (1,2.5) {$\loc_4$};
	\end{scope}
	
	\begin{scope}[xshift=11cm,scale=2,yshift=-2cm]
	\draw[->] (0,0) -- (0,2.5);
	\draw[->] (0,0) -- (2.5,0);
	
	\draw[dashed] (1,0) -- (1,2);
	\draw[dashed] (2,0) -- (2,2);
	
	\draw[dashed] (0,1) -- (2,1);
	\draw[dashed] (0,2) -- (2,2);

	\node at (2.5,-.25) {$x_1$};
	\node at (-.25,2.5) {$x_2$};
	
	\node at (.1,-.15) {\textcolor{red}{$x_1 - 2\,\perturbation$}};
	\node at (2.6,1.25) {\textcolor{red}{$x_1 - 5\,\perturbation - \frac{1}{2}$}};
	\node at (1.9,-.25) {\textcolor{red}{$2 - 2\,\perturbation$}};

	\node at (1,-.25) {$1$};
	\node at (-.35,.9) {\textcolor{red}{$1-2\perturbation$}};
	\node at (-.35,.45) {\textcolor{red}{$\frac{1}{2}-\perturbation$}};
	\node at (-.25,2) {$2$};
	
	\draw[Brown,fill=Tan,opacity=.7] (.75,0) -- (1,0) -- (1,.25) --  (.75,0);
	\draw[RedOrange,fill=Peach,opacity=.7] (.1,0) -- (.75,0) -- (1,.25) --
	(1,.9) -- (.1,0);
	\draw[ForestGreen,fill=LightGreen,opacity=.7] (1,0) -- (1,.45) -- (1.9,.45) --
	(1.9,0) -- (1,0);
	\draw[Purple,fill=Lavender,opacity=.7] (1,.45) -- (1,.9) -- (1.9,.9) -- (1.9,.45) -- (1,.45);
	
	\draw[red,very thick] (1,0) -- (1,2);
	\draw[red,very thick] (0,.45)  -- (.6,.45); % y = 1/2-9p on [0, .6]
	\draw[red,very thick,dotted] (.6,.45)  -- (1,.45); % y = 1/2-p on [.6, 1]
	\draw[red,very thick] (1.9,.45)  -- (2,.45); % y = 1/2-p on [1.9, 2]
	\draw[red,very thick] (1.9,.9) -- (1.9,2); % x= 2 -2p on [0.9, 2]
	\draw[red,very thick] (0,.9)  -- (1,.9); % y = 1-2p
	\draw[red,very thick] (1.9,.9)  -- (2,.9); % y = 1-2p
	\draw[red,very thick] (1,.9) -- (2,1.9);% y = x-2p sur [1, 2]
	\draw[Purple,very thick] (1,.9) -- (1.9,.9); % y = 1-2p
	\draw[Blue,very thick] (1.9,0) -- (1.9,.9); % x= 2 -2p on [0, 0.9]
	\draw[ForestGreen,very thick] (1,.45)  -- (1.9,.45); % y = 1/2-p on [1, 1.9]
	\draw[ForestGreen,very thick] (1,0)  -- (1.9,0);
	\draw[red,very thick,dotted] (1,.25)  -- (1.65,.9);
	\draw[red,very thick] (1.65,.9)  -- (2,1.25);
	\draw[Brown,very thick] (0.75,0)  -- (1,0.25);
	\draw[Brown,very thick] (1,0)  -- (1,0.25);
	\draw[Brown,very thick] (0.75,0)  -- (1,0);
	\draw[RedOrange,very thick] (.1,0) -- (1,.9);% y = x-2p sur [0, 1]
	\draw[RedOrange,very thick] (.1,0) -- (.75,0);
	\draw[RedOrange,very thick] (1,.25) -- (1,.9);

	\node at (1,2.5) {$\loc_0$};
	
	\end{scope}
	
	\begin{scope}[xshift=0cm,yshift=-4.75cm,scale=1.25]
	\draw[->] (0,0) -- (0,2.5);
	\draw[->] (0,0) -- (2.5,0);
	
	\draw[dashed] (1,0) -- (1,2);
	\draw[dashed] (2,0) -- (2,2);
	
	\draw[dashed] (0,1) -- (2,1);
	\draw[dashed] (0,2) -- (2,2);

	\node at (2.5,-.25) {$x_1$};
	\node at (-.25,2.25) {$x_2$};
	
	\node at (.2,-.25) {\textcolor{red}{$x_1 - 4\,\perturbation$}};
	\node at (1,-.25) {$1$};
	\node at (1.8,-.25) {\textcolor{red}{$2-2\,\perturbation$}};
	
	\node at (-.25,0) {$0$};
	\node at (-.5,.8) {\textcolor{red}{$1-4\,\perturbation$}};
	
	\draw[Brown,fill=Tan,opacity=.7] (0.2,0) -- (1,0) -- (1,.8) -- (0.2,0);
	\draw[ForestGreen,fill=LightGreen,opacity=.7] (1,0) -- (1,.8) -- (1.9,.8) -- (1.9,0) -- (1,0);
	
	\draw[red,very thick] (1,0.8) -- (1,2); % x = 1
	\draw[red,very thick] (1,0.8) -- (2,1.8); % x-4p
	\draw[red,very thick] (0,.8) -- (1,.8); % y=1-4p
	\draw[ForestGreen,very thick] (1,.8) -- (2,.8); % y=1-4p
	\draw[Brown,very thick] (0.2,0) -- (1,.8); % x-4p
	\draw[Brown,very thick] (0.2,0) -- (1,0);
	\draw[ForestGreen,very thick] (1,0) -- (1.9,0);
	\draw[red,very thick] (1.9,0) -- (1.9,2); % x = 2-2p
	\draw[Brown,very thick] (1,0) -- (1,0.8); % x = 1
	
	\node at (1,2.5) {$\loc_0 \to \loc_4$};
	
	\end{scope}
	
	\begin{scope}[xshift=5.5cm,yshift=-4.75cm,scale=1.25]
	\draw[->] (0,0) -- (0,2.5);
	\draw[->] (0,0) -- (2.5,0);
	
	\draw[dashed] (1,0) -- (1,2);
	\draw[dashed] (2,0) -- (2,2);
	
	\draw[dashed] (0,1) -- (2,1);
	\draw[dashed] (0,2) -- (2,2);

	\node at (2.75,0) {$x_1$};
	\node at (-.25,2.25) {$x_2$};
	
	\node at (.1,-.25) {\textcolor{red}{$x_1-2\,\perturbation$}};
	\node at (1,-.25) {$1$};
	\node at (1.8,-.25) {\textcolor{red}{$2-2\,\perturbation$}};
	
	\node at (-.25,0) {$0$};
	\node at (-.5,.8) {\textcolor{red}{$1-2\,\perturbation$}};
	
	\draw[ForestGreen,fill=LightGreen,opacity=.7] (0.1,0) -- (1,.9) -- (1.9, .9)
	-- (1.9, 0) -- (0.1,0);
	
	\draw[red,very thick] (1,0.9) -- (2,1.9); % x-2p
	\draw[red,very thick] (0,.9) -- (1,.9); % y=1-2p
	\draw[red,very thick] (1.9,.9) -- (1.9,2); % x=2-2p
	\draw[ForestGreen,very thick] (0.1,0) -- (1,.9); % x-2p
	\draw[ForestGreen,very thick] (1,.9) -- (2,.9); % y=1-2p
	\draw[ForestGreen,very thick] (1.9,0) -- (1.9,.9); % x=2-2p
	\draw[ForestGreen,very thick] (.1,0) -- (1.9,0);
	
	\node at (1,2.5) {$\loc_0 \to \loc_3$};
	
	\end{scope}
	\end{tikzpicture}
	\caption{Cells for the fixed-perturbation robust value functions
		computed for all locations of the acyclic \WTG
		depicted in \figurename{~\ref{fig:rob-valIt_ex-3}} when $\perturbation = \frac{1}{18}$.}
	\label{fig:rob-valIt_ex-3_cons-cell}
\end{figure}

\text{}

\noindent\textbf{The fixed-perturbation robust value function in $\loc_0$}
Finally, we compute the fixed-perturbation robust value function of $\loc_0$
by remarking~that
\begin{displaymath}
\rValue(\loc_0, \val) =
\min\Big(\rValue(\loc_0 \to \loc_3, \val),
\rValue(\loc_0 \to \loc_4, \val)\Big) \,
\end{displaymath}
where $\rValue(\loc_0 \to \loc_3, \val)$ (resp. $\rValue(\loc_0 \to \loc_4, \val)$) denotes 
the robust value from $(\loc_0, \val)$ when \MinPl chooses to go to the location $\loc_3$ (resp. $\loc_4$) 
from $\loc_0$.
To conclude, we distinguish two cases according to the choice of
transition by \MinPl.

\emph{We suppose that \MinPl chooses to go to $\loc_4$}.
Since the clock $x_1$ is reset by the transition between
$\loc_0$ and $\loc_4$ only the green cell on its border $x_1 = 0$
can be reached to obtain a finite value function, i.e.
$\loc_4$ is reached when $\val(x_2) \leq 1 - 2\,\perturbation$.
Now, since the weight of $\loc_0$ is positive and the
fixed-perturbation robust value function increases
when the valuation of $x_2$ increases, \MaxPl wants to maximise
its perturbation, i.e.\ $\textcolor{blue}{2\, \perturbation}$,
whereas \MinPl wants to minimise its delay, i.e. directly leave if
$\val(x_1) \geq 1$, or wait until $1$ for the valuation of $x_1$.
\begin{itemize}
	\item We suppose that $1 < \val(x_1) < 2 - 2\,\perturbation$ and
	\MinPl plays the delay $\textcolor{magenta}{0}$ when $\val(x_2) \leq 1 - 4\,\perturbation$.
	Moreover, the obtained weight~is
	$1 \times \textcolor{blue}{2\perturbation} + 1 -
	(2 - 2\perturbation - (\val(x_2) + 2\perturbation))
	= \val(x_2) + 6\,\perturbation - 1$.
	
	\item We suppose that $0 \leq \val(x_1) \leq 1$ and \MinPl plays the minimal
	delay $\textcolor{magenta}{1 - \val(x_1)}$ when $\val(x_2) \leq \val(x_1) - 4\,\perturbation$.
	In particular, the obtained weight is
	$1 \times (\textcolor{magenta}{1 - \val(x_1)} + \textcolor{blue}{2\,\perturbation}) + 1
	- (2 - 2\perturbation - (\val(x_2) + 1 + 2\perturbation - \val(x_1))) =
	1 + \val(x_2) + 6\perturbation - 2\,\val(x_1)$.
\end{itemize}
Thus, for all valuations in $[0,2]^2$, we have
\begin{displaymath}
\rValue(\loc_0 \to \loc_4, \val) = 
\begin{cases}
\val(x_2) + 1 + 6\,\perturbation - 2\,\val(x_1)
& \text{if \textcolor{Brown}{$x_1 \leq 1$ and $\val(x_2) \leq \val(x_1) - 4\,\perturbation$};} \\
\val(x_2) - 1 + 6\,\perturbation
& \text{if \textcolor{ForestGreen}{$1 < \val(x_1) < 2- 2\,\perturbation$ and $\val(x_2) \leq 1 - 4\,\perturbation$};} \\
+\infty & \text{otherwise.} \\
\end{cases}
\end{displaymath}

\emph{We suppose that \MinPl chooses to go to $\loc_3$}.
According to the guard between $\loc_0$ and $\loc_3$, we know that $\loc_3$
is reached with the green cell, i.e.\ in $\loc_3$ (after the perturbation),
the valuations of clocks are such that $\val(x_1) \leq 2$ and $\val(x_2) \leq 1$.
Again, \MinPl wants to minimise its delay and \MaxPl perturbs with
$\textcolor{blue}{2\,\perturbation} $.
Thus, from the guard, we consider two cases according to the valuation of~$x_1$.
\begin{itemize}
	\item We suppose that $1 \leq \val(x_1) \leq 2 - 2\perturbation$ and
	\MinPl plays with the delay $\textcolor{magenta}{0}$
	when $\val(x_2) \leq 1 - 2\,\perturbation$.
	In particular, the fixed-perturbation robust value function is equal to
	$1 \times \textcolor{blue}{2\,\perturbation} + 4\perturbation -
	(\val(x_2) + 2\,\perturbation) = 4\,\perturbation - \val(x_2)$.
	
	\item We suppose that $\val(x_1) < 1$ and \MinPl plays with the delay
	$\textcolor{magenta}{1 - \val(x_1)}$ when $x_2 \leq \val(x_1) - 2\,\perturbation$.
	Moreover, the fixed-perturbation robust value function is equal to
	$1 \times (\textcolor{magenta}{1 - \val(x_1)} + \textcolor{blue}{2\,\perturbation})
	+ 4\perturbation - (\val(x_2) + 1 + 2\perturbation - \val(x_1))
	= 4\,\perturbation - \val(x_2)$.
\end{itemize}
Thus, for all valuations in $[0,2]^2$, we have
\begin{displaymath}
\rValue(\loc_0 \to \loc_3, \val) = 
\begin{cases}
4\perturbation - \val(x_2)
& \text{if \textcolor{ForestGreen}{$\val(x_1) \leq 2 - 2\perturbation$ and
		$\val(x_2) \leq \min(\val(x_1) - 2\,\perturbation, 1 - 2\,\perturbation)$};} \\
+\infty & \text{otherwise.}
\end{cases}
\end{displaymath}

\emph{Computation of the minimum}.
To conclude, we take the minimum between these two robust values functions:
we look for the minimum according to valuations of clocks.
\begin{itemize}
	\item We suppose that $\val(x_1) \leq 1$ and
	$\val(x_1) - 4\,\perturbation \leq \val(x_2) < \val(x_1) - 2\,\perturbation$.
	In this case, only $\rValue(\loc_0 \to \loc_3, \val)$ is finite.
	Thus, we deduce that $\rValue(\loc_0, \val) = 4\,\perturbation - \val(x_2)$.
	
	\item We suppose that $1 \leq \val(x_1) \leq 2- 2\,\perturbation$ and
	$1 - 4\,\perturbation < \val(x_2) \leq 1 - 2\perturbation$.
	Again, only $\rValue(\loc_0 \to \loc_3, \val)$ is finite,
	i.e.\ $\rValue(\loc_0, \val) = 	4\,\perturbation - \val(x_2)$.
	
	\item We suppose that $\val(x_1) = 2 - 2\,\perturbation$ and $\val(x_2) \leq 1 - 2\perturbation$
	and $\rValue(\loc_0, \val) = 4\,\perturbation - \val(x_2)$ by choosing~$\loc_3$.
	
	\item We suppose that $\val(x_1) \leq 1$ and $\val(x_2) \leq \val(x_1) - 4\,\perturbation$.
	In this case, both fixed-perturbation robust value functions are finite.
	Thus, we want to decide when
	$\rValue(\loc_0 \to \loc_4, \val) \leq \rValue(\loc_0 \to \loc_3, \val)$,
	i.e. when $\val(x_2) + 1 - 6\,\perturbation - 2\,\val(x_1) \leq 4\,\perturbation - \val(x_2)$.
	As this inequality holds when $\val(x_2) \leq \val(x_1) + 5\,\perturbation - 1/2$, we fix
	$\perturbationBound = 1/18$ to guarantee that the cell
	defined by $\val(x_1) + 5\,\perturbation- 1/2 \leq \val(x_1) - 4\,\perturbation$ is not empty.
	Thus, in this case, we obtain that
	\begin{displaymath}
	\rValue(\loc_0, \val) =
	\begin{cases}
	\val(x_2) + 1 + 6\,\perturbation - 2\,\val(x_1)
	& \text{if $\val(x_2) \leq \val(x_1) + 5\,\perturbation - \frac{1}{2}$;} \\
	4\,\perturbation - \val(x_2)
	& \text{if $\val(x_1) + 5\,\perturbation - \frac{1}{2} \leq \val(x_2) \leq \val(x_1) - 4\,\perturbation$;} \\
	+\infty & \text{otherwise.}
	\end{cases}
	\end{displaymath}
	
	\item We suppose that  $1 < \val(x_1) < 2 - 2\perturbation$ and $\val(x_2) \leq \val(x_1) - 4\,\perturbation$.
	In this case both fixed-perturbation robust value functions are finite and we want to
	decide when $\val(x_2) - 1 + 6\,\perturbation \leq 4\,\perturbation - \val(x_2)$.
	Since this inequality holds when $\val(x_2) \leq \frac{1}{2} - \perturbation$ that is possible
	when $\perturbation = 1/2$. In particular, $\perturbationBound$ remains $1/18$.
	Thus, in this case, we obtain that
	\begin{displaymath}
	\rValue(\loc_0, \val) =
	\begin{cases}
	\val(x_2) - 1 + 6\,\perturbation & \text{if $\val(x_2) \leq \frac{1}{2} - \perturbation$;} \\
	4\,\perturbation - \val(x_2)
	& \text{if $\frac{1}{2} - \perturbation < \val(x_2) \leq \val(x_1) - 4 \,\perturbation$;} \\
	+\infty & \text{otherwise.}
	\end{cases}
	\end{displaymath}
\end{itemize}
Thus, for all valuations in $[0,2]^2$, we have
\begin{displaymath}
\rValue(\loc_0, \val) = 
\begin{cases}
\val(x_2) - 1 + 6\,\perturbation
& \text{if }
\begin{cases}
\text{\textcolor{Brown}{$\val(x_1) \leq 1$ and $\val(x_2) \leq \val(x_1) - 5\,\perturbation - \frac{1}{2}$};} \\
\text{\textcolor{ForestGreen}{$1 < \val(x_1) < 2 - 2\,\perturbation$ and
		$\val(x_2) \leq \frac{1}{2} - \perturbation$};}
\end{cases} \\
4\,\perturbation - \val(x_2)
& \text{if }
\begin{cases}
\text{\textcolor{RedOrange}{$\val(x_1) \leq 1$ and
		$\val(x_1) - 5\,\perturbation -\frac{1}{2} < \val(x_2) \leq \val(x_1) - 2\,\perturbation$};} \\
\text{\textcolor{Purple}{$1 < \val(x_1) < 2 - 2\,\perturbation$ and
		$\frac{1}{2} - \perturbation < \val(x_2) \leq 1-2\perturbation$};} \\
\text{\textcolor{Blue}{$\val(x_1) = 2 - 2\,\perturbation$ and $\val(x_2) \leq 1 - 2\,\perturbation$};}
\end{cases} \\
+\infty & \text{otherwise.} \\
\end{cases}
\end{displaymath}

\section{Proof of Lemma~\ref{lem:valIt-robust-sem-F}}
\label{app:valIt-robust-sem-F}

To prove this result, we introduce the fixed-perturbation robust value at horizon $i$,
i.e.\ the fixed-perturbation robust value computed only on plays on length at most $i$.
Formally, given two robust strategies $\robminstrategy \in \rStratMin{\perturbation}$
and $\robmaxstrategy \in \rStratMax{\perturbation}$, we define
$\outcomes^i((\loc, \val), \robminstrategy, \robmaxstrategy)$ be
the maximal play conforming to $\robminstrategy$ and $\robmaxstrategy$ with a length at
most\footnote{If the play reaches a deadlock within $j$ steps ($j < i$), we consider
	that it is an outcome of both strategies within $i$ steps.}
$i$ from $(\loc, \val)$.
In particular, we denote by $\rValue_i$
the \emph{fixed-perturbation robust value at horizon $i$}, defined by
\begin{displaymath}
\rValue_i(\loc, \val) =
\inf_{\robminstrategy \in \rStratMin{\perturbation}}~
\sup_{\robmaxstrategy \in \rStratMax{\perturbation}}
\weightP(\outcomes^i((\loc, \val), \robminstrategy, \robmaxstrategy)) \,.
\end{displaymath}

\lemValItF*
\begin{proof}
	Let $i \in \N$ and $(\loc, \val)$ be a configuration.
	By grouping all infimum/supremum together
	in $(\F_\perturbation)^i(\V)(\loc, \val)$, it can be rewritten~as
	\begin{displaymath}
	\inf_{\big(f_k\colon ((\move[0]),\ldots,(\move[k-1])) \mapsto
		(\move[k])\big)_{\substack{0\leq k \leq i \\ \states_k \in \StatesMin}}} \,
	\sup_{\big(f_k\colon ((\move[0]),\ldots,(\move[k-1])) \mapsto
		(\move[k])\big)_{\substack{0\leq k \leq i \\ \states_k \in \StatesMax}}}
	\weight(\play)
	\end{displaymath}
	where $\play$ is the finite play $\states_0 \moveto{(\move[0])=f_0}
	\states_1 \moveto{(\move[1])=f_1(\move[0])} \cdots
	\moveto{(\move[i])=f_i((\move[0]),\ldots,(\move[i-1]))} \states_i$
	such that $\weight(\play)$ is finite if and only if $\play$ reaches a target location 
	within $i$ steps.
	Since the operator $\F_\perturbation$ only
	used robust decisions for $\perturbation$ for both players,
	we notice that the mapping $f_k$, chosen by the player owning state $\states_k$,
	describes the robust decision for $\perturbation$
	for the $k$-th step as a function of the previously chosen (robust) decisions.
	In particular, for all $i \in \N$ and configurations $(\loc, \val)$,
	we have
	\begin{equation}
	\label{eq:rob-valIt_frob}
	\rValue_i(\loc, \val) = (\F_\perturbation)^i(\V)(\loc, \val) \,.
	\end{equation}
	
	Finally, to conclude the proof, we remark that the hypothesis 
	of acyclicity of $\game$ guarantees that
	$\rValue = \rValue_D$ where $D$ is the length of the longest path in $\game$ 
	(that is finite since $\game$ is acyclic).
	In particular, by~\eqref{eq:rob-valIt_frob},
	we have $\rValue(\loc, \val) =
	\F_\perturbation^D(\V_0)(\loc, \val)$.
	To conclude, we note that
	$\F_\perturbation(\rValue)(\loc, \val) =
	\F_\perturbation^{D+1}(\V_0)(\loc, \val) =
	\rValue(\loc, \val)$.
\end{proof}

\section{Proof of Lemma~\ref{lem:rob-valIt_Frob-cons}}
\label{app:rob-valIt_Frob-cons}

\lemValItFcons*
\begin{proof}
	The case for $\loc \in \LocsT$ is direct from the definition of $\F_\perturbation$. 
	
	Now, we suppose that $\loc \in \LocsMax$ and we distinguish two cases. 
	If $\val \in R_{\loc}$, i.e.~$E(\loc, \val) = \emptyset$, by definition, $\F_\perturbation(\V_{\loc})(\val) = +\infty$.
	Now, we suppose that $\val \notin R_{\loc}$, i.e.~$E(\loc, \val) \neq \emptyset$,
	and by decomposing the supremum, we have
	\begin{align*}
	\F_\perturbation(\V_{\loc})(\val) &=
	\max_{\trans = (\loc, \guard, \reset, \loc')} \big(\weight(\trans) +
	\sup_{(\loc, \val) \moveto{\move} (\loc', \val')}
	[\delay \, \weight(\loc) + \V_{\loc'}(\val')]\big) \\
	&= \max_{\trans = (\loc, \guard, \reset, \loc')} \big(\weight(\trans) +
	\sup_{(\loc, \val) \moveto{\move} (\loc', \val')}
	[\delay \, \weight(\loc) + \unreset{\reset}(\V_{\loc'})(\val+\delay)]\big)
	\end{align*}
	since $\val' = (\val + \delay)[\reset \coloneqq 0]$.
	When we fix a transition, choosing a robust decision (to define an edge in
	the semantic) is equivalent to choosing $\delay$ to be a delay such that
	$(\val + \delay) \models \guard$.
	In particular, we deduce that
	\begin{align*}
	\F_\perturbation(\V_{\loc})(\val)
	&= \max_{\trans = (\loc, \guard, \reset, \loc')} \big(\weight(\trans) +
	\sup_{\delay}
	[\delay \, \weight(\loc) + \fguard{\trans}(\unreset{\reset}(\V_{\loc'}))(\val+\delay)]\big) \\
	&= \max_{\trans = (\loc, \guard, \reset, \loc')} \big(\weight(\trans) +
	\pre{\loc}(\fguard{\trans}(\unreset{\reset}(\V_{\loc'})))(\val)\big) \,.
	\end{align*}
	
	Finally, we suppose that  $\loc \in \LocsMin$ and we distinguish two cases. If $\val \in R_{\loc}$, i.e.~$E(\loc, \val) = \emptyset$, by definition, $\F_\perturbation(\V_{\loc})(\val) = +\infty$, which is 
	consistent with the definition of an infimum/minimum over an empty set. 
	Now, we suppose that $\val \notin R_{\loc}$.
	In particular, by decomposing the infimum,
	we obtain that $\F_\perturbation(\V_{\loc})(\val)$ is equal to
	\begin{displaymath}
	\min_{\trans = (\loc, \guard, \reset, \loc')} \big(\weight(\trans) +
	\inf_{(\loc, \val) \moveto{\move} (\loc, \val, \delay, \trans)}
	\sup_{\substack{\varepsilon \in [0, 2\perturbation] \\
			(\loc, \val, \delay, \trans) \moveto{\trans, \varepsilon} (\loc, \val')}}
	[(\delay+\perturbation)\,\weight(\loc) +
	\unreset{\reset}(\V_{\loc'})(\val+\delay+\varepsilon)]
	\end{displaymath}
	since $\val' = (\val + \delay + \varepsilon)[\reset:= 0]$.
	Then, like in the case of \MaxPl, the operator $\pre{\loc}$ 
	does not distinguish decisions that are robust and those that are not.
	In particular, this operator does not compute, a priori, the set of safe delays 
	(i.e.~the delays such that the guard is satisfied for all perturbations).
	Now, to guarantee that the chosen delay by $\pre{\loc}$ induces an edge 
	in $\sem{\game}_\perturbation$, we use the combination between 
	operators $\perturbs{\loc}$ and $\fguard{\trans}$. 
	Indeed, if the operator $\pre{\loc}$ chooses a delay such that $(\move)$
	is a non-robust decision, then there exists $\varepsilon \in [0, 2\perturbation]$ such that 
	$\fguard{\trans}(\V_{\loc'})(\val+\delay+\varepsilon) = +\infty$. 
	In particular, the computation of the following supremum 
	$\sup_{\varepsilon \leq 2\perturbation} \fguard{\trans}(\V_{\loc'})(\val+\delay+\varepsilon)]$
	guarantee that $\pre{\loc}$ (and more precisely its infimum) compute a 
	delay that induces a robust decision when it is possible.
	Thus, we rewrite the infimum and the supremum over all
	edges in $\sem{\game}_\perturbation$ of $\F_\perturbation(\V_{\loc})(\val)$ 
	by a constraint on the delay and the perturbations that we consider before
	applying the filter defined by the guard,
	i.e.\ $\F_\perturbation(\V)_{\loc}(\val)$ is equal to
	\begin{displaymath}
	\min_{\trans = (\loc, \guard, \reset, \loc')} \Big(\weight(\trans) +
	\inf_{\delay} \big(\delay \, \weight(\loc) +
	\sup_{\varepsilon \leq 2\perturbation}
	[\varepsilon \, \weight(\loc) +
	\fguard{\trans}(\unreset{\reset}(\V_{\loc'}))(\val+\delay+\varepsilon)]\big)\Big)\,.
	\end{displaymath}
	Finally, we conclude by definition of operators $\pre{\loc}$ and $\perturbs{\loc}$:
	\begin{displaymath}
	\F_\perturbation(\V)_{\loc}(\val)
	= \min_{\trans = (\loc, \guard, Y, \loc')} \left[\weight(\trans) +
	\pre{\loc}(\perturbs{\loc}(\fguard{\trans}(\unreset{Y}(\V_ {\loc'}))))(\val)\right] \,.
	\qedhere
	\end{displaymath}
\end{proof}

\section{Proof of Lemma~\ref{lem:rob-valIt_computation-atomic}}
\label{app:rob-valIt_computation-atomic}

\lemComputationAtomic*
\begin{proof}
	To define the atomic parametric piecewise affine function $F'$, we add diagonal 
	inequalities defined from each intersection between two (non-diagonal) affine 
	expressions.
	
	Formally, we consider two (non-diagonal) affine expressions 
	$E: \sum_{\clockx \in \Cl} a_\clockx\,\clockx + b + c\perturbationParam$ 
	and $E': \sum_{\clockx \in \Cl} a_\clockx'\,\clockx + b' + c'\perturbationParam$.
	By letting $A = \sum_{\clockx \in \Cl} a_\clockx$ and 
	$A' = \sum_{\clockx \in \Cl} a_\clockx'$. 
	We fix the diagonal intersection of $E$ and $E'$ (denoted by $E \cap_d E'$), 
	i.e. the hyperplane that contains the intersection of hyperplanes defined by equations $E=0$ and $E'=0$, by
	\begin{displaymath}
	E \cap_d E' = \sum_{\clockx \in \Cl} (Aa_\clockx' - A'a_\clockx)\clockx + 
	(Ab' - A'b) + (Ac' - A'c)\perturbationParam \,.
	\end{displaymath}
	On the right of \figurename{~\ref{fig:valIt-cell}}, the diagonal intersection 
	of equations of~$\Eq$ are depicted in red.

	Finally, we define $F'$ by letting
	$\Eq' = \Eq \cup \{E \cap_d E' ~|~ E, E' \in \Eq\}$, 
	$\perturbationBound' = \min(\eta(\Eq'), \eta)$ and 
	$(f'_c)_{c \in \cells(\Eq')}$ by the restriction of $(f_c)_{c \in \cells(\Eq)}$ 
	(we note that since $\Eq \subseteq \Eq'$, a cell of $\Eq'$ is always 
	contained in a cell of $\Eq$).
	To conclude, we show that $F'$ is atomic as in the non-robust case 
	(see \cite[Lemma 7.7]{BusattoGastonMR23}).
\end{proof}

\section{Proof of Proposition~\ref{prop:Fp_parametric}}
\label{app:Fp_parametric}

In this section, we detailed the proof of Proposition~\ref{prop:Fp_parametric} by given 
the computation of each operator:

\propFpParametric*

We start by considering the minimum (resp.\ the maximum) of two 
parametric value functions $F_1$ and $F_2$. 
To do so, we consider their parametric partitions $\tuple{\Eq_1, \perturbationBound_1}$ 
and $\tuple{\Eq_2, \perturbationBound_2}$, and build their \emph{intersection} 
$\tuple{\Eq_1, \perturbationBound_1}\cap \tuple{\Eq_1, \perturbationBound_1}$ 
that is a parametric partition $\tuple{\Eq, \perturbationBound}$ obtained 
by considering the union of all the expressions as well as an upperbound 
$\perturbationBound= \min( \perturbationBound_1,  \perturbationBound_2,  \perturbationBound(\Eq))$: 
therefore, each cell $c$ of the intersection is such that there exist a cell $c_1$ 
of $\tuple{\Eq_1, \perturbationBound_1}$ and a $c_2$ of $\tuple{\Eq_2, \perturbationBound_2}$ 
such that for all $p\leq \perturbationBound$, $\sem c_p = \sem{c_1}_p \cap \sem{c_2}_p$.

\begin{lemma}%{lemRobComputationMin}
	\label{lem:rob-valIt_computation-min}
	Let $\tuple{\Eq_i, \perturbationBound_i, (f_{i,c})_{c\in \cells(\Eq_i)}}_{1\leq i \leq k}$ be a set of
	parametric value functions.
	Then, there exists a parametric value function 
	$\tuple{\Eq, \perturbationBound, (f_c)_{c\in \cells(\Eq)}}$
	such that $\perturbationBound \leq \min_i \perturbationBound_i$, 
	and $\sem{F}_p = \min_{1\leq i\leq k} \sem{F_i}_p$
	(respectively, $\sem{F}_p = \max_{1\leq i\leq k} \sem{F_i}_p$) 
	for all $p\leq \perturbationBound$.
\end{lemma}
\begin{proof}
	We provide the proof only in the case $k=2$ and for the minimum operation, 
	since we can deduce the general one by induction. 
	We thus consider the intersection $\tuple{\Eq', \perturbationBound'}$
	of the two partitions $\tuple{\Eq_1, \perturbationBound_1}$ and $\tuple{\Eq_2, \perturbationBound_2}$. For a cell $c$ of this partition, we know that it is obtained by the intersection of two cells $c_1$ and $c_2$ of the two original partitions. We consider the parametric affine expressions \[f_{1,c_1} = \sum_{x\in \Cl} \alpha_x x + \beta + \gamma\perturbationParam \quad\text{ and }\quad f_{2,c_2}=\sum_{x\in \Cl} \alpha'_x x + \beta' + \gamma'\perturbationParam\] 
	We need to refine the cell $c$ to take into account the minimum of the two functions. To do so, we consider the equation describing the equality of the two functions: it is of the form $E=0$ with $E=\sum_{x\in\Cl}(\alpha'_x - \alpha_x)x +
	(\beta' - \beta) + (\gamma' - \gamma)\perturbationParam$, and 
	we thus add the expression $E$ in the partition, if this one goes through the cell $c$. In this case, this gives rise to two new cells over which $f_{1,c_1}$ and $f_{2,c_2}$ are respectively the minima: we can decide this by a careful examination of the parametric affine expressions. 
\end{proof}

Thanks to that, we are able to adapt the proofs from \cite{AlurBernadskyMadhusudan-04,BusattoGastonMR23} 
for the operators $\fguard{\trans}$, $\unreset{Y}$ and $\pre{\loc}$, that exist also in the non-robust setting. 

\begin{lemma}
	\label{lem:rob-valIt_computation-guard}
	Let $F = \tuple{\Eq, \perturbationBound, (f_c)_{c\in \cells(\Eq)}}$ 
	be a parametric value function.
	For all transitions $\trans\in\Trans$, we can compute a parametric value 
	function $F' = \tuple{\Eq', \perturbationBound', (f'_c)_{c\in \cells(\Eq')}}$ 
	such that $\perturbationBound' \leq \perturbationBound$, and $\sem{F'}_p = \fguard{\trans}(\sem{F}_p)$
	for all $p \leq \perturbationBound'$.
\end{lemma}
\begin{proof} We let $\tuple{\Eq', \perturbationBound'}$ be the intersection
	of the partition $\tuple{\Eq, \perturbationBound}$ and the partition
	obtained by considering the set of affine expressions that compose the
	guard of the transition (with all coefficients in front of
	$\perturbationParam$ being 0, and thus an upperbound as big as we want).
	For each cell of the obtained partition, either all the valuations in its semantics entirely fulfil the guard (for all values of $p\leq \perturbationBound$), or
	entirely do not fulfil it. Thus, for each cell $c\in \cells(\Eq')$, either we
	let $f'_c$ be equal to the mapping $f_{\bar c}$ with $\bar c$ the unique
	cell of $\cells(\Eq)$ that contains $c$, or we let $f'_c=\pm \infty$ otherwise, according to which player the initial location of $\trans$ belongs to.
\end{proof}

\begin{lemma}
	\label{lem:rob-valIt_computation-unreset}
	Let $F = \tuple{\Eq, \perturbationBound, (f_c)_{c\in \cells(\Eq)}}$ 
	be a parametric value function.
	For all $Y \subseteq \Cl$, we can compute a parametric value function
	$F' = \tuple{\Eq', \perturbationBound', (f'_c)_{c\in \cells(\Eq')}}$
	such that $\perturbationBound' \leq \perturbationBound$, and $\sem{F'}_p = \unreset{Y}(\sem{F}_p)$
	for all $p \leq \perturbationBound'$.
\end{lemma}
\begin{proof}
	Given a parametric affine expression $E$ of the form 
	$\sum_{x\in \Cl} \allowbreak\alpha_x x + \beta + \gamma\perturbationParam$, 
	we let $\unreset{Y}(E)$ be the parametric affine expression 
	$\sum_{x\in \Cl\setminus Y} \alpha_x x + \beta + \gamma\perturbationParam$ 
	(where we have replaced by $0$ every coefficient $\alpha_x$ with $x\in Y$). 
	In particular, we obtain that 
	$\sem E_p (\val[Y:=0]) = \sem {\unreset{Y}(E)}_p (\val)$. 
	We thus let $\Eq'$ be the set of expressions $\unreset{Y}(E)$ for all $E\in \Eq$.
	For all cells $c\in \cells(\Eq')$, let $\tilde c$ be the cell of $\cells(\Eq)$ that 
	contains the part of the cell $c$ where all clocks of $Y$ are equal to $0$.
	Then, we let $f'_{c} = \unreset{Y}(f_{\tilde c})$, so that for all $p \leq
	\min(\perturbationBound, \perturbationBound')$, and all valuations
	$\val\in\sem {c}_p$, we have $\sem{F'}_p(\val) = \sem{f'_{c}}_p(\val)=
	\sem{\unreset{Y}(f_{\tilde c})}_p(\val) =
	\sem{f_{\tilde c}}_p(\val[Y:=0]) = \unreset{Y}(\sem{f_{\tilde c}}_p(\val)) =
	\unreset{Y}(\sem{F}_p(\val))$.
\end{proof}

\begin{lemma}
	\label{lem:rob-valIt_computation-pre}
	Let $F = \tuple{\Eq, \perturbationBound, (f_c)_{c\in \cells(\Eq)}}$ 
	be a parametric value function.
	For all locations~$\loc$, we can compute a parametric value function 
	$F' = \tuple{\Eq', \perturbationBound', (f'_c)_{c\in \cells(\Eq')}}$
	such that $\perturbationBound' \leq \perturbationBound$, and $\sem{F'}_p = \pre{\loc}(\sem{F}_p)$ 
	for all $p \leq \perturbationBound'$.
\end{lemma}
\begin{proof}
	By Lemma~\ref{lem:rob-valIt_computation-atomic}, we can suppose that the 
	partition we start with is atomic.
	Suppose now that $\loc \in \LocsMax$
	(the case for \MinPl is symmetrical by replacing the supremum with an infimum),
	and consider a valuation $\val$, as well as a fixed perturbation $p\leq \perturbationBound$. Then, 
	\begin{displaymath}
	\pre{\loc}(\sem{F}_p)(\val) = \sup_{\delay\geq 0}
	[\delay \, \weight(\loc) + \sem{F}_p(\val + \delay)] \,.
	\end{displaymath}
	For every delay $\delay>0$, the valuation $\val + \delay$ belongs
	to the open diagonal half-line from $\val$, which crosses some of the semantics $\sem{c'}_p$ for $c'$ some cells of the parametric partition (this subset of cells is finite since there are anyway only a finite 
	number of cells in the partition). 
	Since the partition is atomic, this subset of cells 
	neither depends on the choice of $\val$ in a given starting cell $c$, 
	nor on the 
	perturbation $p$ as long as it is at most $\perturbationBound$. 
	For a starting cell $c$, we thus let $D_{c}$ 
	be the set of cells that intersect the open diagonal half-lines starting 
	from $c$.\footnote{This is a subset of what we intuitively called a \emph{tube} in Section~\ref{sec:encoding}.}
	
	Since the function $\sem{F}_p$ is affine in each cell, the above supremum 
	over the possible delays is obtained for a value $\delay$ that is either 
	$0$, or tending to $+\infty$, or on one of the two non-diagonal 
	borders of a cell that intersect the 
	open diagonal half-line from $\val$ (indeed diagonal borders either completely contain this half-line or do not intersect it), and we thus only have to consider those borders. 
	Once again because the partition is atomic (and $\perturbationBound$ 
	is small enough), the set of those borders do not depend on the choice
	of $\val$ in a given starting cell $c$, nor on the perturbation $p$ as long as it is at most~$\perturbationBound$.
	For a starting cell $c$ and a cell $c'\in D_c$, we thus let $B_{c,c'}$ be 
	the non-diagonal borders of $c'$ that intersect the diagonal half-lines from $c$: if $c=c'$, this consists of one of the non diagonal borders of $c'$, 
	otherwise of both non diagonal borders of~$c'$.
	
	Consider again a valuation $\val$, as well as a fixed perturbation $p$, and let 
	$c$ be the cell that contains $\val$ when the perturbation is $p$. 
	For $c'\in D_c$, and $B\in B_{c,c'}$, there exists
	a unique $\delay_{\val, B, p} \in \Rplus$ such that $\val + \delay_{\val, B, p}$ 
	lies on the border $B$: 
	if $B=\sum_{x\in\Cl} \alpha_x \, x + \beta + \gamma\perturbationParam$ with
	$A = \sum_{x\in\Cl} \alpha_x\neq 0$ (since $B$ is non-diagonal), 
	then $\delay_{\val, B, p}  = -\frac{1}{A} \big(\sum_{x\in\Cl} \alpha_x \val(x) + 
	\beta + \gamma\perturbation\big)$.
	The supremum in the definition $\pre{\loc}(\sem{F}_p)(\val)$ 
	can thus be rewritten as
	\begin{displaymath}
	\max
	\begin{cases}
	\sem{f_c}_p(\val) \\
	\max_{c' \in D_c} \max_{B \in B_{c,c'}} [\delay_{\val, B, p} \, \weight(\loc) + 
	\sem{f_{c'}}_p(\val + \delay_{\val, B, p})] \\
	\lim_{t\to +\infty} (t\weight(\loc)+\sem{F}_p(\val+t))
	\end{cases}
	\end{displaymath}
	where the first term of the external maximum corresponds to the delay $0$, 
	the second term corresponds to a jump arbitrarily close to the border $B$, 
	and the last term corresponds to a delay tending to~$+\infty$.
	
	The various finite maxima in this formula can be computed by using 
	Lemma~\ref{lem:rob-valIt_computation-min}. The limit when $t$ tends to $+\infty$ 
	can be computed by using the parametric affine expression $f_{c'}$ in the furthest cell $c'$ in the $D_c$.
	It only remains to explain how to compute the value 
	$\delay_{\val, B, p} \, \weight(\loc) + 
	\sem{f_{c'}}_p(\val + \delay_{\val, B, p})$, in a way that does not depend on 
	$\val$ and $p$. Suppose that $f_{c'} = \sum_{x\in \Cl} \alpha'_x x + \beta' + 
	\gamma'\perturbationParam$, with $A' = \sum_{x\in \Cl} \alpha'_x$. By using the 
	value of $\delay_{\val, B, p}$ computed above, we have 
	\begin{multline*}
	\delay_{\val, B, p} \, \weight(\loc) + 
	\sem{f_{c'}}_p(\val + \delay_{\val, B, p}) \\ 
	\quad\begin{aligned}
	&= \sum_{x\in \Cl} \alpha'_x \val(x) + (A'+\weight(\loc)) \delay_{\val, B, p} + 
	\beta' +\gamma'\perturbation \\
	&= \sum_{x\in \Cl} \Big(\alpha'_x-\frac{A'+\weight(\loc)}A\alpha_x\Big) \val(x) + \beta' - \frac {A'+\weight(\loc)} A \beta
	+ \Big(\gamma'-\frac {A'+\weight(\loc)}A \gamma\Big)\perturbation
	\end{aligned}
	\end{multline*}
	This term can be encoded by the following parametric expression, allowing us to conclude:  
	\begin{displaymath}
	\sum_{x\in \Cl} \Big(\underbrace{\alpha'_x-\frac{A'+\weight(\loc)}A\alpha_x}_{\in \Q}\Big) x + \underbrace{\beta' - \frac {A'+\weight(\loc)} A \beta}_{\in\Q\cup\{-\infty,+\infty\}}  +
	\Big(\underbrace{\gamma'-\frac {A'+\weight(\loc)}A \gamma}_{\in\Q}\Big)\perturbationParam\qedhere
	\end{displaymath}
\end{proof}

Finally, we need to compute $\perturbs{\loc}$ for a location $\loc\in \LocsMin$. We obtain this by a subtle adaptation of the operator $\pre{\loc}$, taking into account the parametric representation of the perturbation $p$. 

\begin{lemma}
	\label{lem:rob-valIt_computation-perturbs}
	Let $F = \tuple{\Eq, \perturbationBound, (f_c)_{c\in \cells(\Eq)}}$ 
	be a parametric value function.
	For all locations $\loc\in\LocsMin$, we can compute a parametric value function 
	$F' = \tuple{\Eq', \perturbationBound', (f'_c)_{c\in \cells(\Eq')}}$
	such that $\perturbationBound' \leq \perturbationBound$, and $\sem{F'}_p = \perturbs{\loc}(\sem{F}_p)$
	for all $p \leq \perturbationBound'$.
\end{lemma}
\begin{proof}
	We suppose that the parametric partition is atomic. 
	Let $\val$ be a valuation. We have:
	\begin{displaymath}
	\perturbs{\loc}(\sem{F}_p)(\val) =
	\sup_{\varepsilon \in [0, 2\perturbation]}
	[\varepsilon\,\weight(\loc) + \sem{F}_p(\val +\varepsilon)] \,.
	\end{displaymath}
	We want to use the same technique as for $\pre{\loc}$, that is to write this supremum as a maximum over a finite number of functions, in particular using the borders of the partition. However, the situation is more complex since $2p$ is now a possible delay, and moreover all delays should be at most $2p$ which suppresses some possible non diagonal borders in the diagonal half-line from $\val$.
	
	We take care of the first issue by considering $\varepsilon=0$ and $\varepsilon=2p$ as special cases. We take care of the second issue by first modifying the parametric partition, in order to incorporate the $2p$ granularity over the delays. Indeed, we let $\tuple{\Eq_1,\perturbationBound_1}$ be the atomic parametric partition obtained by Lemma~\ref{lem:rob-valIt_computation-atomic} from the set of expressions that contains $\Eq$ and all expressions $E-2\perturbationParam$ for $E$ a non-diagonal expression of $\Eq$. We can restrict the affine functions $f_c$ to be also defined on cells of $\Eq_1$.

	We now claim that, similarly to the proof of $\pre{\loc}$, for every cell
	$c\in \cells(\Eq_1)$ and perturbation $p\leq \perturbationBound_1$, if two
	valuations $\val_1$ and $\val_2$ belong to $\sem c_p$, then the open
	diagonal segments from the valuation to the valuation translated by the
	delay $2p$ cross the same cells and borders of $\Eq$. We again let $D_c$
	and $B_{c,c'}$ the set of cells $c'$ of $\Eq$ crossed by the open diagonal
	segments of length $2p$ from valuations in $c$, and the non diagonal
	borders of $c'$ that are indeed crossed. Moreover, for all valuations $\val \in \sem{c}_p$, the valuation $\val+2p$ is always in the same cell $c''$ of $\Eq$, independent on the choice of $\val$ and $p$. Thus, for $c\in \cells(\Eq_1)$ and
	$\val\in \sem{c}_p$, we can write
	$\perturbs{\loc}(\sem{F}_p)(\val)$ as 
	\begin{displaymath}
	\max 
	\begin{cases}
	\sem{f_c}_p(\val) \\
	\max_{c' \in C_c}
	\max_{B \in B_{c,c'}}
	[t_{\val, B, p} \weight(\loc) +
	\sem{f_{c'}}(\val + t_{\val, B, p})] \\
	\sem{f_{c''}}(\val+2p)
	\end{cases}
	\end{displaymath}
	with $t_{\val, B, p}$ defined as in the proof of $\pre{\loc}$.
	The last term can be rewritten as a parametric affine expression, as is the first term. 
	The value $t_{\val, B, p} \weight(\loc) + \sem{f_{c'}}(\val + t_{\val, B, p})$ 
	can be computed in a way that does not depend on $\val$ and $p$, for similar 
	reasons as for $\pre{\loc}$. We thus conclude by repeated applications of 
	Lemma~\ref{lem:rob-valIt_computation-min}.
\end{proof}

This concludes the proof of Proposition~\ref{prop:Fp_parametric}.

\section{Proof of Theorem~\ref{thm:rob-limit_div}}
\label{app:rob-limit_div}

\thmDiv*
We recall that to compute the robust value of a divergent \WTG 
(without configurations with a value equal to $-\infty$), 
we prove that the value iteration algorithm converges in a finite time 
(i.e. the number of iterations does not depend on $\perturbation$) 
by relying on the decomposition into SCC of divergent \WTG. 

Before to prove this result, we provide some preliminary results. 
First, we recall that a strategy for \MinPl, $\robminstrategy \in \rStratMin{\perturbation}$, 
is \emph{$\varepsilon$-optimal strategy w.r.t.~$\rValue$} when, 
for all configurations $(\loc, \val)$, 
$\sup_{\robmaxstrategy \in \rStratMax{\perturbation}}
\weight(\outcomes((\loc, \val), \robminstrategy, \robmaxstrategy)) \leq \rValue(\loc, \val) + \varepsilon$.
Then, we prove that the conservative semantics does not create new cyclic path. 
In particular, if the \WTG is divergent under the exact semantics, then all cyclic play 
following a cyclic region path under the conservative semantics have the same weight.

\paragraph*{The conservative semantics preserves the sign of paths}

Since the conservative semantics only filter plays from the 
exact semantics, we prove  that the 
positivity or the negativity of the paths of $\game$ are preserved by 
the conservative semantics: 
\begin{restatable}{lemma}{lemLimitDiv}
	\label{lem:rob_valPb-gdelta-div}
	Let $\game$ be a \WTG and $\ppath$ be a cyclic path of $\game$.
	If $\ppath$ is positive (resp. negative), then all plays 
	defined with the conservative semantics along $\ppath$ 
	are positive (resp. negative).
\end{restatable}
\begin{proof}
	We suppose that $\ppath$ is a positive cyclic path (the negative case is analogous). 
	Let $\play$ be a play in $\game$ w.r.t. the conservative semantics
	following $\rpath$, and we prove that $\weightC(\play) \geq 1$.	
	To do it, we define a play $\play'$ in $\game$ 
	w.r.t. the exact semantics such that $\play'$ follows $\rpath$ 
	and $\weightC(\play) = \weightC(\play')$.
	Indeed, in this case, by hypothesis on $\ppath$, we know that 
	$\weightC(\play') > 1$, thus $\weightC(\play) > 1$.
	
	To conclude the proof, we define a function to define $\play'$. 
	In particular, we define the function 
	$\inj : \FPlays[\perturbation] \to \FPlays[\exact]$
	by induction on the length of plays such that for all finite plays ending in a configuration of $\game$
	$\play \in \FPlays[\perturbation]$, we~let 
	\begin{displaymath}
	\inj(\play) = 
	\begin{cases}
	(\loc, \val) & \text{if $\play = (\loc, \val)$} \\
	(\loc, \val) \moveto{\move} \inj(\play')
	& \text{if $\play = (\loc, \val) \moveto{\move} \play'$ with $\loc \in \LocsMax$} \\
	(\loc, \val) \moveto{\move + \varepsilon} \inj(\play')
	& \text{if $\play = (\loc, \val) \moveto{\move} (\loc, \val, \move) \moveto{\trans, \varepsilon} \play'$ 
		with $\loc \in \LocsMin$} \\
	(\loc, \val, \move + \varepsilon) \moveto{\trans, 0} \inj(\play')
	& \text{if $\play = (\loc, \val, \move) \moveto{\trans, \varepsilon} \play'$} 
	\end{cases}
	\end{displaymath}
	The function $\inj$ fulfils the following properties:
	given a play w.r.t. the conservative semantics $\play \in \FPlays[\perturbation]$, 
	we have
	\begin{enumerate}
		\item\label{itm:rob_valPb-rob2exact-inj-last}
		$\inj(\play) \in \FPlays[\exact]$
		and $\last(\play) \in \Conf$ if and only if
		$\last(\inj(\play)) \in \Conf$.
		Moreover, if $\last(\play) \in \Conf$, then
		$\last(\play) = \last(\inj(\play))$;
		
		\item\label{itm:rob_valPb-rob2exact-inj-path}
		let $\rpath$ be a region path followed by $\play$, then $\inj(\play)$ follows $\rpath$;
		
		\item\label{itm:rob_valPb-rob2exact-inj-weight}
		if $\play$ ends in a configuration of $\game$, then $\weightC(\play) = \weightC(\inj(\play))$.
	\end{enumerate}
	
	The proof of~\eqref{itm:rob_valPb-rob2exact-inj-last} is given by definitions of the conservative semantics 
	and the function $\inj$. 
	
	Next, we prove~\eqref{itm:rob_valPb-rob2exact-inj-path} by remarking that $\inj$ that does not 
	change the sequence of transitions of the play (that defines a path)
	nor the sequence of configuration by~\eqref{itm:rob_valPb-rob2exact-inj-last} (that defines 
	the sequence of regions in a region path).  
	
	Finally, we prove~\eqref{itm:rob_valPb-rob2exact-inj-weight} by induction on the length of $\play$.
	If $\play = (\loc, \val)$ then the property is trivial.
	Otherwise, we let $\play = (\loc, \val) \moveto{\move} \play'$
	and we distinguish two cases according to $\loc$.
	If $\loc \in \LocsMax$, then we have 
	$\inj(\play) = 	(\loc, \val) \moveto{\move} \inj(\play')$. 
	We conclude by hypothesis of induction applied on $\play'$. 
	Otherwise $\loc \in \LocsMin$ and, since $\play$ ends in a configuration of $\game$, we rewrite
	$\play' = (\loc, \val, \move) \moveto{\trans, \varepsilon} \play''$. 
	Thus,we have
	$\inj(\play) = (\loc, \val) \moveto{\move + \varepsilon} (\loc, \val, \move + \varepsilon) \moveto{\trans, 0} \inj(\play'')$.
	Moreover, since $\play''$ is a play (it starts by a configuration), we have $\weightC(\inj(\play'')) = \weightC(\play'')$
	(by the induction hypothesis applied on $\play''$).
	Now, by definition of the conservative semantics, we deduce that
	\begin{align*}
	\weightC(\inj(\play))
	&= \weightC(\inj(\play'')) + (\delay + \varepsilon) \, \weight(\loc) +
	\weight(\trans) \\
	&= \weightC(\play'') + (\delay + \varepsilon) \, \weight(\loc) + \weight(\trans)
	= \weightC(\play) \,. \qedhere
	\end{align*}
\end{proof}

\paragraph*{The case of positive SCCs}
Let $\game$ be a \WTG that only contains positive cyclic region paths (or just a positive SCC). 
Intuitively, in such a \WTG, the interest of \MinPl is to quickly 
reach a target location of $\game$ to minimise the number of 
positive cyclic region paths followed along the play.
Formally, by letting $A$ be the bound obtained in Proposition~\ref{prop:reach-cons}
on the length of plays to reach a target location and $\QLocs$ be the 
set of region locations of $\game$, we deduce the following lemma:
\begin{lemma}
	\label{lem:rob-limit_div-pos}
	Let $p >0$ be a perturbation, and $(\loc, \val)$ be a configuration 
	from which \MinPl can reach a target location in $\sem[\perturbation]{\game}$. 
	Let $0 < \varepsilon < 1$, $\robminstrategy \in \rStratMin{\perturbation}$ be an
	$\varepsilon$-optimal strategy w.r.t.~$\rValue$, 
	and $\play$ be a play conforming to $\robminstrategy$ from $(\loc, \val)$.
	Then, $|\play| \leq (A\maxWeightEdge + |\QLocs|\maxWeightEdge+1)|\QLocs|$.
\end{lemma}
\begin{proof}
	First, since \MinPl can reach a target location in $\sem[\perturbation]{\game}$ 
	from $(\loc, \val)$, there exists a (robust) strategy $\robminstrategy^*$ 
	for \MinPl such that, for all plays $\play^*$ conforming to $\robminstrategy^*$ 
	in $\sem[\perturbation]{\game}$, $\weight(\play^*) \leq A\maxWeightEdge$ 
	(by Proposition~\ref{prop:reach-cons2}).
	In particular, we deduce that  
	$\rValue(\loc, \val) \leq A\maxWeightEdge$.
	Moreover, since $\robminstrategy$ is an $\varepsilon$-optimal strategy w.r.t.~$\rValue$, 
	we obtain that 
	\begin{displaymath}
	\weight(\play) \leq \rValue(\loc, \val) + \varepsilon \leq A\maxWeightEdge + \varepsilon \,.
	\end{displaymath}
	By contradiction, we suppose that 
	$|\play| >  (A\maxWeightEdge + |\QLocs|\maxWeightEdge+1)|\QLocs|$.
	In particular, $\play$ follows at least $(A\maxWeightEdge + |\QLocs|\maxWeightEdge + 1)$ 
	positive cyclic region paths before reaching a target location within $|\QLocs|$ steps. 
	Thus, we deduce that 
	\begin{displaymath}
	\weight(\play) \geq A\maxWeightEdge + |\QLocs|\maxWeightEdge + 1 -|\QLocs|\maxWeightEdge
	=  A\maxWeightEdge + 1 > A\maxWeightEdge + \varepsilon  \,.
	\end{displaymath}
	Finally, we obtain a contradiction by combining both inequalities.
\end{proof}

Using this property, we obtain a bound $H^+$ (independent of $\perturbation$ and~$\varepsilon$) 
over the length of the plays 
conforming to an $\varepsilon$-optimal strategy of \MinPl 
w.r.t.~$\rValue$.  In particular, since Min can avoid the plays 
longer than $H^+$, 
we deduce that these are irrelevant in the computation of the robust value.
Thus, $H^+$ is a bound on the number of iterations needed for the value iteration to get its fixpoint.
In particular, we use the fixed-perturbation robust value 
at horizon $i$ (introduced in the proof of Lemma~\ref{lem:valIt-robust-sem-F}).
We recall that the fixed-perturbation robust value at horizon $i$, 
denoted by $\rValue_i$, is defined by
\begin{displaymath}
\rValue_i(\loc, \val) =
\inf_{\robminstrategy \in \rStratMin{\perturbation}}~
\sup_{\robmaxstrategy \in \rStratMax{\perturbation}}
\weightP(\outcomes^i((\loc, \val), \robminstrategy, \robmaxstrategy)) 
\end{displaymath}
where $\outcomes^i((\loc, \val), \robminstrategy, \robmaxstrategy)$ denotes the 
plays of length at most $i$ from $(\loc, \val)$ and conforming to 
both strategies. Let $i \in \N$. In particular, by grouping all infimum/supremum together 
along the computation of $\F_\perturbation$, 
we obtain~\eqref{eq:rob-valIt_frob}, 
i.e. $\rValue_i = (\F_\perturbation)^i(\V^0)$.
Moreover, since for all $\robminstrategy \in \rStratMin{\perturbation}$, 
and $\robmaxstrategy \in \rStratMax{\perturbation}$, 
$\weightP(\outcomes^i((\loc, \val), \robminstrategy, \robmaxstrategy)) \in 
\{\weightP(\outcomes((\loc, \val), \robminstrategy, \robmaxstrategy)), +\infty\}$, 
then $\rValue \leq \rValue_i$.

\begin{restatable}{proposition}{robLimDivPos}
	\label{prop:rob-limit_div-pos}
	Let $\game$ be a \WTG that contains only positive cyclic region paths,
	and $p >0$ be a perturbation. Then,
	$\rValue = (\F_\perturbation)^{(A\maxWeightEdge + |\QLocs|\maxWeightEdge+1)|\QLocs|}(\V^0)$.
\end{restatable}
\begin{proof} Let $N = (A\maxWeightEdge + |\QLocs|\maxWeightEdge+1)|\QLocs|$.
	Let $(\loc, \val)$ be a configuration of $\game$. 
	We distinguish two cases according to the existential 
	robust reachability value in $\game$ from $(\loc, \val)$:
	\begin{itemize}
		\item If \MinPl cannot reach a target location in $\sem[\perturbation]{\game}$ 
		from $(\loc, \val)$, then for all $i \in \N$, $(\F_\perturbation)^i(\V^0)(\loc, \val) = +\infty$.
		Thus, we have  
		$(\F_\perturbation)^{N}(\V)(\loc, \val) = 
		+\infty = \rValue(\loc, \val)$.
		
		\item Otherwise, \MinPl can reach a target location in $\sem[\perturbation]{\game}$ 
		from $(\loc, \val)$. In particular, we need to prove that  
		$\rValue(\loc, \val) = \rValue_{N}(\loc, \val)$
		to conclude by~\eqref{eq:rob-valIt_frob}.
		Since, for all $i \in \N$, $\rValue \leq \rValue_i$, we just need to prove that 
		$\rValue_{N}(\loc, \val) \leq \rValue(\loc, \val)$.
		
		To prove this inequality, by letting $0 < \varepsilon < 1$, 
		we remark that all plays conforming to an $\varepsilon$-optimal strategy 
		w.r.t.~$\rValue$ for \MinPl have a length at most $N$ 
		(by Lemma~\ref{lem:rob-limit_div-pos}).
		In particular, we have
		\begin{displaymath}
		\inf_{\robminstrategy \in \rStratMin{N,\perturbation}}~
		\sup_{\robmaxstrategy \in \rStratMax{\perturbation}}
		\weightP(\outcomes((\loc, \val), \robminstrategy, \robmaxstrategy)) 
		\leq \rValue(\loc, \val) + \varepsilon
		\end{displaymath} 
		where $\rStratMin{N,\perturbation}$ denotes the set of strategies 
		of \MinPl such that all plays conforming to them has 
		a length at most $N$, i.e.~for all $\robmaxstrategy \in \rStratMax{\perturbation}$, 
		$\outcomes((\loc, \val), \robminstrategy, \robmaxstrategy) = 
		\outcomes^N((\loc, \val), \robminstrategy, \robmaxstrategy)$.
		In particular, from this definition, we deduce that 
		\begin{displaymath}
		\inf_{\robminstrategy \in \rStratMin{N,\perturbation}}~
		\sup_{\robmaxstrategy \in \rStratMax{\perturbation}}
		\weightP(\outcomes^N((\loc, \val), \robminstrategy, \robmaxstrategy)) 
		\leq \rValue(\loc, \val) + \varepsilon \,.
		\end{displaymath}
		Now, by considering an infimum over all possible strategies for \MinPl, we deduce that  
		\begin{displaymath}
		\rValue_N(\loc, \val) = 
		\inf_{\robminstrategy \in \rStratMin{\perturbation}}~
		\sup_{\robmaxstrategy \in \rStratMax{\perturbation}}
		\weightP(\outcomes^N((\loc, \val), \robminstrategy, \robmaxstrategy)) 
		\leq \rValue(\loc, \val) + \varepsilon \,.
		\end{displaymath}
		Since this holds for all $\varepsilon > 0$, we conclude that $\rValue(\loc, \val) = 
		\rValue_{N}(\loc, \val)$.\qedhere
	\end{itemize}
\end{proof}

\paragraph*{The case of negative SCCs}
Let $\game$ be a \WTG that only contains negative cyclic region paths (or just a negative SCC) and no configurations of value $-\infty$. 
In this case, \MaxPl can always reach a target location 
(otherwise, \MinPl controls a negative cyclic region path, 
and the value of each locations in this cyclic region path is 
$-\infty$~\cite{BusattoGastonMR23}).
In particular, the interest of \MaxPl is to quickly reach a target 
location of $\game$ to minimise the number of negative cyclic region 
paths followed along the play. Formally, by letting 
$B = |\inf_{(\loc, \val) \in \sem{\game}} \dValue(\loc, \val)|$ 
(that is finite by hypothesis over the value in $\game$)
and $\QLocs$ be the set of region locations of $\game$, 
we obtain the following lemma:
\begin{lemma}
	\label{lem:rob-limit_div-neg}
	Let $p >0$ be a perturbation and $(\loc, \val)$ be a 
	configuration from where \MinPl can reach a target location in 
	$\sem[\perturbation]{\game}$. 
	Let $0 < \varepsilon < 1$, $\robmaxstrategy \in \rStratMax{\perturbation}$ 
	be an $\varepsilon$-optimal strategy w.r.t.~$\rValue$, 
	and $\play$ be a play conforming to $\robmaxstrategy$ from $(\loc, \val)$.
	Then, $|\play| \leq (B + |\QLocs|\maxWeightEdge + 1)|\QLocs|$.
\end{lemma}
\begin{proof}
	First, since $\robmaxstrategy$ is an $\varepsilon$-optimal strategy 
	w.r.t.~$\rValue$, we obtain (by Lemma~\ref{lem:rVal-monotony}) that 
	\begin{displaymath}
	\weight(\play) \geq \rValue(\loc, \val) - \varepsilon  \geq \dValue(\loc, \val) - \varepsilon
	>  \dValue(\loc, \val) - 1\,.
	\end{displaymath}
	Now, by contradiction, we suppose $|\play| > (B + |\QLocs|\maxWeightEdge + 1)|\QLocs|$.
	In particular, $\play$ follows at least $(B + |\QLocs|\maxWeightEdge + 1)$ 
	negative cyclic region paths before reaching a target location within
	$|\QLocs|$ steps. Since $-B \leq \dValue(\loc, \val)$, 
	we deduce that 
	\begin{displaymath}
	\weight(\play) \leq |\QLocs|\maxWeightEdge - |\QLocs|\maxWeightEdge - B - 1 
	= - B - 1 \leq \dValue(\loc, \val) - 1
	\end{displaymath}
	Thus, we obtain a contradiction by combining both inequalities.
\end{proof}

Using this property, we obtain a bound $H^-$ (independent of $\perturbation$ and~$\varepsilon$) 
over the length of the plays conforming to an $\varepsilon$-optimal 
strategy of \MaxPl w.r.t.~$\rValue$. In particular, since \MaxPl 
can avoid the plays longer than $H^-$, 
we deduce that these are irrelevant in the 
computation of the robust value. 
Thus, $H^-$ is a bound on the number of iterations needed for the value iteration to get its fixpoint.

\begin{restatable}{proposition}{robLimDivNeg}
	\label{prop:rob-limit_div-neg}
	Let $\game$ be a \WTG that contains only negative cyclic region paths
	and no configuration of value $-\infty$.
	Let $p >0$ be a perturbation, then 
	$\rValue = (\F_\perturbation)^{(B + |\QLocs|\maxWeightEdge + 1)|\QLocs|}(\V^0)$.
\end{restatable}
\begin{proof} 
	Let $N = (B + |\QLocs|\maxWeightEdge + 1)|\QLocs|$.
	We follow the same proof as for Proposition~\ref{prop:rob-limit_div-pos}. 
	Since for all $\robminstrategy \in \rStratMin{\perturbation}$, 
	and $\robmaxstrategy \in \rStratMax{\perturbation}$, 
	$\weightP(\outcomes^i((\loc, \val), \robminstrategy, \robmaxstrategy)) \in 
	\{\weightP(\outcomes((\loc, \val), \robminstrategy, \robmaxstrategy)), +\infty\}$, 
	we also have $\rValue \leq \rValue_i$.
	
	Let $(\loc, \val)$ be a configuration of $\game$. 
	We distinguish two cases according to the existential 
	robust reachability value in $\game$ from $(\loc, \val)$:
	\begin{itemize}
		\item If \MinPl cannot reach a target location in $\sem[\perturbation]{\game}$ 
		from $(\loc, \val)$, then for all $i \in \N$, $(\F_\perturbation)^i(\V^0)(\loc, \val) = +\infty$.
		Thus, we have  
		$(\F_\perturbation)^{N}(\V)(\loc, \val) = 
		+\infty = \rValue(\loc, \val)$.
		
		\item Otherwise, \MinPl can reach a target location in $\sem[\perturbation]{\game}$ 
		from $(\loc, \val)$. In particular, we need to prove that  
		$\rValue(\loc, \val) = \rValue_{N}(\loc, \val)$
		to conclude by~\eqref{eq:rob-valIt_frob}.
		Again we just need to prove that 
		$\rValue_{N}(\loc, \val) \leq \rValue(\loc, \val)$, 
		since, for all $i \in \N$, $\rValue \leq \rValue_i$.
		
		To prove this inequality, let $0 < \varepsilon < 1$, 
		$\robminstrategy \in \rStratMin{\perturbation}$ and 
		$\robmaxstrategy \in \rStratMax{\perturbation}$ be two $\varepsilon$-optimal strategies 
		for \MinPl and \MaxPl respectively.
		Since $\robminstrategy$ is $\varepsilon$-optimal, we have that 
		$\weight(\Plays((\loc, \val), \robminstrategy, \robmaxstrategy)) \leq \rValue(\loc, \val) + \varepsilon$.
		Moreover, since all plays conforming to an $\varepsilon$-optimal strategy 
		w.r.t.~$\rValue$ for \MaxPl have a length at most $(|\QLocs|\maxWeightEdge+B+1)|\QLocs|$ 
		(by Lemma~\ref{lem:rob-limit_div-neg}), we have
		\begin{displaymath}
		\weight(\Plays^N((\loc, \val), \robminstrategy, \robmaxstrategy)) = 
		\weight(\Plays((\loc, \val), \robminstrategy, \robmaxstrategy))  \leq \rValue(\loc, \val) + \varepsilon
		\end{displaymath} 
		Now, by applying the supremum over robust strategies of \MaxPl in this inequality, we have 
		\begin{displaymath}
		\sup_{\robmaxstrategy \in \rStratMax{\perturbation}} 
		\weight(\Plays^N((\loc, \val), \robminstrategy, \robmaxstrategy)) \leq \rValue(\loc, \val) + \varepsilon \,.
		\end{displaymath} 
		By considering an infimum over all possible strategies for \MinPl, we have
		\begin{displaymath}
		\rValue_N(\loc, \val) = 
		\inf_{\robminstrategy \in \rStratMin{\perturbation}}~
		\sup_{\robmaxstrategy \in \rStratMax{\perturbation}}
		\weightP(\outcomes^N((\loc, \val), \robminstrategy, \robmaxstrategy)) 
		\leq \rValue(\loc, \val) + \varepsilon
		\end{displaymath}
		and we conclude since this the inequality holds for all $\varepsilon > 0$.\qedhere
	\end{itemize}
\end{proof}

\paragraph*{The case of a divergent \WTG}
To compute the robust value in a divergent \WTG, 
we combine both approaches by successively computing the robust value 
in each SCC of the graph of the region game. 
Formally, let $\game$ be a divergent \WTG 
without configurations of value $-\infty$. 
We prove that there exists $H \in \N$ such that, 
for all $p > 0$, $\rValue = \F^{H}_p(\V^0)$. 
Then, since we know how to compute one iteration of $\F_{\perturbation}$ 
(from Proposition~\ref{prop:Fp_parametric}), we can compute the robust value 
in $\game$. 

The theoretical existence of the bound $H$ above comes from the decomposition into SCCs of the graph of~$\rgame$, though we do not need to compute this decomposition then to compute the robust value: we will simply stop as soon as we have reached a fixpoint (which is forced to happen by this proof). 
For each SCC, we use the bound ($H^+$ or $H^-$ according to the sign of the SCC) 
over the number of iterations of $\F_{\perturbation}$ given 
by Propositions~\ref{prop:rob-limit_div-pos} or~\ref{prop:rob-limit_div-neg} 
to (only) solve the SCC. 
Then, for each branch along the (acyclic) decomposition into SCCs of the region game, 
we sum the bound of each SCC as well as its length. 
Finally, by letting $H$ be the maximum obtained over each branch, 
we conclude the proof of Theorem~\ref{thm:rob-limit_div}.


\begin{thebibliography}{10}
	
	\bibitem{AlurBernadskyMadhusudan-04}
	Rajeev Alur, Mikhail Bernadsky, and P.~Madhusudan.
	\newblock Optimal reachability for weighted timed games.
	\newblock In {\em Proceedings of the 31st International Colloquium on Automata,
		Languages and Programming (ICALP'04)}, volume 3142 of {\em LNCS}, pages
	122--133. Springer, 2004.
	
	\bibitem{AlurDill-94}
	Rajeev Alur and David~L. Dill.
	\newblock A theory of timed automata.
	\newblock {\em Theoretical Computer Science}, 126(2):183--235, 1994.
	
	\bibitem{AsarinMaler-99}
	Eugene Asarin and Oded Maler.
	\newblock As soon as possible: Time optimal control for timed automata.
	\newblock In {\em Hybrid Systems: Computation and Control}, volume 1569 of {\em
		LNCS}, pages 19--30. Springer, 1999.
	
	\bibitem{BacciBFLMR21}
	Giovanni Bacci, Patricia Bouyer, Uli Fahrenberg, Kim G. Larsen, Nicolas Markey, and Pierre{-}Alain Reynier.
	\newblock Optimal and robust controller synthesis using energy timed automata with uncertainty.
	\newblock In {\em Formal Aspects Comput.}, 2021.
	
	\bibitem{BouyerBrihayeBruyereRaskin-07}
	Patricia Bouyer, Thomas Brihaye, V{\'e}ronique Bruy{\`e}re, and Jean-Fran{\c c}ois Raskin.
	\newblock On the Optimal Reachability Problem of Weighted Timed Automata.
	\newblock In {\em Formal Methods in System Design}, 2007.
	
	\bibitem{BouyerBrihayeMarkey-06}
	Patricia Bouyer, Thomas Brihaye, and Nicolas Markey.
	\newblock Improved Undecidability Results on Weighted Timed Automata.
	\newblock In {\em Information Processing Letters}, 2006.
	
	\bibitem{BouyerCassezFleuryLarsen-04}
	Patricia Bouyer, Franck Cassez, Emmanuel Fleury, and Kim~G. Larsen.
	\newblock Optimal strategies in priced timed game automata.
	\newblock In {\em FSTTCS 2004:
		Foundations of Software Technology and Theoretical Computer Science}, pages
	148--160, Berlin, Heidelberg, 2005. Springer Berlin Heidelberg.
	
	
	\bibitem{BouyerJaziriMarkey-15}
	Patricia Bouyer, Samy Jaziri, and Nicolas Markey.
	\newblock On the value problem in weighted timed games.
	\newblock In {\em {P}roceedings of the 26th {I}nternational {C}onference on
		{C}oncurrency {T}heory ({CONCUR}'15)}, volume~42 of {\em Leibniz
		International Proceedings in Informatics}, pages 311--324. Leibniz-Zentrum
	f{\"u}r Informatik, 2015.
	
	\bibitem{BouyerLarsenMarkeyRasmussen-06}
	Patricia Bouyer, Kim~G. Larsen, Nicolas Markey, and Jacob~Illum Rasmussen.
	\newblock Almost optimal strategies in one clock priced timed games.
	\newblock In {\em FSTTCS 2006:
		Foundations of Software Technology and Theoretical Computer Science}, pages
	345--356, Berlin, Heidelberg, 2006. Springer Berlin Heidelberg.
	
	\bibitem{BouyerMR06}
	Patricia Bouyer, Nicolas Markey, and Pierre{-}Alain Reynier.
	\newblock Robust Model-Checking of Linear-Time Properties in Timed Automata.
	\newblock In {\em {LATIN} 2006: Theoretical Informatics, 7th Latin American Symposium}, 
	Valdivia, Chile, March 20-24, volume 3887 of {\em Lecture Notes in Computer Science}, 
	pages 238--249, 2006. 
	
	\bibitem{BouyerMarkeySankur-13}
	Patricia Bouyer, Nicolas Markey, and Ocan Sankur.
	\newblock Robust Weighted Timed Automata and Games.
	\newblock In {\em Formal Modeling and Analysis of Timed Systems - 11th International Conference, {FORMATS} 2013}, 
	volume 8053 of {\em Lecture Notes in Computer Science}, pages 31--46, 2013.
	
	\bibitem{BouyerMarkeySankur-15}
	Patricia Bouyer, Nicolas Markey, and Ocan Sankur.
	\newblock Robust reachability in timed automata and games: {A} game-based approach.
	\newblock In {\em Theor. Comput. Sci.}, 2015.
	
	\bibitem{BrihayeBruyereRaskin-05}
	Thomas Brihaye, V{\'e}ronique Bruy{\`e}re, and Jean-Fran{\c c}ois Raskin.
	\newblock On optimal timed strategies.
	\newblock In {\em Formal Modeling and
		Analysis of Timed Systems}, pages 49--64, Berlin, Heidelberg, 2005. Springer
	Berlin Heidelberg.
	
	\bibitem{BrihayeGeeraertHaddadLefaucheuxMonmege-15}
	Thomas Brihaye, Gilles Geeraerts, Axel Haddad, Engel Lefaucheux, and Benjamin
	Monmege.
	\newblock Simple priced timed games are not that simple.
	\newblock In {\em Proceedings of the 35th IARCS Annual Conference on
		Foundations of Software Technology and Theoretical Computer Science
		(FSTTCS'15)}, volume~45 of {\em LIPIcs}, pages 278--292. Schloss
	Dagstuhl--Leibniz-Zentrum f{\"u}r Informatik, 2015.
	
	\bibitem{BrihayeGeeraertsNarayananKrishnaManasaMonmegeTrivedi-14}
	Thomas Brihaye, Gilles Geeraerts, Shankara {Narayanan Krishna}, Lakshmi Manasa, Benjamin Monmege, and Ashutosh Trivedi.
	\newblock Adding Negative Prices to Priced Timed Games.
	\newblock In {\em {CONCUR} 2014 - Concurrency Theory - 25th International Conference}, 
	volume 8704 of {\em Lecture Notes in Computer Science} pages 560--575, 2014.
	
	\bibitem{brihaye2021oneclock}
	Thomas Brihaye, Gilles Geeraerts, Axel Haddad, Engel Lefaucheux, and Benjamin
	Monmege.
	\newblock One-clock priced timed games with negative weights.
	\newblock Research Report 2009.03074, arXiv, 2021.
	
	\bibitem{Bus19}
	Damien Busatto{-}Gaston.
	\newblock {\em Symbolic controller synthesis for timed systems: robustness and optimality}.
	\newblock PhD thesis, Aix-Marseille Université, 2019.
		
	
	\bibitem{BusattoGastonMonmegeReynier-17}
	Damien Busatto{-}Gaston, Benjamin Monmege, and Pierre{-}Alain Reynier.
	\newblock Optimal reachability in divergent weighted timed games.
	\newblock In {\em Foundations of Software Science and Computation Structures -
		20th International Conference, {FOSSACS} 2017, Held as Part of the European
		Joint Conferences on Theory and Practice of Software, {ETAPS} 2017}, pages 162--178, 2017.
	
	\bibitem{BusattoGastonMonmegeReynier-18}
	Damien Busatto{-}Gaston, Benjamin Monmege, and Pierre{-}Alain Reynier.
	\newblock Symbolic Approximation of Weighted Timed Games.
	\newblock In {\em 38th {IARCS} Annual Conference on Foundations of Software Technology 
		and Theoretical Computer Science, {FSTTCS} 2018}, 2018.
	
	
	\bibitem{Busatto-GastonM19}
	Damien Busatto{-}Gaston, Benjamin Monmege, Pierre{-}Alain Reynier, and Ocan Sankur.
	\newblock Robust Controller Synthesis in Timed B{\"{u}}chi Automata: {A} Symbolic Approach.
	\newblock In {\em Computer Aided Verification - 31st International Conference, {CAV} 2019}, 
	volume 11561 of {\em Lecture Notes in Computer Science} pages 572--590, 2019.
	
	\bibitem{BusattoGastonMR23}
	Damien Busatto{-}Gaston, Benjamin Monmege, and Pierre{-}Alain Reynier.
	\newblock Optimal controller synthesis for timed systems.
	\newblock In {\em Log. Methods Comput. Sci.}, 2023.
	
	\bibitem{GuhaKrishnaManasaTrivedi-15}
	Shibashis Guha, Shankara Narayanan Krishna, Lakshmi Manasa, and Ashutosh Trivedi.
	\newblock Revisiting Robustness in Priced Timed Games.
	\newblock In {\em 35th {IARCS} Annual Conference on Foundation of Software Technology 
		and Theoretical Computer Science, {FSTTCS} 2015}, 2015.
	
	
	\bibitem{JurdzinskiTrivedi-07}
	Marcin Jurdzi{\'n}ski and Ashutosh Trivedi.
	\newblock Reachability-time games on timed automata.
	\newblock In {\em Proceedings of the 34th International Colloquium on Automata,
		Languages and Programming (ICALP'07)}, volume 4596 of {\em LNCS}, pages
	838--849. Springer, 2007.
	
	\bibitem{LarsenLTW14}
	Kim G. Larsen, Axel Legay, Louis{-}Marie Traonouez, and Andrzej Wasowski.
	\newblock Robust synthesis for real-time systems.
	\newblock In {\em Theor. Comput. Sci.}, 2014.
	
	
	\bibitem{MonmegeParreauxReynier-ICALP21}
	Benjamin Monmege, Julie Parreaux, and Pierre-Alain Reynier.
	\newblock {Playing Stochastically in Weighted Timed Games to Emulate Memory}.
	\newblock In {\em 48th International Colloquium on Automata, Languages, and Programming 
		(ICALP 2021)}, volume 198 of {\em Leibniz International Proceedings in Informatics
		(LIPIcs)}, pages 137:1--137:17, 2021.
	
	\bibitem{MonPar22}
	Benjamin Monmege, Julie Parreaux, and Pierre-Alain Reynier.
	\newblock {Decidability of One-Clock Weighted Timed Games with Arbitrary Weights}.
	\newblock In {\em 33rd International Conference on Concurrency Theory, {CONCUR} 2022,}, 
	volume 243 of {\em Leibniz International Proceedings in Informatics
		(LIPIcs)}, pages 15:1--15:22, 2022.
	
	
	
	
	\bibitem{OualhadjReynierSankur-14}
	Youssouf Oualhadj, Pierre{-}Alain Reynier, and Ocan Sankur.
	\newblock {Probabilistic Robust Timed Games}.
	\newblock In {\em {CONCUR} 2014 - Concurrency Theory - 25th International Conference}, 
	volume 8704 of {\em Lecture Notes in Computer Science}, pages 203--217, 2014.
	
	\bibitem{Puri98}
	Anuj Puri.
	\newblock {Dynamical Properties of Timed Automata}.
	\newblock In {\em Formal Techniques in Real-Time and Fault-Tolerant Systems, 5th International Symposium, FTRTFT'98}, 
	volume 1486 of {\em Lecture Notes in Computer Science}, pages 210--227, 1998.
	
	
	\bibitem{SankurBM-11}
	Ocan Sankur, Patricia Bouyer, and Nicolas Markey.
	\newblock {Shrinking Timed Automata}.
	\newblock In {\em IARCS Annual Conference on Foundations of Software Technology and Theoretical Computer Science (FSTTCS 2011)}, 
	volume 13 of {\em Leibniz International Proceedings in Informatics
		(LIPIcs)}, pages 90--1027, Dagstuhl, Germany, 2011.
	
	\bibitem{Sankur2013}
	Ocan Sankur, Patricia Bouyer, Nicolas Markey, and Pierre{-}Alain Reynier.
	\newblock {Robust Controller Synthesis in Timed Automata}.
	\newblock In {\em {CONCUR} 2013 - Concurrency Theory - 24th International Conference}, 2013.
	
	\bibitem{Sankur-13}
	Ocan Sankur.
	\newblock {\em Robustness in timed automata : analysis, synthesis, implementation.
		(Robustesse dans les automates temporisés : analyse, synthèse, implémentation)}.
	\newblock PhD thesis, École normale supérieure de Cachan, Paris, France, 2013.
	
	\bibitem{WulfDMR04}
	Martin De Wulf, Laurent Doyen, Nicolas Markey, and Jean{-}Fran{\c{c}}ois Raskin.
	\newblock {Robustness and Implementability of Timed Automata}.
	\newblock In {\em Formal Techniques, Modelling and Analysis of Timed and Fault-Tolerant Systems, Joint International 
		Conferences on Formal Modelling and Analysis of Timed Systems, {FORMATS} 2004 and Formal Techniques in Real-Time 
		and Fault-Tolerant Systems, {FTRTFT} 2004}, 
	volume 3253 of {\em Lecture Notes in Computer Science}, pages 118--133, 2013.
	
\end{thebibliography}
\end{document}